
\input harvmac
\input amssym.def
\input amssym
\def \bR{{\Bbb R}}
\baselineskip 15pt
\magnification\magstep1
\parskip 6pt
\newdimen\itemindent \itemindent=32pt
\def\textindent#1{\parindent=\itemindent\let\par=\resetpar%
\indent\llap{#1\enspace}\ignorespaces}

\let\oldpar=\par
\def\resetpar{\oldpar\parindent=20pt\let\par=\oldpar}

\font\ninerm=cmr9 \font\ninesy=cmsy9
\font\eightrm=cmr8 \font\sixrm=cmr6
\font\eighti=cmmi8 \font\sixi=cmmi6
\font\eightsy=cmsy8 \font\sixsy=cmsy6
\font\eightbf=cmbx8 \font\sixbf=cmbx6
\font\eightit=cmti8
\def\eightpoint{\def\rm{\fam0\eightrm}
  \textfont0=\eightrm \scriptfont0=\sixrm \scriptscriptfont0=\fiverm
  \textfont1=\eighti  \scriptfont1=\sixi  \scriptscriptfont1=\fivei
  \textfont2=\eightsy \scriptfont2=\sixsy \scriptscriptfont2=\fivesy
  \textfont3=\tenex   \scriptfont3=\tenex \scriptscriptfont3=\tenex
  \textfont\itfam=\eightit  \def\it{\fam\itfam\eightit}%
  \textfont\bffam=\eightbf  \scriptfont\bffam=\sixbf
  \scriptscriptfont\bffam=\fivebf  \def\bf{\fam\bffam\eightbf}%
  \normalbaselineskip=9pt
  \setbox\strutbox=\hbox{\vrule height7pt depth2pt width0pt}%
  \let\big=\eightbig  \normalbaselines\rm}
\catcode`@=11 %
\def\eightbig#1{{\hbox{$\textfont0=\ninerm\textfont2=\ninesy
  \left#1\vbox to6.5pt{}\right.\n@space$}}}
\def\vfootnote#1{\insert\footins\bgroup\eightpoint
  \interlinepenalty=\interfootnotelinepenalty
  \splittopskip=\ht\strutbox %
  \splitmaxdepth=\dp\strutbox %
  \leftskip=0pt \rightskip=0pt \spaceskip=0pt \xspaceskip=0pt
  \textindent{#1}\footstrut\futurelet\next\fo@t}
\catcode`@=12 %
\def \up#1{{\raise2pt\hbox{$\scriptstyle #1$}}}
\def \olr{{\raise7pt\hbox{$\leftrightarrow  \! \! \! \! \!$}}}
\def \limsub#1{\mathrel{\smash{\mathop{\longrightarrow}\limits_{#1}}}}
\def \l{\langle}
\def \r{\rangle}
\def \de{\delta}
\def \si{\sigma}

\def \bsi{\bar \sigma}

\def \pr{\partial}

\def \te{{\tilde e}}

\def \tD{{\tilde D}}
\def \teta{{\tilde \eta}}
\def \tsi{{\tilde \sigma}}
\def \tth{{\tilde \theta}}
\def \tTh{{\tilde \Theta}}
\def \bTh{{\bar \Theta}}

\def \d{{\rm d}}

\def \tr{{\rm tr \, }}
\def \bth{{\bar\theta}}
\def \bph{{\bar\phi}}
\def \bla{{\bar\lambda}}
\def \bom{{\bar{\omega}}}
\def \bkap{{\bar{\kappa}}}
\def \btau{{\bar{\tau}}}
\def \bi{{\bar \imath}}
\def \bj{{\bar \jmath}}
\def \ba{{\bar a}}
\def \bv{{\bar v}}
\def \bg{{\bar g}}
\def \bh{{\bar h}}
\def \bs{{\bar s}}
\def \bt{{\bar t}}
\def \bq{{\bar q}}
\def \1{{\bar 1}}
\def \2{{\bar 2}}
\def \3{{\bar 3}}
\def \br{{\bar r}}
\def \bs{{\bar s}}
\def \bt{{\bar t}}
\def \bep{{\bar \epsilon}}
\def \bta{{\bar \eta}}
\def \hom{{\hat{\omega}}}
\def \tom{{\tilde{\omega}}}
\def \hbom{{\hat{\bom}}}
\def \h{{\rm h}}

\def \b{{\rm b}}
\def \e{{\rm e}}
\def \x{{\rm x}}

\def \J{{\rm J}}
\def \P{{\rm P}}
\def \T{{\rm T}}
\def \X{{\rm X}}
\def \bD{{\bar D}}
\def \bJ{{\bar J}}
\def \bK{{\bar K}}
\def \bL{{\bar L}}
\def \bO{{\bar O}}
\def \bU{{\bar U}}
\def \Xb{{\bar X}} 
\def \bZ{{\bar Z}}
\def \bW{{\overline W}}
\def \bI{{\bar I}}
\def \bX{{\bar{\rm X}}}
\def \tX{{\tilde{\rm X}}}
\def \ta{{\tilde {\rm a}}}
\def \tb{{\tilde {\rm b}}}
\def \te{{\tilde {\rm e}}}
\def \th{{\tilde {\rm h}}}
\def \tx{{\tilde {\rm x}}}
\def \ty{{\tilde {\rm y}}}
\def \tb{{\tilde {\rm b}}}
\def \brh{{\bar{\rm h}}}
\def \dal{{\dot \alpha}}
\def \dbe{{\dot \beta}}
\def \dga{{\dot \gamma}}
\def \dde{{\dot \delta}}
\def \dep{{\dot \epsilon}}
\def \dta{{\dot \eta}}

\def \vep{\varepsilon}
\def \half{{\textstyle {1 \over 2}}}
\def \thir{{\textstyle {1 \over 3}}}
\def \quar{{\textstyle {1 \over 4}}}
\def \hh{{1\over 2}}
\def \ts{ \textstyle}
\def \L{{\cal L}}
\def \cJ{{\cal J}}
\def \cT{{\cal T}}

\def \D{{\cal D}}
\def \E{{\cal E}}
\def \H{{\cal H}}
\def \G{{\cal G}}

\def \S{{\cal S}}
\def \N{{\cal N}}
\def \V{{\cal V}}
\def \R{{\cal R}}
\def \bcD{{\bar{\cal D}}}
\def \bG{{\bar{\cal G}}}
\def \bH{{\bar{\cal H}}}
\def \bV{{\bar{\cal V}}}
\def \vphi{\varphi}

\def \ep{\epsilon}

\font \bigbf=cmbx10 scaled \magstep1

\lref\hughtwo{J. Erdmenger and H. Osborn, Nucl. Phys. {B483} (1997)
431; hep-th/9605009.}
\lref\hughone{H. Osborn and A. Petkou,
    Ann. Phys. {231} (1994) 311; hep-th/9307010.}
\lref\Gris{D. Anselmi, M.T. Grisaru and A.A. Johansen, Nucl. Phys. B491 (1997)
221\semi
D. Anselmi, D.Z. Freedman, M.T. Grisaru and A.A. Johansen, Phys. Lett. B394
(1997) 329\semi
D. Anselmi, D.Z. Freedman, M.T. Grisaru and A.A. Johansen, Nucl. Phys. B526
(1998) 543, hep-th/9708042\semi
D. Anselmi, J. Erlich,  D.Z. Freedman and A.A. Johansen, Phys. Rev.
D57 (1998) 7570, hep-th/9711035.}
\lref\TMP{B.L. Aneva, S.G. Mikhov and D.Ts. Stoyanov, Theoretical and
Mathematical Physics, 27 (1976) 502; 31 (1977) 394.}
\lref\TMPf{V.V. Molotkov, S.G. Petrova and D.Ts. Stoyanov, Theoretical and
Mathematical Physics, 26 (1976) 125\semi
B.L. Aneva, S.G. Mikhov and D.Ts. Stoyanov, Theoretical and Mathematical
Physics, 35 (1978) 383.}
\lref\Park{J-H. Park, Int. J. Mod. Phys., 13 (1998) 1743, hep-th/9703191.}
\lref\Wess{J. Wess and J. Bagger, {\it Supersymmetry and Supergravity}
(Princeton University Press, Princeton, 1983).}
\lref\HW{P. Howe and P. West, preprint KCL-TH-95-9, hep-th/9509140\semi
P.S. Howe and P.C. West, Phys. Lett. B389 (1996) 273, hep-th/9607060\semi
P.S. Howe and P.C. West, Nucl. Phys. B488 (1997) 425, hep-th/9607239\semi
P.S. Howe and P.C. West, Phys. Lett. B400 (1997) 305, hep-th/9611075\semi
P.S. Howe and P.C. West, preprint KCL-TH-96-18, hep-th/9611074.}
\lref\Con{B.P. Conlong, Kings College London PhD thesis, 1993\semi
B.P. Conlong and P.C. West, unpublished\semi
P.C. West, {\it Introduction to Rigid Supersymmetric Theories}, hep-th/9805055.}
\lref\Pick{A. Pickering,  Kings College London PhD thesis, 1997.}
\lref\Buch{I.L. Buchbinder and S.M. Kuzenko, {\it Ideas and Methods of 
Supersymmetry and Supergravity} (IOP Publishing Ltd., Bristol, 1995).}
\lref\Howe{P.S. Howe and G.G. Hartwell, Class. Quantum Grav. 12 (1995) 1823.}
\lref\Sei{N. Seiberg, Nucl. Phys. B435 (1995) 129.}
\lref\WCon{B.P. Conlong and P.C. West, J. Phys. A 26 (1993) 3325.}
\lref\Unit{M. Flato and C. Fronsdal, Lett. in Math. Phys. 8 (1984) 159\semi
V.K. Dobrev and V.B. Petkova, Phys. Lett. 162B (1985) 127.}
\lref\Pet{V.K. Dobrev and V.B. Petkova, Lett. in Math. Phys. 9 (1985) 287;
Fortschr. Phys. 35 (1987) 537.}
\lref\Witten{P.C. Argyres, M.R. Plesser, N. Seiberg and E. Witten, Nucl. Phys.
{B461} (1996) 71, hep-th/9511154.}
\lref\Dondi{P.H. Dondi and M. Sohnius,  Nucl. Phys. B81 (1974) 317.}
\lref\Cod{C. Codirla and H. Osborn, Ann. Phys. 260 (1997) 91.}
\lref\Og{V. Ogievetsky and E. Sokatchev, Dubna preprint E2-11528 (1978),
Yad. Phys. 28 (1978) 825.}
\lref\Scurr{S. Ferrara and B. Zumino, Nucl. Phys. B87 (1975) 207.}
\lref\Johanna{J. Erdmenger, C. Rupp and K. Sibold, {\it Conformal Transformation
Properties of the Supercurrent in Four Dimensional Supersymmetric Theories},
Nucl. Phys. B530 (1998) 501, hep-th/9804053.}
\lref\DEPP{M. D'Eramo, L. Petiti and G. Parisi, Lett. al Nuovo Cimento
2 (1971) 878.}
\lref\Symanzik{K. Symanzik, Lett. al Nuovo Cimento 3 (1972) 734.}
\lref\DZ{D.Z. Freedman, private communication.}
\lref\Bon{L. Bonora, P. Pasti and M. Tonin, Nucl. Phys. B252 (1985) 458.}
{\nopagenumbers
\rightline{DAMTP/98-101}
\rightline{hep-th/9808041}
\vskip 2truecm
\centerline {\bigbf N=1 Superconformal Symmetry in Four Dimensional} 
\vskip 5pt
\centerline {\bigbf Quantum Field Theory}
\vskip 2.0 true cm
\centerline {Hugh Osborn\footnote{}{email:
{\tt ho@damtp.cam.ac.uk}}}
\vskip 12pt
\vskip 8pt
\centerline {\ Department of Applied Mathematics and Theoretical Physics,}
\centerline {Silver Street, Cambridge, CB3 9EW, England}
\vskip 2.0 true cm

{\eightpoint
\parindent 1.5cm{

{\narrower\smallskip\parindent 0pt
The implications of $\N=1$ superconformal symmetry for four dimensional
quantum field theories are studied. Superconformal covariant expressions 
for two and three point functions of quasi-primary superfields of arbitrary
spin are found and connected with the operator product expansion. The general
formulae are specialised to cases involving a scalar superfield $L$, which
contains global symmetry currents, and the supercurrent, which contains
the energy momentum tensor, and the consequences of superconformal Ward
identities are analysed. The three point function of $L$ is shown to have unique
completely antisymmetric or symmetric forms. In the latter case the superspace
version of the axial anomaly equation is obtained.  The three point function 
for the supercurrent is shown to have two linearly independent forms. A 
linear combination of the associated coefficients for the general expression is
shown to be related to the scale of the supercurrent two point function through
Ward identities. The coefficients are given for the two free field
superconformal theories and are also connected with the parameters present in
the supercurrent anomaly for supergravity backgrounds. Superconformal
invariants, which are possible even in three point functions, are discussed.

\narrower}}

\vfill
\eject}}
\pageno=1

\newsec{Introduction}

The very first paper \ref\one{J. Wess and B. Zumino, Nucl. Phys. B70 (1974) 39.}
in the western literature on supersymmetry in four space-time dimensions
in fact introduced the $\N=1$ superconformal group. Nevertheless, given the
inevitable breakdown of conformal invariance in perturbative treatments of
non trivial quantum field theories, most subsequent discussions of supersymmetric
theories were concerned with theories invariant under just the restricted 
supersymmetry group which is the minimal extension of the Poincar\'e group,
whose elements are standard Lorentz and translation transformations,
and for which there are no perturbative quantum anomalies.
However in the last few years the work of Seiberg \Sei\ and 
others have shown that there should exist a host of non trivial
superconformal field theories which in many cases can be identified with
renormalisation group fixed points where the $\beta$-function vanishes. Just
as two-dimensional conformal field theories have a very rich mathematical
structure, with applications in string theories and statistical physics,
it is now possible to hope for similar elegant exact results in the as yet
relatively unexplored case of four dimension field theories. Virtually all the
new results depend essentially on constraints imposed by supersymmetry and
in consequence superconformal field theories are the most promising
immediate candidates for potential extension of some of the two-dimensional 
results for conformal field theories to higher dimensions.

Although $\N=2$ and $4$ superconformal theories in four space-time dimensions
have considerable interest, and also remarkably there may also be possible
superconformal field theories in five and six dimensions, we here consider
just the $\N=1$ case in four dimensions using standard superspace formalism. 
Some relevant results were obtained long ago \refs{\TMP,\TMPf}
and recently there has been extensive work by Howe and West \HW, on which we
attempt to build (although much of their discussion was concerned with $\N>1$).
Furthermore Anselmi and co-workers \Gris\ have undertaken specific
calculations in $\N=1$
supersymmetric Yang Mills theory, exploring analogues in four-dimensional
superconformal theories of the two-dimensional Virasoro
central charge $c$. Here we extend previous results \refs{\hughone,\hughtwo}
which give explicit forms for two and three point functions in conformal
field theories for operators of arbitrary spin to the $\N=1$ superconformal
case. The resulting expressions determine the forms of the operator
product expansions for quasi-primary operators and although essentially
kinematic are, in our view, a necessary precursor to dynamical investigations.
We also analyse the supercurrent \Scurr, which contains the energy momentum 
tensor amongst its component fields, and its Ward identities which reflect
superconformal invariance. One of the main concerns of this paper is to analyse
in detail the two and three point functions of the supercurrent and their
relation through Ward identities and also their connection to anomalies
present on curved supergravity backgrounds. We also discuss possible
superconformal invariants which can appear in the general expression
for four-point functions. A similar analysis was described in \Park\ but
the present discussion is perhaps more complete and differs in some details.

In the next section we establish notation and
review $\N=1$ superconformal transformations on
superspace in four space-time dimensions. We discuss primarily infinitesimal
superconformal transformations which are super-diffeomorphisms restricted
by a natural condition playing a similar role to the conformal Killing
equation for ordinary conformal transformations. We identify the associated
Lie algebra with that for the supergroup $Sl(4|1)$ with suitable reality
conditions. From the discussion of the action
of superconformal transformations on superspace we construct variables which
transform homogeneously and may be used in a simple construction of two and
three point functions. Unlike the non supersymmetric case there is even
a superconformal invariant for three points.
In section 3 we describe  how
quasi-primary superfields may be defined in general by a simple
transformation rule, depending on the scale dimension and $U(1)$ $R$-symmetry
charge, as well as its particular spin representation. We further show how 
derivatives  of quasi-primary superfields in particular special cases
are also quasi-primary. Such results demonstrate the consistency of conservation
conditions on the supercurrent with superconformal invariance and are
important in the subsequent analysis. A corollary of these results is that
the Bianchi identity for a $\N=1$ supersymmetric gauge theory is consistent
with superconformal invariance only if the scale dimension and $R$-charge
are those of the free abelian theory. We also describe, in terms of the results
of section 2, general superconformal covariant constructions for two and 
three point functions of quasi-primary superfields. The result for the three
point function depends on a homogeneous function on superspace coordinates
which can be directly related to the leading coefficient of the term in 
the operator product expansion associated with the operators appearing in the
three point function. In section 4 we introduce the supercurrent by a variant 
of Noether's construction which is used to find the corresponding Ward 
identities. In section 5 we consider first the application of the general 
formalism to the simple cases involving chiral scalar superfields. 
After showing how conditions flowing from the
conservation equations may be imposed on the general form for the superfield
three point function in section 6 we consider the three point function
for superfields containing an internal symmetry current. We discuss the
associated Ward identities and also the anomalies which are supersymmetric
extension of the usual axial current anomaly. 
In section 7 we consider the supercurrent and obtain results for the three point
functions involving two supercurrents and a scalar superfield and also 
three supercurrents. For the latter case we show that there
are two possible linearly independent forms although one linear combination
is shown to be related to the coefficient of the two point function through
a Ward Identity. In section 8 general results are then restricted to the
case of free fields which give two different trivial superconformal theories
in four dimensions. In section 9 we show how superconformal invariance
allows for the evaluation of integrals, generalising old results in the
non supersymmetric case, and in section 10 we discuss possible superconformal
invariants that may be present in higher point correlation functions which
include generalisations of the usual invariant cross ratios as well as
the Grassmann valued invariants present for just three points. Finally
in a conclusion we relate the coefficients which are present in the general
supercurrent three point function to the coefficients $c,a$ appearing
in the supergravity extension of the energy momentum tensor trace for
a curved space background. Both  $c,a$ are possible generalisations of the
Virasoro central charge to four dimensional superconformal theories. An
appendix contains some details concerning the transformation properties of
the supercurrent in free theories.

\newsec{Superconformal Transformations}

The conformal group is defined by the intersection of the group of
diffeomorphisms with local rescalings of the metric. The superconformal
group can be obtained in a variety of equivalent ways but here we consider
it as a reduction of those super-diffeomorphisms which leave the chiral
subspaces of superspace invariant. With standard superspace coordinates
$z^A = (x^a,\theta^\alpha, {\tilde\bth}_\dal) \in \bR^{4|4}$
the chiral restrictions are given by 
\eqn\ione{
z_+^{\, A_+} = (x_+^a , \theta^\alpha) \, , \qquad
z_-^{\, A_-} = (x_-^a , {\tilde\bth}_\dal) \, , \qquad 
x_\pm^a = x^a \pm i \theta \si^a \bth \, ,
}
so that $\bD_\dal z_+ = 0 , \ D_\alpha z_- =0$.\foot{We use the notation
of Wess and Bagger \Wess, with minor emendations, thus $ \theta^\alpha, \,
\bth^\dal$ are regarded as row, column vectors and we let 
$\tth_\alpha = \ep_{\alpha\beta}\theta^\beta, \, {\tilde \bth}{}_\dal
= \ep_{\smash {\dal \dbe}} \bth{}^\dbe$ form associated column, row vectors,
$\theta^2 = \theta \tth, \, \bth^2 = {\tilde \bth}\bth$, as usual $4$-vectors
are identified with $2\times 2$-matrices using the hermitian $\si$-matrices
$\si_a, \, \tsi_a, \, \si_{(a} \tsi_{b)} = - \eta_{ab}1$, 
$x^a \to \x_{\alpha\dal} =
x^a (\si_a)_{\alpha\dal}, \ \tx^{\dal\alpha} = x^a (\tsi_a)^{\dal\alpha}
= \ep^{\alpha\beta}\ep^{\smash {\dal \dbe}} \x_{\smash {\beta \dbe}}$,
with inverse $x^a = - {1\over 2}{\rm tr}(\si^a \tx)$. 
Hence \ione\ gives $\tx_\pm = \tx \pm 2i \bth \theta, \, \x_\pm = \x \mp
2i \tth {\tilde \bth}$. The associated spinor derivatives satisfy
$\{D_\alpha, \bD_\dal\} = - 2i(\si^a \pr_a)_{\alpha\dal}$.} 
For a general diffeomorphism preserving the chiral decomposition of superspace
we may write
\eqn\ch{
\de x_+^a = v^a(z_+) \, , \quad \de \theta^\alpha = \lambda^\alpha
(z_+) \, , \qquad
\de x_-^a = \bv^a(z_-) \, , \quad \de \bth^\dal = \bla^\dal (z_-) \, .
}
The corresponding differential generators acting in chiral superfields are
\eqn\chL{
\L_+ = h^a \pr_a + \lambda^\alpha D_\alpha \, , \qquad
\L_- = \bh^a \pr_a + {\tilde\bla}_\dal {\tilde\bD}{}^\dal \, , 
}
where
\eqn\defh{
h^a(z) = v^a(z_+)- 2i \lambda(z_+) \si^a \bth \, , \quad
{\bh}{}^a (z) = \bv^a(z_-) + 2i \theta  \si^a \bla (z_-) \, .
}
Alternatively we may define $\L_\pm$ by the requirements $[\bD_\dal , \L_+]
= 0 , \ [D_\alpha , \L_- ] =0$ which, assuming just the form \chL, leads to
\eqn\Dh{
\bD_\dal h^b = - 2i \,  (\lambda \si^b)_\dal \, , \quad
\bD_\dal \lambda^\beta = 0 \, , \qquad
D_\alpha \bh^b = 2i \, (\si^b \bla)_\alpha \, ,
\quad D_\alpha \bla{}^\dbe = 0 \, ,
}
which are easily seen to be solved by \defh. $\lambda, \, \bla$ are thus
determined in terms of $h, \, \bh$ with the remaining constraints
on $h, \, \bh$ which follow from \Dh\ may be written as
\eqn\Dhh{
{\tilde\bD}{}^{(\dal} {\tilde\h}{}^{\dbe)\beta} = 0 \, , \qquad \quad
D_{(\alpha} {\bar \h}{}_{\smash{\beta)\dbe}}^{\vphantom g} = 0 \, .
}

We here define infinitesimal superconformal transformations as those
diffeomorphisms of the form \ch\ where with \defh\  we have\foot
{Alternatively \Dhh\ with $h=\bh$ can be regarded as the superconformal Killing
equations, they  were obtained previously by Conlong and West \Con.}
\eqn\hh{
h^a = \bh^a \, ,
}
With this restriction it is easy to see from \Dh\ that
\eqn\Kill{
\pr_a h_b + \pr_b h_a = 2\rho \, \eta_{ab} \, , \qquad
\rho = \half ( D_\alpha \lambda^\alpha + {\tilde\bD}{}^\dal 
{\tilde\bla}_\dal ) \, ,
}
which is just the standard conformal Killing equation. The solution of
\Kill\ is well known and for dimensions $d>2$ gives the standard result
for an infinitesimal conformal transformation in terms of a translation
$a^a$, rotation $\omega^{ab}= - \omega^{ba}$, special conformal transformation
$b^a$ and scale parameter $\lambda$. In the present case the solution 
becomes \Buch, with $\lambda = \kappa+\bkap$,
\eqn\sol{ \!\!\!\!\!\!\eqalign{ 
v^a(y,\theta) ={}& a^a + \omega^a{}_b y^{\, b} + (\kappa+\bkap) y^a
+ b^a  y^{2} - 2 y^{a} y{\cdot b} + 2i \, \theta \si^a \bep - 2\, \theta \si^a
\ty \eta \, , \cr
\lambda^\alpha (y,\theta) ={}& \ep^\alpha - \theta^\beta \omega_\beta{}^\alpha
+ \kappa \, \theta^\alpha + (\theta \b \ty)^\alpha -i ( \bta \,\ty )^\alpha
+ 2\teta^\alpha \theta^2 \, , \quad \omega_\beta{}^\alpha = -\quar \omega^{ab}
(\si_a \tsi_b)_\beta{}^\alpha \, , \cr
\bv^a(y,\bth) ={}& a^a + \omega^a{}_b y^{\, b} + (\kappa+\bkap) y^a + 
b^a  y^{2} - 2 y^{a} y{\cdot b} -2i \, \bep \si^a \bth 
- 2\, \bta \,\ty \si^a \bth \, , \cr
\bla^\dal (y,\bth) ={}& \bep^\dal + \bom{}^\dal{}_{\smash\dbe} \bth^\dbe
+ \bkap \, \bth^\dal + (\ty \b \bth)^\dal +i ( \ty \eta )^\dal
+ 2 {\tilde \bta}{}^\dal \bth^2 \, , \quad 
\bom{}^\dal{}_{\smash\dbe} = 
-\quar \omega^{ab} (\tsi_a \si_b)^\dal{}_{\smash\dbe}\, , \cr}
}
Superconformal transformations are parameterised additionally by the
supertranslations $\ep^\alpha, \, \bep^\dal$ and an extra Grassmann spinor
$\eta_\alpha, \, \bta_\dal$ (in our notational conventions
$\teta^\alpha = \ep^{\alpha\beta}\eta_\beta, \, {\tilde \bta}{}^\dal
= \ep^{\dal\dbe} \bta_{\smash\dbe})$, as well as, if 
$\kappa=\half(\lambda+i\alpha), \,
\bkap =\half(\lambda-i\alpha)$ the  angle $\alpha$ which  corresponds to the
$U(1)$ $R$-symmetry acting on $\theta, \, \bth$.

The action of infinitesimal  superconformal transformations on fields
defined on superspace is then generated by
\eqn\Ls{
\L = h^a \pr_a + \lambda^\alpha D_\alpha  + 
{\tilde \bla}_\dal {\tilde\bD}{}^\dal  \, . 
}
{}From \Dh\ it is easy to see that
\eqn\LD{
[D_\alpha , \L ] = (D_\alpha \lambda^\beta) D_\beta \, , \qquad
[{\tilde \bD}{}^\dal , \L ] = ( {\tilde \bD}{}^\dal {\tilde \bla}_{\smash\dbe})
{\tilde \bD}{}^\dbe \, .
}
Writing the superspace exterior derivative $\d = \d z^A \pr_A = e^A D_A$,
which requires
\eqn\defe{
e^a = \d x^a + i \theta \si^a \d \bth - i \d \theta \si^a \bth \, ,
}
then
if $[\L, D_A] = - R_A{}^{\! B} D_B$ the associated variation of one-forms
is given by $\delta e^A = e^B R_B{}^{\! A}$. The
results in \LD\ imply that for superconformal transformations
$R_a{}^{\! \beta} = R_a{}^{\! \dbe} = 0$ and that in this case the variation of $e^a$
is homogeneous 
\eqn\vare{
\de e^a = e^b \pr_b h^a \, ,
}
since $R_a{}^{\! b} = \pr_a h^b$. Using \Kill\ for $e^2=e^a e_a$ therefore,
\eqn\varesq{
\de e^2 = 2\rho \, e^2 \, .
}
The invariance of the square of the superspace interval, $e^2$ with
$e^a$ given by \defe, up to a
local rescaling can be regarded as an  alternative basic characterisation
of superconformal transformations.

Besides transformations connected to the identity it is natural to
extend the superconformal group by inversions $z\longrightarrow z'$ where
\Buch,
\eqn\invert{
\tx'{}_{\!\! -} = - \x_+{}^{\!\! -1} \, , \quad \bth{}' = -i \x_+{}^{\!\! -1}
\tth \, , \quad  \theta' = i {\tilde \bth}\,  \x_-{}^{\!\!\! -1} \, , 
\quad \Rightarrow \quad \tx'{}_{\!\! +} = - \x_-{}^{\!\!\! -1} \, ,
}
with $\x_\pm{}^{\!\! -1} = - \tx_\pm/ x_\pm{}^{\! 2}$. From \defe\ it is
straightforward to find, as in \varesq, that $e^2$ is invariant up to a
rescaling,
\eqn\inverte{
\te' = \x_+{}^{\!\! -1} \e \, \x_-{}^{\!\!\! -1} \, , \qquad e'{}^2 = 
{e^2 \over x_+{}^{\! 2}x_-{}^{\! 2}} \, .
}
It is easy to verify from \invert\ that inversions are idempotent, $(z')'=z$.

The Lie algebra of the differential generators $\L$, given by \Ls, is easily
calculated
\eqn\alg{
\L' = [ \L_2 , \L_1 ] \, ,
}
so that, for instance,
\eqn\hprime{
h^{\prime a} = [h_2, h_1]^a + 2i ( \lambda_2 \si^a \bla_1 - 
\lambda_1 \si^a \bla_2 ) \, .
}
It is straightforward to check that $h'=\bh{}', \lambda', \bla{}'$ satisfy \Dh.
The superconformal algebra may be identified with that of supermatrices, in
terms of the parameters in \sol,
\eqn\MM{
M = \pmatrix{\omega - {\ts{1\over3}}(\kappa + 2\bkap)1 &
- i \b & 2\eta \cr -i \ta & \bom + {\ts{1\over3}}(2\kappa + \bkap ) 1& 2\bep \cr
 2 \ep & 2 \bta & {\ts{2\over3}} ( \kappa - \bkap ) } \, ,
}
since \alg\ corresponds exactly to
\eqn\algm{
M' = [M_1,M_2] \, .
}
It is easy to see that ${\rm str} M =0$ so that $M$ belongs to Lie algebra
$sl(4|1)$ which is restricted to $su(2,2|1)$ by the reality condition,
\eqn\real{
M = - B M^\dagger B \, , \qquad B = \pmatrix{0&1&0\cr 1&0&0\cr 0&0& -1} \, .
}

The non-linear realisation of the superconformal group on superspace may
be recovered by regarding $z=(x,\theta,{\tilde\bth})$ as coordinates on the
coset $SU(2,2|1)/G_0$, where $G_0\in SU(2,2|1)$ is the stability group of the 
point
$z=0$ under superconformal transformations, generated by matrices $M_0$ of the
same form as in \MM\ with $a=\ep=\bep=0$.\foot{For a general review of such 
constructions see \Howe.} To describe the coset explicitly it is 
convenient to define
\eqn\VV{
\V(z) = \pmatrix{1&0\cr -i\tx_+ & 2 \bth\cr 2\theta & 1} \, ,\qquad
\bV(z) = \V(z)^\dagger B = \pmatrix { i \tx_- & 1 & - 2\bth\cr
2 \theta& 0 & -1} \, ,
}
which are constrained by $\bV(z)\V(z)=\pmatrix{0&0\cr0&-1}$.
For $M$ given by \MM\ the associated $\de x_\pm, \de \theta, \de \bth$ 
as given by \ch\ are then obtained by
\eqn\inf{\eqalign{
M \V(z) = {}&\V(\de z) + \V(z) \H(z) \, , \qquad  \V(\de z) = \L \V(z) \, ,\cr
- \bV(z) M = {}& \bV(\de z) + \bH(z) \bV(z) \, , \qquad  
\bV(\de z) = \L \bV(z) \, , \cr}
}
if $\H, \, \bH$ are matrices of the form
\eqn\HH{
\H = \pmatrix{\hom - \si \, 1 & \tau\cr 0 & 2(\bsi - \si)} \, , \qquad
\bH = \pmatrix{- \hbom - \bsi \, 1 & 0 \cr \btau & - 2(\bsi - \si)} \, ,
}
which are determined by requiring the structure of $\V, \, \bV$ in \VV\ to be
preserved. The elements
of $\H, \, \bH$ are given by
\eqnn\el
$$\eqalignno{ 
D_\alpha \lambda^\beta = {}& - \hom_\alpha{}^\beta + \delta_\alpha{}^{\!\beta}
(2\bsi - \si) \, , \quad
{\tilde\bD}{}^\dal {\tilde\bla}_{\smash\dbe} = - \hbom{}^\dal{}_{\smash\dbe} + 
\de^\dal{}_{\!\smash\dbe}(2\si - \bsi)\, , 
\quad  \hom_\alpha{}^\alpha = \hbom{}^\dal{}_\dal = 0 \, , \cr
& D_\alpha \si = - \thir D_\beta \hom_\alpha{}^\beta = \tau_\alpha \, , \qquad
\bD_\dal \bsi = \thir \bD_{\smash\dbe} \hbom{}^\dbe{}_\dal = \btau_\dal \, ,
& \el \cr}
$$
or explicitly, independent of $a,\ep,\bep$,
\eqn\form{\eqalign{
\hom_\alpha{}^\beta(z_+) = {}& \omega_\alpha{}^\beta + \half 
(\x_+ \tb - \b \, \tx_+ )_\alpha{}^\beta + 4\, \eta_\alpha \theta^\beta + 2
\delta_\alpha{}^{\!\beta} \, \theta \eta \, , \cr
\si(z_+) = {}&
{\ts{1\over 3}}(\kappa+2\bkap) + 2\, \theta \eta - b{\cdot x_+} \, , \qquad
\btau_\dal(z_+) = 2(\bta+ i \theta \b)_\dal \, , \cr
\hbom{}^\dal{}_{\smash\dbe}(z_-) = {}& \bom{}^\dal{}_{\smash\dbe} + \half
( \tx_- \b - \tb \, \x_-)
{}^\dal{}_{\smash\dbe} - 4 \, \bth{}^\dal \bta_{\smash\dbe} - 
2 \de^\dal{}_{\!\smash\dbe} \,
\bta \bth \, , \cr
\bsi(z_-) = {}&{\ts{1\over 3}}(2\kappa + \bkap) + 2\, \bta \bth - b
{\cdot x_-} \, , \qquad \tau_\alpha(z_-) = 2 ( \eta - i \b \bth)_\alpha \, .
\cr}
}
It is easy to check that $M_0 \V(0) = \V(0) \H(0), \, \bH(0)\bV(0) = - \bV(0)
M_0$. For general $z$
the important result that $\hom, \si$ depend only on $z_+$ follows directly
from \Dh\ by using $\bD_\dal D_\alpha \lambda^\beta = \de_\alpha{}^{\!\beta}
\bD_\dal \bD_{\smash\dbe} \bla^\dbe$ which from \el\ gives 
$\bD_\dal \hom_\alpha{}^\beta = \bD_\dal \si =0$, and similarly for 
the dependence of $\hbom,\bsi$ on $z_-$. We may also note that
${\rm str}\, \H(z) = - 2\bsi(z_-), \, {\rm str}\, \bH(z) = - 2\si(z_+)$.
{}From \Kill\ we find
\eqn\rh{
\rho = \si + \bsi \, ,
}
and using \Dh\
\eqn\rot{
\hom = \quar \pr^{[a} h^{b]} \si_a \tsi_b \, , \qquad 
\hbom = \quar \pr^{[a} h^{b]} \tsi_a \si_b \, .
}
By using  \alg\ and \algm\  in \inf\ we may find
\eqn\algh{\eqalign{
\L_2 \H_1 - \L_1 \H_2 + [\H_1 , \, \H_2 ] = {}& \H'\, ,\cr
\L_2 \bH_1 - \L_1 \bH_2 - [\bH_1 , \, \bH_2 ] = {}& \bH'\ . \cr}
}

It is straightforward to rewrite \inf\ for the non infinitesimal case.
The element of the superconformal group formed by
exponentiating $M$ corresponds to a finite superconformal transformation
$z \longrightarrow z'$ given by
\eqn\fin{
e^M \V(z) = \V(z')\G(z) \, \qquad \bV(z)e^{-M} = \bG(z) \bV(z') \, ,
}
where $\G(z), \, \bG(z)$ are matrices of the same form as $\H, \, \bH$ in \HH,
with the group property that if $z\ {\buildrel e^{M_1} \over \longrightarrow}\
z'\ {\buildrel e^{M_2} \over \longrightarrow} \ z''$ and $e^{M_2} e^{M_1} =
e^M$, so that $z\ {\buildrel e^{M} \over \longrightarrow}\ z''$,
then $\G_2(z')\G_1(z) = \G(z)$ while $\bG_1(z)\bG_2(z') = \bG(z)$.
The action of finite conformal transformations on coordinates $x^a \in \bR^4$
is globally well defined 
on a compactification of Minkowski space, $\bR^4 \to S^3\times S^1$, or some
multiple covering\foot{For a discussion of some global issues see \Cod.}
and similar considerations apply in the superconformal case
when the transformations act on $SU(2,2|1)/G_0$.
However such issues are not relevant for the considerations of this paper. 

The usefulness of the coset construction becomes manifest if we consider
\eqn\Vtwo{
\bV(z_1) \V(z_2) = \pmatrix{ i\, \tx_{\1 2} & - 2\, \bth_{12}\cr 
2\, \theta_{12}& -1 } \, ,
}
where
\eqn\defvar{
\tx_{\1 2} = \tx_{1 -} - \tx_{2 +} + 4i \, \bth_1 \theta_2 \, , \qquad
\theta_{12} = \theta_1 - \theta_2 \, , \quad
\bth_{12} = \bth_1 - \bth_2 \, .
}
The expression \Vtwo\ is a function of the supertranslation invariant
interval given by
\eqn\yint{
z_{12}^{\, A} = (x_{12}^{\, a},\theta_{12}^{\, \alpha}, 
{\tilde\bth}^{\vphantom g}_{12\dal}) \, , \qquad 
y_{12} = x_1 - x_2 -i \theta_1 \si \bth_2 + i \theta_2 \si \bth_1 =-y_{21}\, ,
}
since $x_{\1 2} = y_{12} - i \theta_{12} \si \bth_{12}$.  From \inf\ \Vtwo\
transforms according to $\de \big ( \bV(z_1) \V(z_2)\big ) = - \bH(z_1)
\bV(z_1) \V(z_2) - \bV(z_1) \V(z_2) \H(z_2)$ and using the form
for $\H, \, \bH$ given in \HH\ this gives
\eqn\varx{
\de \tx_{\1 2} = \big ( \hbom(z_{1-}) + \bsi(z_{1-}) 1 \big ) \tx_{\1 2}
+ \tx_{\1 2}  \big ( - \hom(z_{2+}) + \si(z_{2+})1 \big ) \, ,
}
as well as
\eqn\varth{\eqalign{
\de \theta_{12} = {}&\theta_{12} \big ( - \hom(z_{2+}) + \si(z_{2+}) 1 \big ) 
+ 2 \big ( \bsi(z_{1-}) - \si(z_{1+}) \big ) \theta_{12} - \half i \, \btau
(z_{1+}) \tx_{\1 2} \, , \cr
\de \bth_{12} = {}& \big ( \hbom(z_{1-}) + \bsi(z_{1-}) 1 \big ) \bth_{12}
- 2  \big ( \bsi(z_{2-}) - \si(z_{2+}) \big ) \bth_{12} + \half i \,
\tx_{\1 2} \tau(z_{2-}) \, . \cr}
}
Combining this result for $\de \theta_{12}$ with \varx\ gives a simpler
expression for the variation in the form
\eqn\varthx{
\de \big ( \theta_{12} \tx_{\1 2}{}^{\! -1} \big )  = 
\big ( \theta_{12} \tx_{\1 2}{}^{\! -1} \big ) \big ( - \hbom(z_{1-}) 
+ \si(z_{1+}) 1 - 2 \bsi(z_{1-})1 \big )  - \half i \, \btau (z_{1+}) \, ,
}
and similarly from the result \varth\ for $\de \bth_{12}$ with 
$1\leftrightarrow 2$ 
\eqn\varbth{
\de \big (  \tx_{\2 1}{}^{\! -1} \bth_{12}\big ) =  
\big ( - \hom(z_{1+}) + \si(z_{1+}) 1 - 2 \bsi(z_{1-})1 \big ) 
\big ( \tx_{\2 1}{}^{\! -1} \bth_{12}\big ) - \half i \,\tau(z_{1-})  \, .
}
If we let
\eqn\defxx{
(\x_{2\1})_{\alpha\dal} = - \ep_{\alpha\beta}\ep_{\smash{\dal\dbe}}
(\tx_{\1 2})^{\dbe\beta}\, , \quad \x_{2\1} = \x_{2+} - \x_{1-} + 4i\,
\tth_2 {\tilde \bth}_1 \, ,
}
then the inverse of $\tx_{\1 2}$ is given explicitly by
\eqn\inv{
\tx_{\1 2}{}^{\! -1} = {1\over x_{\1 2}{}^{\! 2}} \, \x_{2\1} \, ,
} 
since $\tx_{\1 2}  \x_{2\1} =  x_{\1 2}{}^{\! 2}1$. 
It is also useful to note that from \varx
\eqn\varxx{
\de \x_{2\1} = \big ( \hom(z_{2+}) + \si(z_{2+})1 \big )  \x_{2\1} 
+ \x_{2\1} \big ( - \hbom(z_{1-}) + \bsi(z_{1-}) 1 \big ) \, .
}
Under inversions as in \invert\ then with \defvar\ and \defxx\
\eqn\invxx{
\tx_{\12} \longrightarrow \x_{1+}{}^{\!\!\! -1} \x_{1\2}\,
\x_{2-}{}^{\!\!\! -1} \, .
}

For two points in superspace $z_1,z_2$ then
$\tx_{\12}= x_{\1 2}{}^{\!\! a}\tsi_a$, 
defined in \defvar\ such that it depends just on $z_{1-}$ and $z_{2+}$ and
satisfies  $\tx_{\12}{}^\dagger = - \tx_{\21}$,
or equivalently $\x_{2\1}$, play a crucial role in the subsequent
construction of superconformally covariant two and higher point amplitudes
as a consequence of their homogeneous transformation properties in
\varx\ or \varxx. Given the result \Vtwo\ we may also define a scalar by
\eqn\sdet{
{\rm sdet} \bV(z_1) \V(z_2) = - 
\big (x_{\1 2} + 2i\, \theta_{12}\si \bth_{12} \big )^2 
= - x_{\2 1}{}^{\! 2}\, .
}
Under a finite transformation $z \longrightarrow z'$ as in \fin\ we
may find directly
\eqn\vars{
x'{}_{\!\2 1}{}^{\! 2} = {x_{\2 1}{}^{\! 2}\over {\bar \Omega}(z_{2-})
\Omega(z_{1+})} \, , \qquad {\bar \Omega}(z_- ) = {\rm sdet}\, \G(z), \,
\quad \Omega(z_+ ) = {\rm sdet}\, \bG(z)\, .
}
In constructing conformally covariant two point functions both $x_{\1 2}$
and $x_{\2 1}$, which are related by $z_1\leftrightarrow z_2$,
are necessary.  A symmetric scalar is given by, with $y_{12}$ defined in \yint,
\eqn\twop{
x_{\1 2}{}^{\! 2} x_{\2 1}{}^{\! 2} = \big ( y_{12} {}^{\! 2} + 
\theta_{12}{}^{\! 2}\bth_{12}{}^{\! 2} \big )^2 \, .
}

For three points $z_1,z_2,z_3$ we may define 4-vectors $X_1{}^{\! a},
{\bar X}_1{}^{\! a}$ by
\eqn\defX{
\X_1 = {\x_{1\2}\, \tx_{\2 3}\, \x_{3\1}\over x_{\2 1}{}^{\! 2}
\, x_{\1 3}{}^{\! 2}} \, , \qquad
\bX_1 = - {\x_{1\3}\, \tx_{\3 2}\, \x_{2\1}\over x_{\3 1}{}^{\! 2}
\, x_{\1 2}{}^{\! 2}} = \X_1{}^{\!\dagger} \, ,
}
so that $\X_1$ transforms homogeneously at $z_1$  according to
\eqn\varX{
\de \X_1 = \big ( \hom(z_{1+}) - \si(z_{1+})1 \big ) \X_1 +
\X_1 \big ( - \hbom(z_{1-}) - \bsi(z_{1-})1 \big )  \, ,
}
and similarly for $\bX_1$. To achieve analogous results for the Grassmann 
variables we consider
\eqn\defTh{
\tTh_1 = i\bigg ( {1\over x_{\2 1}{}^{\! 2}}\, \x_{1\2} \bth_{12} -
{1\over x_{\3 1}{}^{\! 2}}\, \x_{1\3} \bth_{13}\bigg ) \, , \quad
{\tilde \bTh}_1 = i\bigg ( {1\over x_{\1 2}{}^{\! 2}}\, \theta_{12} \x_{2\1} -
{1\over x_{\1 3}{}^{\! 2}}\, \theta_{13} \x_{3\1} \bigg )\, ,
}
since then the inhomogeneous terms involving 
$\btau, \, \tau$ in \varthx\ and \varbth\ cancel and
$\Theta_1{}^{\!\alpha} , \, \bTh_1{}^{\!\dal}$ transform 
homogeneously as chiral, anti-chiral spinors at $z_1$,
\eqn\varTh{\eqalign{
\de \Theta_1 = {}&\Theta_1 \big ( - \hom(z_{1+}) + \si(z_{1+})1 - 2\bsi
(z_{1-})1 \big ) \, , \cr
\de \bTh_1 = {}& \big ( \hbom(z_{1-}) + \bsi(z_{1-}) 1 - 2 \si(z_{1+}) 1 \big )
\bTh_1 \, . \cr}
}
Using formulae such as
\eqn\xxx{
\tx_{\2 3} = \tx_{\2 1} + \tx_{\1 3} + 4i\, \bth_{12} \theta_{13} \, ,
}
we may find the relation
\eqn\XTh{
\X_1 - \bX_1 = 4i\, \tTh_1 {\tilde \bTh}_1 \, ,
}
and for future reference we may note
\eqn\Xsq{
X_1{}^{\! 2} = {x_{\2 3}{}^{\! 2} \over x_{\2 1}{}^{\! 2}
x_{\1 3}{}^{\! 2}} \, , \qquad
{\bar X}_1{}^{\! 2} = {x_{\3 2}{}^{\! 2} \over x_{\3 1}{}^{\! 2}
x_{\1 2}{}^{\! 2}} \, .
}

With obvious cyclic permutation of indices in \defX\ and \defTh\ we may
similarly define $X_2 , \bar X_2 , \Theta_2 , \bTh_2$ and 
$X_3 , \bar X_3 , \Theta_3 , \bTh_3$
which transform homogeneously at $z_2$ and $z_3$ respectively. It is easy
to see that
\eqn\Xtran{
\tx_{\2 1} \X_1 \tx_{\1 2} = {1\over \bar X_2{}^{\up{\! 2}}}\, 
{\tilde \bX}_2 \, ,
\qquad
\tx_{\2 1} \bX_1 \tx_{\1 2} = {1\over X_2{}^{\up{\! 2}}}\, \tX_2 \, ,
}
and
\eqn\Thtran{
{x_{\1 2}{}^{\! 2} \over x_{\2 1}{}^{\! 2}} \, \tx_{\2 1} \tTh_1 =
{1\over X_2{}^{\up{\! 2}}}\, \tX_2 \tTh_2 \, , \qquad
{x_{\2 1}{}^{\! 2} \over x_{\1 2}{}^{\! 2}} \, {\tilde \bTh}_1 \tx_{\1 2}
= - {1\over \bar X_2{}^{\up{\! 2}}}\, {\tilde \bTh}_2 {\tilde \bX}_2 \, ,
}
with similar results giving $X_3 , \bar X_3 , \Theta_3 , \bTh_3$. From
\Thtran\ we have
\eqn\Thsqtran{
\Theta_1{}^{\!2}= \bigg({x_{\2 1}{}^{\! 2} \over x_{\1 2}{}^{\! 2}}\bigg )^{\!2}
{x_{\3 2}{}^{\! 2} \over x_{\3 1}{}^{\! 2}} \, \Theta_2{}^{\!2} \, , \qquad
\bTh_1{}^{\! 2} = \bigg({x_{\1 2}{}^{\! 2} \over x_{\2 1}{}^{\! 2}}\bigg )^{\!2}
{x_{\2 3}{}^{\! 2} \over x_{\1 3}{}^{\! 2}} \, \bTh_2{}^{\! 2} \, .
}
A straightforward check on \Xtran\ and \Thtran\ are that the
conformal transformation properties of both sides are consistent. From
\Xtran\
\eqn\XX{
{X_1{}^{\! 2} \over \bar X_1{}^{\up{\! 2}}} = 
{X_2{}^{\! 2} \over \bar X_2{}^{\up{\! 2}}} =
{X_3{}^{\! 2} \over \bar X_3{}^{\up{\! 2}}} \, ,
}
and from \varX\ this is a superconformal invariant. Under 
$z_2 \leftrightarrow z_3$, $X_1 \leftrightarrow -{\bar X}_1$, while
$X_2 \leftrightarrow - {\bar X}_3$, so that
\eqn\inv{
I = \half \bigg ( {X_1{}^{\! 2} \over \bar X_1{}^{\up{\! 2}}} +
{\bar X_1{}^{\! 2} \over X_1{}^{\up{\! 2}}} \bigg )  - 1 = 
4 \, {\Theta_1{}^{\!2} \bTh_1{}^{\! 2} \over  X_1{}^{\up{\! 2}}} \, ,
}
using \XTh, is a completely symmetric superconformal invariant.\foot{This
invariant was found by Park \Park.} We may also construct an invariant
which is completely antisymmetric by \eqn\invJ{
J = \half \bigg ( {X_1{}^{\! 2} \over \bar X_1{}^{\up{\! 2}}} -
{\bar X_1{}^{\! 2} \over X_1{}^{\up{\! 2}}} \bigg ) = - 2i \Theta_1
 \bigg ( {\bX_1 \over \bar X_1{}^{\up{\! 2}}} + 
{\X_1 \over X_1{}^{\up{\! 2}}} \bigg )\bTh_1 \, .
}
Such invariants, depending only on three points $z_1,z_2,z_3$, do not exist 
with ordinary conformal symmetry, it is immediately evident from \inv\ 
and \invJ\ that $J^2 = 2 I$, $I^2=0$.

\newsec{Superfield Transformations}

A quasi-primary superfield $O^i(z)$, with $i$ denoting vector or spinor
indices, is here defined by requiring that it forms a representation under
superconformal transformations induced from a finite dimension irreducible
representation of $G_0$. Under an infinitesimal
superconformal transformation, as described in the previous section,
\eqn\supertr{
\de O^i(z) = - \L O^i(z) + \half \pr^a h^b(z) (s_{ab})^i{}_{\! i'} O^{i'}(z)
- 2q\, \si(z_+) O^i(z) - 2\bq\, \bsi(z_-) O^i(z) \, ,
}
where $s_{ab}= - s_{ba}$ are the generators of $O(3,1)$, or the associated
spin group, for the representation $(j,\bj)$, $2j, 2{\bj} = 0,1,2,\dots$,
defined by $O^i$, and $q,\bq$ are
parameters such that $q+\bq$ is the scale dimension and $3(q-\bq)$ is
the $U(1)$ $R$-symmetry charge of the field $O^i$. Thus the superfield
representation may be labelled $(j,{\bar \jmath},q,\bq)$. Using \rot\ we may
write
\eqn\spin{
\half \pr^a h^b(z)\, s_{ab} =  \hom_\alpha{}^\beta(z_+)\, s_\beta{}^\alpha
+ \hbom{}^\dal{}_{\smash\dbe}(z_-)\, \bs{}^\dbe {}_{\dal} \, ,
}
where $s, \bs$, $s_\alpha{}^\alpha= \bs{}^\dal {}_{\dal}=0$,
act on undotted, dotted spinor indices and form spin $j, \, {\bar \jmath}$
representations of the algebra,
\eqn\algspin{
[ s_\alpha{}^\beta , s_\gamma{}^\delta ] =  \de_\alpha{}^{\!\delta}
s_\gamma{}^\beta - \de_{\gamma}{}^{\!\beta} s_\alpha{}^\delta \, ,\qquad
[\bs{}^\dal {}_{\smash\dbe} , \bs{}^\dga{}_{\smash\dde} ] = \de^\dal{}_{\smash{\!\dde}} \,
\bs^\dga{}_{\smash\dbe} - \de^\dga{}_{\smash{\!\dbe}} \,\bs^\dal{}_{\smash\dde} \, ,
\quad [ s_\alpha{}^\beta , \bs{}^\dal {}_{\smash\dbe}] = 0 \, ,
}
which leads to $[\half \pr^a h_1^b\, s_{ab} ,\,\half \pr^c h_2^d\, s_{cd}]
= [\hom_1, \hom_2]_\alpha{}^\beta\, s_\beta{}^\alpha 
+ [\hbom_1, \hbom_2]{}^\dal{}_{\smash\dbe}\, \bs{}^\dbe {}_{\dal}$.
Using this and  \algh\ it is straightforward
to check that the form \supertr\ is consistent with the
algebra in \alg, $[\de_1,\de_2]O = - \de' O$.
If the field is chiral, depending only on $z_+$, then there must be no
terms in \supertr\ corresponding to $\hbom$, so that only $(j,0)$
representations, when $O^i(z) \to O_{\alpha_1 \dots \alpha_{2j}}(z_+)$
totally symmetric in $\{\alpha_1 \dots \alpha_{2j}\}$, are
possible, or to $\bsi$ which requires $\bq=0$. For such chiral
superfields, for which the representation may therefore be denoted $(j,q)_+$, 
and also anti-chiral superfields, labelled by $(\bj,\bq)_-$, the scale
dimension is therefore related to the $R$-charge \refs{\Dondi,\WCon}.

The field transformation defined by \supertr\ in general defines an
irreducible representation. However for particular  $(j,{\bar \jmath},q,\bq)$
the superfield representation is reducible since then suitable 
derivatives also transform as quasi-primary fields \Pet. These cases are
physically significant in application to superconformal covariant
conservation equations. As an illustration which is relevant
subsequently we may consider a $\bj=0$ representation for which \supertr\
becomes
\eqn\superj{
\de \phi_{\alpha_1 \dots \alpha_{2j}} = - \L
\phi_{\alpha_1 \dots \alpha_{2j}} + 2j \, \hom_{(\alpha_1}{}^{\!\beta}
\phi_{\alpha_2 \dots \alpha_{2j})\beta}  - (2q \, \si+ 2\bq \, \bsi )
\phi_{\alpha_1 \dots \alpha_{2j}} \,.
}
{}From \LD\ we have
\eqn\LDalpha{
[ D_\alpha, \L ] = - \hom_\alpha{}^\beta D_\beta + (2\bsi - \si) D_\alpha \, ,
}
and using
\eqn\Domega{
D_\gamma \hom_\alpha{}^\beta = - 2 \de_\gamma{}^{\! \beta} D_\alpha \si
+  \de_\alpha{}^{\! \beta} D_\gamma \si \, ,
}
we may find
\eqn\Dphi{\eqalign{
D_\alpha \de \phi_{\alpha_1 \dots \alpha_{2j}} = {}& - \L \big (
D_\alpha \phi_{\alpha_1 \dots \alpha_{2j}} \big )
+  2j \,   \hom_{(\alpha_1}{}^{\!\beta}  D_{|\alpha|}
\phi_{\alpha_2 \dots \alpha_{2j})\beta} + \hom_\alpha{}^\beta D_\beta
 \phi_{\alpha_1 \dots \alpha_{2j}} \cr
{}& - \big ( (2q-1)\si + 2(\bq+1) \bsi \big ) 
D_\alpha \phi_{\alpha_1 \dots \alpha_{2j}} \cr
{}& - 4j \, D_{(\alpha_1}\!\si \,\phi_{\alpha_2 \dots \alpha_{2j})\alpha}
+ 2(j-q) \, D_\alpha \si\, \phi_{\alpha_1 \dots \alpha_{2j}} \, . \cr}
}
{}From \Dphi\ we may easily see, by requiring cancellation of the 
$D_\alpha \si$ terms, that
\eqn\deriv{
{\tD^\alpha \phi_{\alpha_1 \dots \alpha_{2j-1}\alpha}\atop
D_{(\alpha} \phi_{\alpha_1 \dots \alpha_{2j})}}\bigg \} \  \ 
\hbox{are quasi-primary if} \ \ \cases { q=j+1\, , \cr q =-j \, .}
}
The conditions in \deriv\ for $D$ derivatives to be quasi-primary in fact 
apply, without any modification of the argument since $D\bsi=D\hbom=0$, to any
$\bj\ge 0$ representation. Hence in general the derivatives
defined in \deriv\  give
\eqna\inv$$\eqalignno{
(j,{\bar \jmath},j+1,\bq) \longrightarrow {}&
(j-\half,{\bar\jmath},j+\half,\bq+1) \, , & \inv a \cr
(j,{\bar \jmath},-j,\bq) \longrightarrow {}&
(j+\half,{\bar\jmath},-j-\half,\bq+1) \, , & \inv b  \cr
}$$
and the kernels of these maps are invariant subspaces of the superfield
representation spaces for these cases, if $j=0$ in \inv{b}\ the kernel
is just the space of $(\bj,\bq)_-$ anti-chiral superfields.
Such results also obviously apply for $\bD$ derivatives 
if $\bq = {\bar \jmath}+1$ or $\bq = - {\bar \jmath}$. Similar
arguments show that $D^2 O^i$ is quasi-primary only for $(0,\bj,1,\bq)$
representations, as expected since $D^2 O^i$ is an anti-chiral $(\bj,\bq+2)_-$ 
superfield, and conversely for $\bD^2 O^i$ if $\bj=0,\, \bq=1$.\foot{In 
\TMP\ the invariant
subspaces defined by the kernel of $D^2$ were considered, in these papers
a representation of the superconformal generators acting on superfields
based on the little group generated by matrices $M$ in \MM\ with
$a=\ep=\bta=0$ is considered. This is related to the representation here
by a Fourier transform with respect to $\bth$.}

The representation defined by \supertr\ is unitary and has positive energy
when $q, \, \bq$ are real with the following restrictions \Unit,
\eqn\unit{\eqalign{
&j, {\bar \jmath} \ge 0 \,  \quad q\ge j+1, \, \bq \ge {\bar \jmath}+1 \, ; \cr
&j=q=0 \,  , \ \bq \ge {\bar \jmath}+1 \, ; \quad {\bar \jmath}= \bq = 0 \, ,
\ q \ge j+1 \, ; \cr
&j= {\bar \jmath} = q = \bq = 0 \, . \cr}
}

The results \deriv\ are relevant when derivative constraints are imposed
on superfields if they are to be superconformal covariant. 
For the vector supercurrent $T_a(z)\to\T_{\alpha\dal}(z)$ then
\eqn\cons{
\tD^\alpha \T_{\alpha\dal} = {\tilde \bD}{}^\dal \T_{\alpha\dal} = 0  \, ,
}
is superconformally covariant as a consequence of \inv{a}\ and its
conjugate only for $T$ 
belonging to the $(\half,\half,{3\over 2},{\ts{3\over 2}})$ representation.
Similarly for the scalar superfield $L(z)$ containing a current 
amongst its components the supersymmetric conservation equations
\eqn\consL{
D^2 L = \bD^2 L = 0  \, ,
}
require $L$ to belong to the  $(0,0,1,1)$ representation.
We may also note that for chiral, anti-chiral spinor
superfields $W_\alpha, \, \bW_\dal$ then
\eqn\conW{
\tD^\alpha W_\alpha = \bD_\dal {\tilde \bW}{}^\dal \, ,
}
is consistent with superconformal invariance only if $W, \bW$ belong to
$(\half,{3\over 2})_+, \, (\half,{3\over 2})_-$ representations respectively 
(from \inv{a}\ both sides of \conW\ are then $(0,0,2,2)$ superfields). 
Eq. \conW\ is just the Bianchi identity for abelian supergauge
theories so this result shows that it can be maintained at a superconformal
point only if the scale dimensions of $W,\bW$
remain equal to their free field values. 
In consequence a non trivial superconformal gauge theory must violate the
Bianchi identity which requires there to be both massless magnetically and
electrically charged fields \Witten.\foot{Similar results hold in  any
dimension for purely bosonic gauge fields. 
If $F_{ab}$ is the field strength, with scale dimension
$\eta$, and which is supposed to transform under
conformal transformations according to $\de F_{ab} = -(h{\cdot \pr} +
\eta \rho) F_{ab} + \tom_{a}{}^c F_{cb}+\tom_{b}{}^c F_{ac}$ 
where $\pr_a h_b = \eta_{ab}\rho - \tom_{ab}$, $\tom_{ab}= - \tom_{ba}$,
then $\pr_{[c}\de F_{ab]} = - (h{\cdot \pr} + (\eta+1)\rho)\pr_{[c} F_{ab]} 
+ \tom_c{}^d \pr_{[d} F_{ab]} + \tom_{a}{}^d \pr_{[c} F_{db]} +
\tom_{b}{}^d \pr_{[c} F_{ad]} + 2 (\eta-2) b_{[c} F_{ab]}$. The 
inhomogeneous terms in the transformation of the Bianchi identity
$\pr_{[c} F_{ab]}=0$, involving the parameter $b$ for special conformal
transformations, vanish only if $\eta=2$ which is the free field case.}

Using \cons\ and the conditions \Dhh\ and \hh\ we may define a scalar
superfield
\eqn\LTh{
L_h^{\vphantom g} = \th^{\dal\alpha}\T_{\alpha\dal} \, ,
}
which satisfies the conservation equations \consL. Under a superconformal
transformation we have
\eqn\LTtr{
\de_1 L_{h_2}^{\vphantom g} = \th_2^{\dal\alpha} \de_1
\T_{\alpha\dal} = - \big (\L_1 + 2 \si_1 + 2 \bsi_1  \big ) L_{h_2}^{\vphantom g}
- L_{h'}^{\vphantom g} \, ,
}
where $h'$ is as in \hprime.

An expression for the superfield $L$ which trivially satisfies the conservation
equations in \consL\ is given by
\eqn\trivL{
L = \tD{}^\alpha F_\alpha \, , \qquad \bD_\dal F_\alpha = 0 \, ,
}
and for this  to be superconformally covariant $F_\alpha$ must be
a $(\half,{3\over 2})_+$ chiral superfield.

To construct an expression for the supercurrent $T$
in which the conservation equations
\cons\ are identically satisfied we consider a $({3\over 2},q)_+$ 
chiral superfield $C^{\alpha\beta\gamma}(z_+)$ which transforms as
\eqn\tranC{
\de C^{\alpha\beta\gamma} = - \L C^{\alpha\beta\gamma} - 3
\, C^{\alpha'(\beta\gamma}\hom_{\alpha'}{}^{\! \alpha)} - 2q\, \si \,
C^{\alpha\beta\gamma} \, ,
}
and using \LDalpha\ and
\eqn\Lpr{
[\pr_{\alpha\dal},\L] = - \hom_\alpha{}^\beta \pr_{\beta\dal}
+ \hbom{}^\dbe{}_\dal \pr_{\smash{\alpha\dbe}} + (\si + \bsi) \pr_{\alpha\dal}
+ i \big ( D_\alpha \si \bD_\dal + \bD_\dal \bsi  D_\alpha \big ) \, ,
}
we may find
\eqn\tranDC{\eqalign{
D_\beta \pr_{\gamma\dal} \de C^{\alpha\beta\gamma} = {}&
- \L \big ( D_\beta \pr_{\gamma\dal}  C^{\alpha\beta\gamma} \big )
- D_\beta \pr_{\gamma\dal}  C^{\alpha'\beta\gamma}\, \hom_{\alpha'}{}^\alpha
- D_\beta \pr_{\smash{\gamma\dbe}}  C^{\alpha\beta\gamma}\, \hbom{}^\dbe{}_\dal\cr
{}& - (2q\si + 3 \bsi)D_\beta \pr_{\gamma\dal}  C^{\alpha\beta\gamma}
+ (3-2q) \big ( D_\beta \si \, \pr_{\gamma\dal} C^{\alpha\beta\gamma}
- D_\beta C^{\alpha\beta\gamma} \, \b_{\gamma\dal}\big )  \, . \cr}
}
In consequence $D_\beta \pr_{\gamma\dal} C^{\alpha\beta\gamma}$ is
quasi-primary if $q={3\over 2}$ and it is easy to see that
\eqn\TC{
\T_{\alpha\dal} = \ep_{\alpha\alpha'}D_\beta \pr_{\gamma\dal} 
C^{\alpha'\beta\gamma} \, , \qquad \bD_\dal C^{\alpha\beta\gamma}=0 \, ,
}
then gives an expression for the supercurrent which satisfies \cons\ 
identically, and also has the correct
superconformal representation $(\half,\half,{3\over 2},{3\over 2})$ for $T$.
Such a $({3\over 2},{3\over 2})_+$ superfield falls outside the restrictions
given in \unit\ so $C^{\alpha\beta\gamma}$ cannot exist in a unitary theory
but the representation \TC\ is important in various cases subsequently.

For any quasi-primary superfield general constructions for two and three
point functions, consistent with superconformal invariance, 
are possible using the results of section 2. For
a superfield $O^i$ we also consider its conjugate $\bO^\bi$ when 
$(j,{\bar \jmath},q,\bq) \to ({\bar \jmath},j,\bq,q)$. To define the two-point
function  for superfields at $z_1,z_2$ we introduce $I^{i\bi}(x_{1\2}, x_{\1 2})$ 
which transforms as
a bilocal invariant tensor for the corresponding superfield representations
at $z_1,z_2$ with $q=\bq=0$,
\eqn\Itran{ \eqalign{
\big ( \L_{z_1} +{}& \L_{z_2} \big ) I^{i\bi}(x_{1\2}, x_{\1 2})\cr
&{} - \half \pr^a h^b(z_1) (s_{ab})^i{}_{\! i'} I^{i'\bi}(x_{1\2}, x_{\1 2}) 
+ I^{i\bi'}(x_{1\2}, x_{\1 2}) 
(\bs_{ab})_{\bi'}{}^{\!\bi}\half \pr^a h^b(z_2) = 0 \, ,\cr}
}
with $\bs_{ab}= - s_{ab}{}^\dagger$ the generators of the conjugate 
representation. With $g_{ii'},
\, \bg_{\bi\bi'}$ group invariant tensors for these representations
then we may also define
\eqn\Ibar{
\bI_{\bi i}(x_{\2 1},x_{2\1}) = \bg_{\bi\bi'} g_{ii'} 
I^{i'\bi'}(x_{1\2}, x_{\1 2}) \, , 
}
and the normalisation of $I,\bI$ is fixed by requiring
\eqn\II{
I^{i\bi}(x_{1\2}, x_{\1 2}) \bI_{\bi i'}(x_{\2 1},x_{2\1}) = \de^i{}_{\! i'} \, , 
\qquad
\bI_{\bi i}(x_{\2 1},x_{2\1}) I^{i\bi'}(x_{1\2}, x_{\1 2}) = \de_\bi{}^{\! \bi'} \, .
}
For the fundamental spinor representation then from \varxx\ and \defxx\
expressions satisfying \Itran\ and also \Ibar\ may be obtained which
depend just on $x_{1\2}=-x_{\21}$,
\eqn\Ispin{
I_{\alpha\dal}(x_{1\2}) = i \, {(\x_{1\2})_{\alpha\dal}\over 
\sqrt{x_{\2 1}{}^{\! 2}}} \, , \qquad 
\bI^{\dal\alpha}(x_{\2 1}) = \ep^{\dal\dbe}\epsilon^{\alpha\beta}
I_{\beta\dbe}(x_{1\2}) = - i\,  {(\tx_{\2 1})^{\dal \alpha} \over
\sqrt{x_{\2 1}{}^{\! 2}}} \, ,
}
while for general representations explicit results are easily obtained
by reduction of tensor products. Thus for 4-vectors, when
$g_{ii'}, \bg_{\bi\bi'} \to \eta_{ab}$,
\eqn\Ivector{
I_{ab}(x_{1\2}, x_{\1 2}) = \bI_{ba}(x_{\2 1},x_{2\1}) = 
{{\rm tr}( \si_a\, \tx_{\1 2}\, \si_b \,\tx_{\2 1}) 
\over 2\sqrt{x_{\1 2}{}^{\! 2}\, x_{\2 1}{}^{\! 2}} } =
{{\rm tr}( \tsi_a \,\x_{1 \2}\, \tsi_b\, \x_{2 \1}) 
\over 2\sqrt{x_{\1 2}{}^{\! 2}\, x_{\2 1}{}^{\! 2}}  } \, . 
}
Here the denominator may be simplified by use of \twop. 

With the definitions \Ivector\ and \Ispin\ we may rewrite the transformations 
\Xtran\ and \Thtran\ as
\eqn\IX{\!\!\!\!
I_{ab}(x_{2\1}, x_{\2 1}) X_1{}^{\! b} = {1\over \big (
X_2{}^{\up{\! 2}}\,  {\bar X}_2{}^{\up{\! 2}}\, x_{\1 2}{}^{\! 2}\, x_{\2 1}{}^{\! 2} 
\big )^{1\over 2} } \, X_{2 a}{}^{\!\!\! I} \, , \ \
X_a{}^{\!\! I} \equiv I_{ab}({\bar X} , X)  X^{b} = - \bigg ( { X^{2} \over
{\bar X}^{\up{2}}} \bigg )^{\!\!{1\over 2}}\! {\bar X}_a  \, ,
}
and
\eqn\ITh{\eqalign{
I_{\alpha\dal}(x_{2\1}) \bTh_1{}^{\!\dal} = {}&
\bigg( {x_{\1 2}{}^{\! 2}\over{\bar X}_2{}^{\up{\! 2}}}\bigg)^{\! {1\over 2}} \!
{1\over x_{\2 1}{}^{\! 2}} \, \bTh_{2\alpha}{}^{\!\!\! I} \, , \quad
\bTh_\alpha{}^{\!\! I} \equiv I_{\alpha\dal} ({\bar X} )\bTh^{\dal}  =
i {1\over ({\bar X}^2)^{1\over 2}}\, (\bX \bTh)_ \alpha \, ,\cr
\Theta_1{}^{\!\alpha} I_{\alpha\dal}( x_{1\2}) = {}& 
\bigg( {x_{\2 1}{}^{\! 2} \over {X}_2{}^{\up{\! 2}}} \bigg)^{\! {1\over 2}} \!
{1\over x_{\1 2}{}^{\! 2}} \,  \Theta_{2\dal}{}^{\!\!\! I}  \, , \quad
\Theta_\dal{}^{\!\! I} \equiv \Theta^{\alpha} I_{\alpha\dal}(-X) = 
- i {1\over ({X}^2)^{1\over 2}}\, (\Theta \X)_\dal \, .\cr}
}
{}From \defX\ we may also obtain
\eqn\III{
\bI_{\bi_1 i_2}(x_{\1 2},x_{1\2}) I^{i_2\bi_3}(x_{2\3},x_{\2 3}) 
\bI_{\bi_3 i_1}(x_{\3 1},x_{3\1}) = \bI_{\bi_1 i_1}({\bar X}_1, X_1) \, .
}

The significance of $I, \, \bI$ becomes
more apparent on considering the transformation of superfields under inversions,
for which $z\longrightarrow z'$ as in \invert, when we require
\eqn\invertO{\eqalign{
O^i(z) \longrightarrow  O^{\prime i}(z) = {}&
{1\over x'_{\, +}{}^{\!\! 2\bq} \, x'_{\, -}{}^{\!\! 2q}}\,
I^{i\bi}(-x'{}_{\! -},  - x'{}_{\! +}) \, \bO_\bi(z') \, , \cr
\bO_\bi(z) \longrightarrow \bO'{}_{\! \bi}(z) =  {}&
{1\over x'_{\, +}{}^{\! 2q} \, x'_{\, -}{}^{\! 2\bq}}\,
\bI_{\bi i}(-x'{}_{\! +},  - x'{}_{\! -}) \, O^i(z') \, . \cr}
}
Since all superconformal transformations can be generated by combining
inversions with ordinary supersymmetry the transformations in \invertO\
are sufficient to obtain any superfield superconformal transformation.
As a consequence of \II\ $I^{i\bi}(-x{}_{ -},  - x{}_{ +})
\bI_{\bi i'}(x_-, x_+) = \de^i{}_{\!i'}$
and using \invert\ $x'_{\,\pm} = - x_{\mp}/ x_\mp{}^{\! 2}$
it is easy to verify from \invertO\ that two inversions leave the superfields
 invariant. In the purely bosonic case $I_{ab}(x_{1\2}, x_{\1 2})$
reduces to the inversion tensor $\eta_{ab} - 2 x_{12a} x_{12b}/x_{12}{}^{\! 2}$
which played a crucial role in the discussion of conformal invariance in
arbitrary dimensions \refs{\hughone,\hughtwo}.

A general superconformal covariant expression for the two point function
of the superfield $O$ and its conjugate $\bO$ is given in terms of $I$ by
\eqn\OO{
\l O^i(z_1) \bO^\bi(z_2) \r = C_O \,
{I^{i\bi}(x_{1\2}, x_{\1 2}) \over x_{\1 2}{}^{\! 2\bq} \, x_{\2 1}{}^{\! 2q}} \, ,
}
with $C_O$ an overall normalisation constant. In the denominator of \OO, and
in other expressions subsequently, the singular behaviour at $x_{\1 2}{}^{\! 2},
x_{\2 1}{}^{\! 2}=0$ has be modified in accord with the standard lore
of quantum field theory, thus for a product of field operators $\phi(x)\phi(0)$
the singular functions should depend on
$x^2 + i\ep x^0$ while for  time ordered products on $x^2+i\ep$. However in
\OO\ and subsequently we leave such resolutions of light cone singularities
implicit. The conditions \unit\ are necessary and sufficient for \OO\ to be
expressible in terms of a sum over intermediate states of positive norm, as in
any unitary theory.

For the three point function we may write a general form as
\eqn\OOO{
\l O_1^{i_1}(z_1) O_2^{i_2}(z_2) O_3^{i_3}(z_3) \r = 
{I_1^{i_1\bi_1}(x_{1\3}, x_{\1 3}) I_2^{i_2\bi_2}(x_{2\3}, x_{\2 3}) \over 
x_{\1 3}{}^{\! 2\bq_1} \, x_{\3 1}{}^{\! 2q_1}\,
x_{\2 3}{}^{\! 2\bq_2} \, x_{\3 2}{}^{\! 2q_2} } \, 
t_{\bi_1 \bi_2}{}^{i_3}(X_3,\Theta_3, \bTh_3) \, ,
}
where $X_3,\Theta_3,\bTh_3$ are defined by appropriate modification of \defX\ and \defTh,
and $I_1 , I_2$ are the bilocal tensors introduced above for the representations
defined by the quasi-primary superfields $O_1,O_2$. The
expression \OOO\ automatically has the correct transformation properties at $z_1, \, z_2$ and
also at $z_3$ if $t_{\bi_1 \bi_2}{}^{i_3}$ has the homogeneity properties
\eqn\homt{
t_{\bi_1 \bi_2}{}^{i_3}( \lambda\bla X , \lambda \Theta, \bla \bTh )
= \lambda^{2a} \bla^{2\ba} \, t_{\bi_1 \bi_2}{}^{i_3}(  X , \Theta,\bTh ) \, ,
}
where
\eqn\aa{
a - 2 \ba = \bq_1 + \bq_2 - q_3  \, , \qquad \ba - 2a = q_1+q_2- \bq_3  \, .
}
Since $\Theta, \bTh$ are two-component Grassmann spinors we must have
\eqn\aba{
2(a- \ba) = {\ts{2\over 3}}\, {\ts \sum_i} \big ( \bq_i - q_i \big ) = 0, \pm 1 , \pm 2 \, ,
}
and if $a=\ba$ we may equivalently write, as a consequence of \XTh, 
$ t_{\bi_1 \bi_2}{}^{i_3}(  X , {\bar X})$. 
Otherwise $t_{\bi_1 \bi_2}{}^{i_3}(  X , \Theta,\bTh )$ is required to transform
according to the appropriate spin representations when $X, \Theta,\bTh $ transform
infinitesimally as $\de \X = \omega \X - \X \bom , \, \de \Theta = - \Theta \omega , \,
\de \bTh = \bom \bTh$.

Using \IX, \ITh, for $2\to 1, 1\to 3$, and \homt\ we may use the invariance
properties of $t_{\bi_1 \bi_2}{}^{i_3}(  X , \Theta,\bTh )$ to obtain the
transformation formula
\eqn\invt{\eqalign{\!\!\!
I_1^{i_1\bi_1} (x_{1\3},x_{\1 3})& I_2^{i_2\bi_2}(x_{1\3}, x_{\1 3})
\bI_{3\, \bi_3 i_3}(x_{\1 3}, x_{1\3})\,
t_{\bi_1 \bi_2}{}^{i_3}(X_3,\Theta_3,\bTh_3) \cr
{}& = {x_{\1 3}{}^{\! 2(a - 2\ba)} x_{\3 1}{}^{\! 2(\ba - 2a)}\over
X_1{}^{\up{\! 2a}}\,  {\bar X}_1{}^{\up{\! 2\ba} } }\, \bt^{i_1 i_2}{}_{\bi_3}
(X_1{}^{\! I} , \Theta_1{}^{\! I}, \bTh_1{}^{\! I}) \, , \cr
\bt^{i_1 i_2}{}_{\bi_3} (X^I , \Theta^I, \bTh^I)
={}& I_1^{i_1\bi_1}({\bar X},X) I_2^{i_2\bi_2}({\bar X},X)\bI_{3\, \bi_3 i_3}
(-{\bar X},-X) \,t_{\bi_1 \bi_2}{}^{i_3}(X,\Theta,\bTh) \, , \cr}
}
with $X^I , \Theta^I, \bTh^I$ defined in \IX\ and \ITh.
This result, with the aid of \III\ and \II, allows us to rewrite \OOO\ in the
equivalent form
\eqn\OOOt{
\l O_1^{i_1}(z_1) O_2^{i_2}(z_2) O_3^{i_3}(z_3) \r = 
{I_2^{i_2\bi_1}(x_{2\1}, x_{\2 1}) I_3^{i_3\bi_3}(x_{3\1}, x_{\3 1}) \over 
x_{\2 1}{}^{\! 2\bq_2} \, x_{\1 2}{}^{\! 2q_2}\,
x_{\3 1}{}^{\! 2\bq_3} \, x_{\1 3}{}^{\! 2q_3} } \, 
{\tilde t}^{i_1}{}_{\bi_2 \bi_3}(X_1,\Theta_1, \bTh_1) \, ,
}
where
\eqn\IIt{
{\tilde t}^{i_1}{}_{\bi_2 \bi_3}(X,\Theta, \bTh) = {1 \over
X^{\up{2(a+\bq_2)}} {\bar X}{}^{\up{2(\ba + q_2)}}} \, 
\bI_{2\, \bi_2 i_2}({\bar X},X) \, \bt^{i_1 i_2}{}_{\bi_3} (X^I , \Theta^I, \bTh^I)
\, .
}
It is clearly possible to obtain a third representation in which we have a 
function ${\tilde{\tilde t}}_{\bi_1}{}^{\! i_2}{}_{\bi_3}(X_2,\Theta_2, \bTh_2)$.
It is straightforward to verify that the result \OOOt\ satisfies the equivalent 
homogeneity properties to \homt\ and \aa.

The relevance of these results becomes more apparent 
if we consider the short distance limit $z_1\to z_2$.
In this limit it is easy to see that
\eqn\limx{
X_1{}^{\! I} \sim - {1\over (x_{\2 1}{}^{\! 2} \, x_{\1 2}{}^{\! 2})^{1\over 2}}
\, x_{1\2} \, , \quad \Theta_1{}^{\! I} \sim - {1\over (x_{\2 1}{}^{\! 2})^{1\over 2}}
\, {\tilde \bth}_{12} \, , \quad \bTh_1{}^{\! I} \sim - 
{1\over (x_{\1 2}{}^{\! 2})^{1\over 2}}\, \tth_{12} \, .
}
Using these limiting expressions in \OOOt\ with \IIt, and applying the homogeneity 
relation \homt\ once more, the leading behaviour has the form
\eqn\short{
\l O_1^{i_1}(z_1) O_2^{i_2}(z_2) O_3^{i_3}(z_3) \r \sim
I_3^{i_3\bi_3}(x_{3\1}, x_{\3 1}) \, \bt^{i_1 i_2}{}_{\bi_3} \big
(x_{2\1}, {\tilde \bth}_{21}, \tth_{21}\big ) \quad \hbox{for} \quad z_1 \sim z_2 \, .
}
Given the result \OO\ for the two point function we may therefore obtain
for the contribution to the operator product expansion of $O_1, O_2$ involving
$\bO_3$,
\eqn\OPE{
 O_1^{i_1}(z_1) O_2^{i_2}(z_2) \sim {1\over C_{O_3}}\, \bt^{i_1 i_2}{}_{\bi_3} \big
(x_{2\1}, {\tilde \bth}_{21}, \tth_{21}\big ) \, \bO_3^{\bi_3}(z_2) \, .
}
This demonstrates how in the operator
product expansion of two quasi-primary fields the most singular coefficient
in the expansion involving a third operator, without any derivatives acting
on it, determines completely the corresponding superconformally
covariant three point function.

In the rest of this paper we apply the general results to various particular
cases, mainly involving the supercurrent. It is important to recognise that
the representations \trivL\ and \TC\ may be used to provide alternative
less singular forms. For the superfield $L$, which contains a conserved
current, then the general formula \OO\ gives simply
\eqn\LL{
\l L(z_1) L(z_2) \r = C_L \, {1\over x_{\2 1}{}^{\! 2} \, x_{\1 2}{}^{\! 2}} \, .
}
Making use of \trivL\ an expression  which satisfies \consL\ identically
and reduces to \LL\ for $z_1 \ne z_2$ is then
\eqn\LLt{
\l L(z_1) L(z_2) \r =  \quar C_L \, \tD_2{}^{\! \alpha} D_{1 \alpha}
\bigg ( {\theta_{12}^{\,\, 2} \over (x_{12+}^{\,\,\, 2})^{\up 2}} \bigg ) \, , 
\qquad x_{12+}= x_{1+}- x_{2+}  \, .
}
In a similar fashion the general form for the two point function of the
supercurrent is from \OO,\foot{This result was essentially given in the first paper
in ref.\Gris.}
\eqn\TT{
\l T_a(z_1) T_b(z_2) \r = C_T \, {I_{ab}(x_{1\2}, x_{\1 2}) \over
\big (x_{\2 1}{}^{\! 2} \, x_{\1 2}{}^{\! 2} \big )^{3\over 2}} \ \ \hbox{or} \ \
\l \T_{\alpha\dal}(z_1) \T_{\smash{\beta\dbe}}(z_2) \r = 2C_T \,
{(\x_{1\2})_{\smash{\alpha\dbe}} \, (\x_{2\1})_{\beta\dal} \over \big (
x_{\2 1}{}^{\! 2} \, x_{\1 2}{}^{\! 2} \big )^2 } \, ,
}
and with the aid of \TC\ this can be rewritten as
\eqn\TTt{
\l \T_{\alpha\dal}(z_1) \T_{\smash{\beta\dbe}}(z_2) \r = - {1\over 16}C_T \,
\pr_{1\gamma\dal}\pr_{\smash{ 2\delta\dbe}} D_{2\ep}D_{1\eta} 
\bigg ( \E_\alpha{}^{\gamma\ep}{}_{,\beta}{}^{\de\eta}\,
{\theta_{12}^{\,\, 2} \over (x_{12+}^{\,\,\, 2})^{\up 2}} \bigg ) \, , 
}
where $\E$ is the projector for symmetric three index spinors
\eqn\proj{
\E_{\alpha\gamma\ep,}{}^{\beta\de\eta} = \de_{(\alpha}{}^{\!\beta}
\de_{\gamma}{}^{\!\de}\de_{\ep)}{}^{\!\eta} \, .
}
The expressions \LLt\ and \TTt\ are no longer manifestly superconformally
covariant but they allow the non integrable singularity at $z_1=z_2$
to be easily regularised, consistent with the conservation equations,
using the method of differential regularisation by replacing in \LLt\ and \TTt\
\eqn\reg{
\R\bigg({1 \over (x_{12+}^{\,\,\, 2})^{\up 2}}\bigg ) 
= - {1\over 4}\, \pr^2 \bigg (
{1\over x_{12+}^{\,\,\, 2}}\ln \big ( \mu^2 x_{12+}^{\,\,\, 2} \big )\bigg ) \, ,
}
with $\mu$ an arbitrary mass scale.

\newsec{Supercurrent and Ward Identities}

In order to derive the Ward identities which constrain correlation functions
and operator product expansions involving the supercurrent we use a variant
of the standard Noether construction of the supercurrent from symmetries of
the action. For orientation we first consider a global continuous symmetry
acting on the basic fields of the theory
under which the action is invariant $\de_\ep S = 0$, for $\ep_i$ the parameter
representing an infinitesimal group transformation. To define the associated
conserved current in terms of a superfield we allow $\ep_i$ to be extended
to independent $\ep_i(z_+), \, \bep_i(z_-)$ which are defined locally on chiral,
anti-chiral superspace (thus chiral fields transform according to $\ep_i$
while anti-chiral fields according to $\bep_i$). The action then transforms 
generally as
\eqn\deS{
\de_{\ep,\bep} S = i \int \! \d^8 z \, \big ( \ep_i K_i - \bep_i \bK_i \big ) \, .
}
For $\ep_i = \bep_i$, a constant, the variation must vanish and this requires
that $K_i -\bK_i$ must be a total derivative so that in general we may therefore
write $K_i -\bK_i = \tD^\alpha U_{i\alpha} + \bD_\dal \bU_i{}^{\!\dal}$ for
some $U_{i\alpha}, \, \bU_i{}^{\!\dal}$. Furthermore in the definition
\deS\ $K_i$ is arbitrary up to $K_i \sim K_i - \bD_\dal \bU_i{}^{\!\dal}$
and similarly $\bK_i \sim \bK_i + \tD^\alpha U_{i\alpha} $. Using this freedom
allows us to set $K_i -\bK_i $ to zero so that \deS\ then becomes
\eqn\deSL{
\de_{\ep,\bep} S = i \int \! \d^8 z \, \big ( \ep_i- \bep_i \big ) L_i \, .
}
Since $\de S=0$ for arbitrary variations of the fields defines the equations
of motion then, so long as these are satisfied, \deSL\ must vanish for any 
$ \ep_i, \, \bep_i$ which leads to $L_i$ being required to obey \consL. 
In a quantum field theory, with $L_i$ an operator superfield,
the Ward identities for correlation functions involving other superfields $O$
may be formally derived from the functional integral using \deSL, 
assuming invariance of the measure or equivalently no anomalies, in the form
\eqn\WardL{
- \int \! \d^8 z \, \big ( \ep_i(z_+) - \bep_i(z_-) \big ) \big \l L_i(z) \dots 
O(z_r) \dots \big \r + 
\sum_O \big \l  \dots \de_{\ep,\bep}O(z_r) \dots \big \r = 0 \, ,
}
which leads to differential relations on taking functional derivatives with
respect to $\ep_i$ or $\bep_i$.

For the supercurrent we follow a similar analysis\foot{For a very different
application of Noether's theorem to a derivation of the supercurrent see
\Og.} considering now the
response of the action to local superspace diffeomorphisms preserving chiral
superspace so that coordinates transform as in \ch\ with $h,\bh$ given by \defh.
For general $h,\bh$ satisfying \Dhh\ it is convenient to define
\eqn\sih{
\omega_{h}^{\vphantom g}{}_\alpha{}^\beta - 3\si_h^{\vphantom g} \, 
\de_\alpha{}^{\! \beta} =
D_\alpha \lambda^\beta + \half\, \pr_{\alpha \dal}\th{}^{\dal\beta} \, , \quad
\bom_{\bh}{}^{\! \dbe}{}_{\dal} + 3 \bsi_{\bar h} \, \de^\dbe{}_{\! \dal}
= \bD_\dal \bla^\dbe - \half\, \pr_{\alpha\dal} {\tilde \bh}{}^{\dbe\alpha} \, ,
}
where $\omega_{h}^{\vphantom g}{}_\alpha{}^\alpha 
= \bom_{\bh}{}^{\! \dal}{}_{\dal}=0$ and these satisfy the chirality conditions
\eqn\chsi{
\bD_\dal \omega_{h}^{\vphantom g}{}_\alpha{}^\beta 
= \bD_\dal \si_h^{\vphantom g} = 0 \, , 
\qquad D_\alpha \bom_{\bh}{}^{\! \dbe}{}_{\dal} = D_\alpha  \bsi_{\bar h} = 0 \, .
} 
Analogous to \deS\ we then assume that under transformations
on the basic fields induced by such diffeomorphisms the action is assumed to
transform as
\eqn\deSh{
\de_{h,\bh} S = -\half i \int \! \d^8 z \, \big ( h^a J_a - \bh^a \bJ_a  \big ) 
+ \int \! \d^6 z_+ \, \si_h^{\vphantom g} \cJ 
+ \int \! \d^6 z_- \, \bsi_{\bar h} {\bar \cJ} \, ,
}
where  $\cJ, \, {\bar \cJ}$ are chiral, anti-chiral superfields, satisfying
$\bD_\dal \cJ = 0, \, D_\alpha {\bar \cJ} = 0$.
As a consequence also of \Dhh\ $J, \, \bJ $ are arbitrary up to
\eqn\arbJ{
\J_{\alpha\dal} \sim \J_{\alpha\dal} + \tD^\beta K_{(\beta\alpha)\dal} \, , \qquad
{\bar \J}_{\alpha\dal} \sim {\bar \J}_{\alpha\dal} +
{\tilde \bD}{}^\dbe \bK{}_{\smash{\alpha(\dal\dbe)}} \, .
}
Using this freedom we show that improvement terms may be added to $J, \, \bJ$
so that their difference can be transformed to zero.\foot{This is analogous
to the usual introduction of improvement terms to obtain a symmetric
traceless energy momentum tensor.
In $d$-dimensions then for an action $S$ depending on fields $\phi$ then if
these transform under translations, Lorentz and scale transformations
according to $\de \phi = - h^a \pr_a \phi - \half \omega^{ab} s_{ab}\phi
- \eta \rho \phi$, with $s_{ab}$ the spin generators and $\eta$ the scale
dimension, then for arbitrary local $h^a(x), \, \omega^{ab}(x) = -
\omega^{ba}(x) , \, \rho(x)$ the action $S$ may be supposed to transform as
$\de S = - \int \! \d^d x \, \big \{ (\pr_a h_b + \omega_{ab} - \rho\, \eta_{ab})
T^{ab} + \pr_c\omega_{ab}X^{cab} + \pr_a\pr_b \rho\, S^{ab}\big \}$, where 
the further assumption that terms involving only single derivatives of
$\rho$ can be removed is also made. With this restriction if
$T^{ab}_{\rm{imp.}} = T^{ab}-\pr_c \big ( X^{cab} - X^{acb} + X^{bac})
+ \pr_c\pr_d \big ( \eta^{ac} J^{db} + \eta^{ad} J^{cb} + \eta^{bc} J^{da}
+ \eta^{bd} J^{ca} - 2 \eta^{ab} J^{cd}- 2\eta^{cd} J^{ab} \big)$  for, if $d>2$,
$J^{ab} = {1\over 2(d-2)} \big ( S^{ab} - {1\over 2(d-1)} \eta^{ab}\, \eta_{cd}
S^{cd} \big )$ then $\de S = - \int \! \d^d x \, 
(\pr_a h_b + \omega_{ab} - \rho\, \eta_{ab})T^{ab}_{\rm{imp.}}$ and since
the variation withs respect to $h_a, \, \omega_{ab},\, \rho$ must vanish
independently, subject to the equations of motion, $T^{ab}_{\rm{imp.}}$ is
conserved, symmetric and traceless.}

To demonstrate this we make essential use of the invariance of $S$ under 
the usual Poincar\'e and supersymmetry transformations on the fields. 
This requires that \deSh\ should vanish for $h^a = \bh^a = h_L{}^{\! a}$, with
$h_L{}^{\! a}$ given by the restriction of \sol\ to the non superconformal case
when $\si_{h_L}^{\vphantom g} = \bsi_{{\bar h}_L}=0$, so that
\eqn\inv{
\int \d^8 z \, h_L{}^{\!\! a} \big (J_a - \bJ_a  \big ) = 0 \, , \qquad
{\tilde \h}_L{}^{\!\! \dal \alpha} 
= \ta^{\dal \alpha} + \bom^\dal{}_{\smash\dbe} \tx_+{}^{\!\!\dbe\alpha}
- \tx_-{}^{\!\!\dal\beta}\omega_\beta{}^\alpha - 4i\, \bth{}^\dal \ep^\alpha
+ 4i\, \bep{}^\dal \theta^\alpha \, .
} 
In consequence $J_a - \bJ_a$ must be expressible as a total derivative which,
in order to cancel the $a,\ep,\bep$ terms in $h_L$, should be of the form
\eqn\diffJ{\eqalign{\!\!\!\!
\J_{\alpha\dal} -{}& {\bar \J}_{\alpha\dal} = \tD^\beta Z_{(\beta\alpha)\dal}
+ {\tilde \bD}{}^\dbe \bZ_{\smash{\alpha(\dal\dbe)}}
+ \tD^\beta {\tilde \bD}{}^\dbe X_{\smash{\beta\alpha\dal\dbe}}
+ {\tilde \bD}{}^\dbe \tD^\beta {\bar X}_{\smash{\beta\alpha\dal\dbe}}  \cr
={}& \tD^\beta Z'{}_{\!(\beta\alpha)\dal} 
+ {\tilde \bD}{}^\dbe \bZ'{}_{\!\smash{\alpha(\dal\dbe)}} + D_\alpha {\tilde \bD}{}^\dbe
X_{\smash{(\dal\dbe)}} + \bD_\dal \tD{}^\beta {\bar X}_{(\beta\alpha)}
+ D_\alpha \bD_\dal X + \bD_\dal D_\alpha {\bar X} \, , \cr}
}
where the second line is obtained by decomposing $X_{\smash{\beta\alpha\dal\dbe}}$ and
${\bar X}_{\smash{\beta\alpha\dal\dbe}}$ into irreducible components
and wherever possible absorbing terms into a redefinition of $Z,\bZ$. 
With this particular form in \inv\ we find
\eqn\diffJS{
{} - i \int \! \d^8 z \,  h_L{}^{\!\! a} \big (J_a - \bJ_a  \big ) = 
4 \int \! \d^8 z \, \big ( \bom^{\dal\dbe} X_{\smash{(\dal\dbe)}}
- \omega^{\beta\alpha} {\bar X}_{\smash{(\beta\alpha)}} \big ) \, .
}
In order to ensure such terms are absent we must further require
\eqn\Xc{
{\bar X}_{\smash{(\beta\alpha)}} = \tD{}^\gamma Y_{(\gamma\beta\alpha)}
+ D_{(\beta}Y_{\alpha)} + {\tilde \bD}{}^\dbe Y_{\smash{\beta\alpha\dbe}}  \, , \quad
Y_{\smash{\alpha\beta\dbe}} = Y_{\smash{\beta\alpha\dbe}}  \, ,
}
and similarly for $X_{\smash{(\dal\dbe)}}$. However we then find
\eqn\XY{
\bD_\dal \tD{}^\beta {\bar X}_{(\beta\alpha)} = - {\ts{3\over 2}} \,
\bD_\dal  D_\alpha \big ( \tD{}^\beta Y_\beta\big ) 
- \tD{}^\beta \bD^2 Y_{\smash{\beta\alpha\dal}}
+ {\tilde \bD}{}^\dbe \big ( {\ts{4\over 3}} \bD_{(\dal} \tD^\beta 
Y_{\smash{\beta\alpha|\dbe)}} + 2 \tD^\beta \bD_{(\dal} Y_{\smash{\beta\alpha|\dbe)}}
\big ) \, .
}
Thus these terms may be removed by a further redefinition of $Z, \, \bZ$ and also
$\bar X$ in \diffJ. Hence, taking account of the freedom in \arbJ, we may therefore
in general write, for suitable $X,{\bar X}$,
\eqn\diffJX{
\J_{\alpha\dal} - {\bar \J}_{\alpha\dal} \sim 
\half [D_\alpha, \bD_\dal]( X - {\bar X}) -i\, \pr_{\alpha\dal}(X+ {\bar X}) \, .
}
{}From the trace of the formulae in \sih\ we may find
\eqn\diff{\eqalign{
i \pr_{\alpha\dal}\big ( \th{}^{\dal\alpha} + {\tilde \brh}{}^{\dal\alpha} \big )
= {}& - {\ts{1\over 6}} [D_\alpha, \bD_\dal] \big ( \th{}^{\dal\alpha} -
{\tilde \brh}{}^{\dal\alpha} \big ) - 16i \big ( \si_h^{\vphantom g} + \bsi_\bh
\big ) \, , \cr
\half  [D_\alpha, \bD_\dal] \big ( \th{}^{\dal\alpha} + 
{\tilde \brh}{}^{\dal\alpha} \big ) = {}& - 3 i \pr_{\alpha\dal}\big ( 
\th{}^{\dal\alpha} - {\tilde \brh}{}^{\dal\alpha} \big ) - 48 i 
\big ( \si_h^{\vphantom g} - \bsi_\bh \big ) \, , \cr}
}
and then using this with \diffJX\ allows us to obtain finally
\eqnn\deTh
$$\eqalignno{
\de_{h,\bh} S = {}&\quar i \int \! \d^8 z \, \big ( \th{}^{\dal\alpha} -
{\tilde \brh}{}^{\dal\alpha} \big ) \T_{\alpha\dal}
+ 4 \int \! \d^8 z \, 
\big ( \si_h^{\vphantom g} (2X- {\bar X} ) - \bsi_\bh (X - 2{\bar X} ) \big ) \cr
{}& \ + \int \! \d^6 z_+ \, \si_h^{\vphantom g} \cJ 
+ \int \! \d^6 z_- \, \bsi_{\bar h} {\bar \cJ} \cr
= {}& - \half i \int \! \d^8 z \, \big ( h^a - {\bar h}^a \big ) T_a
+ \int \! \d^6 z_+ \, \si_h^{\vphantom g} \cT 
+ \int \! \d^6 z_- \, \bsi_{\bar h} {\bar \cT} \, , &
\deTh \cr}
$$
where the supercurrent is now given by
\eqn\dT{
\T_{\alpha\dal} = \half \big ( \J_{\alpha\dal} + {\bar \J}_{\alpha\dal} \big )
- {\ts {1\over 12}} [D_\alpha, \bD_\dal] ( X + {\bar X}) 
+ 3i \, \pr_{\alpha\dal}(X - {\bar X}) \, ,
}
and, using the chiral properties \chsi\ of $\si_h^{\vphantom g} , \, \bsi_\bh$,
\eqn\chT{
\cT = \cJ - \bD{}^2 (2X- {\bar X} ) \, , \qquad {\bar \cT} = {\bar \cJ} +
D^2 (X - 2{\bar X} ) \, .
}
The implicit definition \deTh\ does not determine the supercurrent uniquely since
if
\eqn\varT{
\T_{\alpha\dal} \to \T_{\alpha\dal} + D_\alpha \bD_\dal {\bar \S} -
\bD_\dal  D_\alpha  \S \, , \qquad D_\alpha  {\bar \S} = 0 \, , \quad \bD_\dal
\S = 0 \, ,
}
then this can be compensated in  \deTh\ by taking
\eqn\varTT{
\cT \to \cT + {\ts {3\over 2}} \bD^2 {\bar \S} \, , \qquad
 {\bar \cT} \to  {\bar \cT} + {\ts {3\over 2}} D^2 \S \, .
}

To obtain the conservation equations it is convenient as usual to solve \Dhh\ in
terms of unconstrained prepotentials $L^\alpha, \, \bL^\dal$ where
\eqn\defL{
\th{}^{\dal\alpha} = 2{\tilde \bD}{}^\dal L^\alpha \, , \ \
{\tilde \brh}{}^{\dal\alpha} = 2\tD{}^\alpha \bL^\dal \quad \Rightarrow \quad 
\si_h^{\vphantom g} = {\ts{1\over 24}} i \, \bD{}^2 D_\alpha  L^\alpha \, , \ \
\bsi_{\bar h} = - {\ts{1\over 24}} i \, D^2 \bD_\dal  \bL^\dal \, .
}
Varying $L^\alpha, \, \bL^\dal$ in \deTh\ now gives
\eqn\consT{
\tD^\alpha \T_{\alpha\dal} =   {\ts{1\over 3}}\, \bD_\dal {\bar \cT} \, , \qquad
{\tilde \bD}{}^\dal \T_{\alpha\dal} =  {\ts{1\over 3}}\, D_\alpha \cT  \, .
}
For superconformal invariance the variation in \deTh\ must vanish when \hh\
is satisfied, when $\si_h^{\vphantom g} = \si , \, \bsi_{\bar h} = \bsi$
as given by \el\ and \form. Thus \deTh\ should be independent of $\si_h^{\vphantom g}, \,
\bsi_{\bar h}$ or $\cT = {\bar \cT} = 0$, for a suitable choice of $\S,{\bar \S}$
in \varTT, in this case and then \consT\ reduces to the
superconformal covariant equation \cons.

Although the above considerations are classical we can use them as previously in
\WardL\ to obtain the corresponding quantum field theory Ward identity, assuming
superconformal invariance extends to the quantum theory, in the form
\eqn\WardT{
\half  \int \! \d^8 z \, \big ( h^a(z) - {\bar h}^a(z) \big ) \big \l T_a(z) 
\dots O(z_r) \dots \big \r +
\sum_O \big \l  \dots \de_{h,{\bar h}}O(z_r) \dots \big \r = 0 \, .
}
For quasi-primary superfields the definition of $\de_{h,{\bar h}}O$ need not be
unique but it should reduce to \supertr\ in the superconformal limit, $h={\bar h}$,
when \WardT\ becomes to just the requirement of superconformal covariance of the
correlation function $\l  \dots O(z_r) \dots \big \r$.

For subsequent applications we apply these results to the trivial
cases of free field theories. For chiral scalar fields $\phi(z_+),\,
\bph(z_-)$ we take
\eqn\Sphi{
S = \int \! \d^8 z \, \bph \phi \, ,
}
and the fields are supposed to transform as
\eqn\tranphi{
\de_h^{\vphantom g}\phi = - \L_+ \phi - 2q\,  \si_h^{\vphantom g} \phi \, , \qquad
\de_\bh \bph = - \L_- \bph - 2 \bq \, \bsi_\bh \, \bph \, ,
}
with $\L_\pm$ given in \chL. The variation of \Sphi\ can then be written as in 
\deSh\ with
\eqn\JJ{\eqalign{
\J_{\alpha\dal} ={}& \half D_\alpha \phi \, \bD_\dal \bph - i \pr_{\alpha\dal} \phi \,
\bph \, , \qquad {\bar\J}_{\alpha\dal}  = \half D_\alpha \phi \, \bD_\dal \bph + i 
\phi \, \pr_{\alpha\dal} \bph \, , \cr 
\cJ ={}& \half q \, D^2(\phi \bph) \, , \qquad \qquad \qquad \quad \ \
{\bar \cJ} = \half \bq \, \bD^2 (\phi \bph) \, . }
}
Clearly the difference is of the required form given by \diffJX\ with 
$X= {\bar X} = \half \phi \bph$.
Hence from \chT\ we have $\cT = {\bar \cT}=0$ if $q=\bq=1$ and the supercurrent
for this theory becomes
\eqn\Tphi{
\T_{\alpha\dal} = {\ts{1\over 3}} \big (  D_\alpha \phi \, \bD_\dal \bph
+ 2i \, \phi \, \olr {\pr}_{\alpha\dal} \bph \big ) \, .
}

The other example of a trivial superconformal theory is formed by the abelian
gauge theory with action
\eqn\SV{
S = {1\over 4} \int \! \d^6 z_+ \, W^2 +{1\over 4} \int \! \d^6 z_- \bW{}^2 \, ,
\qquad W_\alpha = - \quar \bD^2 D_\alpha V \, , \quad
\bW_\dal = - \quar D^2 \bD_\dal V \, .
}
The action of superdiffeomorphisms on the scalar gauge superfield $V$ is taken as
\eqn\tranV{
\de_{h,\bh} V = - \big ( \half (h^a + \bh^a ) \pr_a + \lambda^\alpha D_\alpha 
+ {\tilde\bla}_\dal {\tilde\bD}{}^\dal\big ) V  - {\ts {1\over 8}}i
\big ( \th{}^{\dal\alpha} - {\tilde \brh}{}^{\dal\alpha} \big ) 
[D_\alpha, \bD_\dal] V \, .
}
This clearly reduces to the general form for a superconformal transformation
\supertr\ when $h={\bar h}$ and has the important property that it preserves
gauge transformations,
\eqn\gaugeV{
\de V = -i\half (\ep - \bep) \ \ \Rightarrow \ \
\de_{h,\bh} \de V =  i\half ( \L_+ \ep - \L_- \bep) \, ,
\qquad \bD_\dal \ep =0 \, , \ D_\alpha \bep = 0 \, .
}
{}From this and using \sih\ the chiral fields $W, \bW$ transform as
\eqn\tranW{\eqalign{
\de_{h,\bh} W_\alpha = {}& - \L_+  W_\alpha +
\omega_{h}^{\vphantom g}{}_\alpha{}^\beta W_\beta 
- 3\si_h^{\vphantom g} W_\alpha
- {\ts {1\over 8}}i \ep_{\alpha\beta}\bD^2 \big ( \bW_{\smash\dbe}
( \th{}^{\dbe\beta} - {\tilde \brh}{}^{\dbe\beta}) \big ) \, ,\cr
\de_{h,\bh} \bW_\dal = {}& - \L_- \bW_\dal  - \bW_{\smash\dbe}
\bom_{\bh}{}^{\! \dbe}{}_{\dal} - 3 \bsi_{\bar h}  \bW_{\smash\dbe}
- {\ts {1\over 8}}i \ep_{{\smash{\dal\dbe}}} D^2  \big ( (
\th{}^{\dbe\beta} - {\tilde \brh}{}^{\dbe\beta}) W_\beta \big ) \, , \cr}
}
which for $h={\bar h}$ automatically give that $W, \, \bW$ are 
$(\half,{3\over 2})_+, \, (\half,{3\over 2})_-$ superconformal superfields. From
\tranW\ we easily find
\eqn\tranWW{\eqalign{
\de_{h,\bh} W^2 = {}& - \big ( \L_+ + 6 \si_h^{\vphantom g} \big ) W^2 
+ \quar i \bD^2  \big ( ( \th{}^{\dbe\beta} - {\tilde \brh}{}^{\dbe\beta}) W_\beta
\bW_{\smash\dbe} \big ) \, , \cr
\de_{h,\bh} \bW{}^2 = {}& - \big ( \L_- + 6 \bsi_{\bar h} \big ) \bW{}^2
+ \quar i D^2  \big ( ( \th{}^{\dbe\beta} - {\tilde \brh}{}^{\dbe\beta}) W_\beta
\bW_{\smash\dbe} \big ) \, ,\cr}
}
and applying this in \SV\ gives the required form in \deTh, with $\cT = {\bar \cT}
=0$, where
\eqn\TV{
\T_{\alpha\dal} = - 2 W_\alpha \bW_\dal \, .
}

\newsec{Chiral Superfields}

The general results in section 3 simplify significantly if they are applied
for cases involving chiral superfields so we consider these first. We denote
a $(0,q)_+$ chiral scalar by $\phi(z_+)$ and its anti-chiral $(0,\bq)_-$ partner,
where $q=\bq$, by $\bph(z_-)$. From \OO\ the associated two point function
is simply
\eqn\phiphi{
\l \phi(z_{1+}) \bph(z_{2-}) \r = C_\phi \, {1\over x_{\2 1}{}^{\! 2q}} \, .
}
For two chiral scalar superfields and an anti-chiral superfield we may write
the three point function from \OOO\ as
\eqn\phithree{
\l \phi_1(z_{1+}) \phi_2(z_{2+})\bph_3(z_{3-}) \r  = C_{12\3} \,
{1\over x_{\3 1}{}^{\! 2q_1} x_{\3 2}{}^{\! 2q_2}} \, , \qquad q_1+q_2 = \bq_3 \, .
}
As in \OPE\ this leads to the operator product
\eqn\OPEphi{
\phi_1(z_{1+}) \phi_2(z_{2+}) \sim {1\over C_\phi}C_{12\3} \, \phi_3(z_{2+}) \, ,
\qquad q_1+q_2 = q_3 \, ,
}
without any singularities as $z_1\to z_2$ so that the chiral scalar fields
form a closed algebraic ring.

Results for two or three chiral fields are only possible in special
cases. For the two point function we may write the conformally invariant
form \TMPf,
\eqn\phph{
\l \phi_1(z_{1+}) \phi_2(z_{2+}) \r = C_{12} \, \de^4(x_{1+} - x_{2+}) \,
\theta_{12}{}^{\! 2} \, , \qquad q_1 + q_2 = 3 \, ,
}
but this is a pure contact term and should be removable by suitable counterterms
in the effective action. For three chiral scalar fields we may write from \OOO\
(in \homt\ $a=\ba-1 = q_3 -2$)
\eqn\phphph{\eqalign{
\l \phi_1(z_{1+}) \phi_2(z_{2+})  \phi_3(z_{3+})\r ={}&  C_{123} \,
{1\over x_{\3 1}{}^{\! 2q_1} x_{\3 2}{}^{\! 2q_2}} \, X_3{}^{\!2(q_3-2)} 
\bTh_3{}^{\! 2} \, , \cr
={}&  C_{123} \,
{1\over x_{\1 2}{}^{\! 2q_2} x_{\1 3}{}^{\! 2q_3}} \, X_1{}^{\!2(q_1-2)}
\bTh_1{}^{\! 2} \, , \quad q_1+q_2+q_3=3\, ,\cr}
}
where consistency depends on the condition $\sum_i q_i =3$ 
and in the second line we have transformed to the alternative form given by 
\OOOt\ using \IIt. The chirality properties
are not manifest in \phphph\ but in the second line the form of $X_1{}^{\! 2}$
in \Xsq\ and of $\bTh_1$ in \defTh\ demonstrate that this expression depends
only on $z_{3+}$ and also since $f(X_1) \bTh_1{}^{\! 2} = 
f(\Xb_1) \bTh_1{}^{\! 2} $ on $z_{2+}$ whereas similar arguments
from the first line of \phphph\ demonstrate that it also depends only on
$z_{1+}$. We later give an equivalent expression in which the chirality
properties are obvious but the conformal properties are less evident.\foot{A
similar but not apparently identical form was given in \Con, see also \Pick.}
Corresponding to \phphph\ we have an operator product,
\eqn\OPEch{
\phi_1(z_{1+}) \phi_2(z_{2+}) \sim - {C_{123} \over C_{\phi_3}} \,
{\theta_{12}^{\vphantom g}{}^{\! 2} \over
\big ( (x_{1+} - x_{2+})^2 \big )^{2 - \bq_3}} \, \bph_3 (z_{2-}) \, , \quad
\bq_3 = 3 -q_1 - q_2 \, .
}

For three point functions involving the supercurrent and chiral scalar fields
we may write from \OOO\ the unique expression
\eqna\phiT$$\eqalignno{
\l \T_{\alpha\dal}(z_1) \phi(z_{2+}) \bph(z_{3-}) \r = {}& - i A \,
{(\x_{1\3})_{\smash{\alpha\dbe}} \, (\x_{3\1})_{\beta\dal} \over \big (
x_{\3 1}{}^{\! 2}\, x_{\1 3}{}^{\! 2}\big )^2 }\, {1\over x_{\3 2}{}^{\! 2q}} \,
{\tX_3^{\, \dbe \beta} \over \big ( X_3^{\, 2} \big )^2 }  & \phiT a \cr
= {}&  - i A \, {1 \over  x_{\1 2}{}^{\! 2q}  x_{\3 1}{}^{\! 2q}} \,  
{ \bX_{1\alpha \dal} \over {\bar X}_1{}^{\up{\! 2(q-1)}}} \, . & \phiT b\cr }
$$
where \phiT{b}\ follows directly or from \OOOt\  with the general result
\IIt\ using $a=\ba = - {3\over 2}$
and taking $t^{\dal\alpha}(X) = i A \, \tX^{\dal\alpha} /(X^2)^2$. This
satisfies the hermeticity condition $t^{\dal\alpha}(X)^\dagger =
t^{\dal\alpha}(-\Xb)$. From the definition of ${\bar X}_1$ in \defX, \phiT{b}\ 
clearly depends only on $z_{2+}, z_{3-}$ as required from the form of the 
l.h.s. With the aid of \phiT{a,b}\ we may easily find the leading
contribution to the operator product expansion for the supercurrent and a chiral
scalar field,
\eqn\OphiT{
 \T_{\alpha\dal}(z_1) \phi(z_{2+}) \sim i {A\over C_\phi} \, 
{(\x_{2\1})_{\alpha\dal} \over \big( x_{\1 2}{}^{\! 2} \big )^2} \, \phi(z_{2+}) \, .
}
To obtain the corresponding Ward identity we may take from \tranphi\ with \defL
\eqn\tranphiL{
\de_h \phi = \quar i \, \bD^2 \big ( L^\alpha D_\alpha \phi \big ) - 
q \, {\ts{1\over 12}} i \, ( \bD^2  D_\alpha L^\alpha ) \, \phi \, ,
}
and using this in \WardT\ gives
\eqn\WardTphi{\eqalign{
{\tilde \bD}_1{}^{\!\dal}& \l  \T_{\alpha\dal}(z_1) \phi(z_{2+}) \bph(z_{3-}) \r \cr
& {} = 
{\ts {2\over 3}}i q \, D_{1\alpha} \de_+^6(z_1-z_2)\, \l  \phi(z_{2+}) \bph(z_{3-}) \r
+ 2i \, \de_+^6(z_1-z_2) \, D_{2\alpha}  \l  \phi(z_{2+}) \bph(z_{3-}) \r \, , \cr}
}
where the chiral delta function is 
\eqn\chde{
\de_+^6(z_1-z_2) = \de^4( x_{1+} - x_{2+} )\,  
\theta_{12}^{\vphantom g}{}^{\! 2} \, .
}
In the next section we show how \phiT{a}\ satisfies the condition that the r.h.s.
of \WardTphi\ is zero for $z_1 \ne z_2$. The delta functions appearing in
the Ward identity \WardTphi\ arise from the singularities in \phiT{a,b}\ for
$z_1 \sim z_2$. The first term on the r.h.s. of \WardTphi\ is thus generated
from the leading singular term in the operator product expansion \OphiT.
The action of the derivative may be calculated by
\eqn\derivS{
{\tilde \bD}_1{}^{\!\dal} {(\x_{2\1})_{\alpha\dal} \over x_{\1 2}{}^{\! 2\lambda}}
= 4i(2-\lambda){1\over x_{\1 2}{}^{\! 2\lambda}} (\tth_{12})_\alpha 
\limsub{{\lambda\to 2}} 4\pi^2 \de^4(x_{\1 2})  (\tth_{12})_\alpha 
= 2\pi^2  D_{1\alpha} \de_+^6(z_1-z_2) \, ,
}
using the result that, as as distribution on $\bR^4$,
$(x^2)^{-\lambda}$ has a pole as $\lambda \to 2$ with a residue which is
proportional to $\de^4(x)$. Using \derivS\ with \OphiT\ in  \WardTphi\ we must 
then require for consistency
\eqn\AW{
{A\over C_\phi} = {1\over 3\pi^2}\,  q \, .
}
Thus the Ward identity determines completely the overall coefficient of the three
point function \phiT{a,b}\ involving chiral scalar fields and the supercurrent.

\newsec{Ward Identities and Correlation Functions}

If the general results of section 3 are applied to correlation functions
involving the current superfield $L$ or the supercurrent $T_a$ then it
is in general necessary to impose restrictions in order to satisfy the
conservation equations \consL\ and \consT\ at non coincident points.
Furthermore the Ward identities \WardL\ and \WardT\ lead to relations
between a three point function containing $L$ or $T_a$ and the associated
two point function without them, as exemplified in \AW.

To obtain simple results for the action of derivatives on three point functions
we first exhibit how  covariant spinor derivatives act on functions of 
$X_3 , \Theta_3 , \bTh_3$ by writing the conformally covariant formulae
\eqn\DX{\eqalign{
{\tilde \bD}_1{}^{\!\dal} f(X_3 , \Theta_3 , \bTh_3) = {}&
-i{1 \over x_{\3 1}{}^{\! 2}}\,
(\tx_{\1 3})^{\dal \alpha} \D_{3 \alpha}  f(X_3 , \Theta_3 , \bTh_3) \, , \cr
D_{1 \alpha}  f(X_3 , \Theta_3 , \bTh_3) = {}& - i{1 \over x_{\1 3}{}^{\! 2}}\,
(\x_{1\3})_{\alpha \dal} {\tilde \bcD}_3 {}^{\!\dal} f(X_3 ,\Theta_3 ,\bTh_3) \, ,\cr}
}
where, for $X_3 , \Theta_3 , \bTh_3 \longrightarrow X, \Theta, \bTh$ and
$\D_3 , \bcD_3 \longrightarrow \D , \bcD$,
\eqn\calD{
\D_\alpha = {\pr \over \pr \Theta^\alpha} - 2i (\si^a \bTh)_\alpha
{\pr \over \pr X^a} \, , \qquad \bcD_\dal  = - {\pr \over \pr \bTh^\dal} \, .
}
With these definitions and from the relation \XTh\ 
$\bar X = X + 2i\Theta \si \bTh$ it is easy to verify that 
$\D_\alpha {\bar X}{}^a=0$ which is in accord with \DX\ 
since $\bD_{1 \dal} {\bar X}_3 = 0$. From \DX\ we may then find
\eqn\Dcon{\eqalign{
{\tilde \bD}_1{}^{\!\dal}\bigg ( {1 \over ( x_{\1 3}{}^{\! 2} )^2 } 
(\x_{3\1})_{\alpha\dal}F^\alpha(X_3 , \Theta_3 , \bTh_3) \bigg )
= {}& - i {1\over x_{\3 1}{}^{\! 2} \, x_{\1 3}{}^{\! 2}} \,  \D_{3 \alpha}
F^\alpha(X_3 , \Theta_3 , \bTh_3) \, , \cr
{\tD}_1{}^{\!\alpha} \bigg ( {1 \over ( x_{\3 1}{}^{\! 2} )^2 }
(\x_{1\3})_{\alpha\dal} {\bar F}{}^\dal (X_3 , \Theta_3 , \bTh_3) \bigg ) = {}&
- i {1\over x_{\3 1}{}^{\! 2} \, x_{\1 3}{}^{\! 2}} \, \bcD_{3 \dal}
{\bar F}{}^\dal (X_3 , \Theta_3 , \bTh_3) \, , \cr}
}
and also
\eqn\Dsqcon{
\bD_1{}^{\!2} \bigg ( {1\over  x_{\1 3}{}^{\! 2} } f(X_3 , \Theta_3 , \bTh_3)
\bigg ) =  {1\over (x_{\3 1}{}^{\! 2})^2} \, \D_3{}^{\!2} f(X_3 , \Theta_3 , 
\bTh_3) \, ,
}
with a similar formula involving $D_1{}^{\!2}$.

With these results it is straightforward to check that \phiT{a}\ satisfies the 
requirement from \WardTphi\ that
${\tilde \bD}_1{}^{\!\dal},\tD_1{}^{\!\alpha}
 \l  \T_{\alpha\dal}(z_1) \phi(z_{2+}) \bph(z_{3-}) \r=0$
at least for $z_1 \ne z_2,z_3$ since applying \Dcon\ in this case requires only
that
\eqn\Dconphi{
\D_\alpha \, {\tX^{\dal\alpha} \over (X^2)^2} = 0 \, , \qquad
\bcD_\dal \, {\tX^{\dal\alpha} \over (X^2)^2} = 0 \, ,
}
which are easily verified.

We now apply the general result to the three point function for the internal 
symmetry current scalar superfield $L_i$, where $i$ is a group index.
Applying \OOO, with $q_L = \bq_L = 1$, gives
\eqn\LLL{
\l L_i(z_1) L_j(z_2) L_k (z_3) \r = {1\over 
x_{\3 1}{}^{\! 2} \, x_{\1 3}{}^{\! 2}\, x_{\3 2}{}^{\! 2} \, x_{\2 3}{}^{\! 2}}
\, t_{ijk}(X_3, {\bar X_3}) \, ,
}
and $t_{ijk}(X,{\bar X})$ is homogeneous
\eqn\homtL{
t_{ijk}(\rho X,\rho {\bar X}) = \rho^{-2} t_{ijk}(X,{\bar X}) \, , 
}
and satisfies the symmetry relations
\eqn\symtL{
t_{ijk}(X,{\bar X}) = t_{jik}(-{\bar X}, -X) = t_{jki}(X^I, {\bar X}^I) \, . 
}
Applying the conservation equation \consL\ leads from \Dsqcon\ to
\eqn\constL{
\D^2 t_{ijk}(X,{\bar X}) = {\bar \D}{}^2 t_{ijk}(X,{\bar X}) = 0 \, .
}
The solution of these conditions is straightforward\foot{This demonstrates
that $\N=1$ superconformal invariance leads to unique totally symmetric
or antisymmetric expressions for the three point functions of conserved 
currents, as was conjectured earlier \DZ.}
\eqn\soltL{
t_{ijk}(X,{\bar X}) = C_f \, if_{ijk}\bigg ( {1\over X^2} - {1\over {\bar X}^2}
\bigg ) + C_d \, d_{ijk} \bigg ( {1\over X^2} + {1\over {\bar X}^2} \bigg ) \, ,
}
where $f_{ijk},\, d_{ijk}$ are totally antisymmetric, symmetric group tensors.

We analyse first the contribution involving $f_{ijk}$ when \LLL\ and \soltL\
give
\eqn\LLLf{
\l L_i(z_1) L_j(z_2) L_k (z_3) \r_f = C_f \, if_{ijk}\bigg ( 
{1\over x_{\1 3}{}^{\! 2}\, x_{\3 2}{}^{\! 2} \, x_{\2 1}{}^{\! 2} }
- {1\over x_{\3 1}{}^{\! 2} \,  x_{\2 3}{}^{\! 2} \,  x_{\1 2}{}^{\! 2}}
\bigg ) \, .
}
To obtain Ward identities we assume that under infinitesimal group
transformations as considered in \deSL,\WardL\ 
\eqn\varL{
\de_{\ep,\bep} L_i = - f_{ijk} \half (\ep_j + \bep_j ) L_k + 
i(\ep_j -  \bep_j ) K_{ij} \, .
}
and then \WardL, assuming $\l K_{ij} L_k \r = 0$, gives
\eqn\WardLLL{\eqalign{
\quar {\bar D}_1{}^{\! 2} & \l L_i(z_1) L_j(z_2) L_k (z_3) \r \cr
&{}+ \half f_{ij\ell}\de_+^6(z_1-z_2) \l L_\ell(z_2) L_k(z_3) \r  + \half 
f_{ik\ell}\de_+^6(z_1-z_3) \l L_j(z_2) L_\ell (z_3) \r  =0 \, , \cr}
}
with a similar equation involving $D_1{}^{\! 2}$. From \LL\ we may take
\eqn\LLij{
\l L_i(z_1) L_j(z_2) \r = C_L \, \de_{ij}
{1\over x_{\2 1}{}^{\! 2} \, x_{\1 2}{}^{\! 2}} \, ,
}
and using the counterpart to \derivS,
\eqn\dist{
{\bar D}_1{}^{\! 2}{1\over x_{\1 2}{}^{\! 2}}= -i16\pi^2 \de_+^6(z_1-z_2) \, ,
}
it is easy to see that \LLLf\ is compatible with \WardLLL\ and \LLij\ if
\eqn\Cf{
8\pi^2 C_f = C_L \, .
}

The part of the three point function involving $d_{ijk}$ may be written as
\eqn\LLLd{\eqalign{ \!\!\!\!
& \l L_i(z_1)  L_j(z_2) L_k (z_3) \r_d = C_d \, d_{ijk}\bigg (
{1\over x_{\1 3}{}^{\! 2}\, x_{\3 2}{}^{\! 2} \, x_{\2 1}{}^{\! 2} }
+ {1\over x_{\3 1}{}^{\! 2} \,  x_{\2 3}{}^{\! 2} \,  x_{\1 2}{}^{\! 2}}
\bigg ) \cr
&\!\!\!\!\! - 4\pi^2 i C_d \, d_{ijk} \bigg ( \de^8(z_1 - z_2) 
{1\over x_{\3 2}{}^{\! 2}  x_{\2 3}{}^{\! 2}} + \de^8(z_2 - z_3) 
{1\over x_{\1 3}{}^{\! 2}  x_{\3 1}{}^{\! 2}} + \de^8(z_3 - z_1) 
{1\over x_{\2 1}{}^{\! 2}  x_{\1 2}{}^{\! 2}} \bigg ) \, , \cr}
}
with
\eqn\deltafu{
\de^8(z_1 - z_2) = \de^4(x_1 - x_2)\,  \theta_{12}^{\vphantom g}{}^{\! 2} \,
\bth_{12}^{\vphantom g}{}^{\! 2} \, .
}
Since $-\quar {\bar D}_1{}^{\! 2} \de^8(z_1 - z_2) = \de_+^6(z_1-z_2)$, the
chiral $\de$-function defined in \chde, the
second line of \LLLd\ removes terms involving a single delta function
from ${\bar D}_1{}^{\! 2} \l L_i(z_1)  L_j(z_2) L_k (z_3) \r_d$, such
as were present in \WardLLL.
The potentiality of introducing such contact terms to impose
the conservation equations is a reflection of the ambiguities in the
precise definition of this three point function as a distribution
arising from the singular behaviour at coincident points. 

A more careful consideration reveals the supersymmetric
counterpart of the well known axial anomalies when calculating the action
of ${\bar D}_1{}^{\! 2}$ on \LLLd.
To demonstrate the necessary presence of such anomalies
in the present formalism and to determine their form we make use of the
representation \trivL\ to impose the conservation equations \consL\
trivially at $z_1,z_2$. Thus, suppressing group indices so that $d_{ijk}\to 1$,
\eqn\FFL{
\l L(z_1)  L(z_2) L(z_3) \r = \tD_2{}^{\!\beta}\tD_1{}^{\!\alpha}
\Gamma_{\alpha\beta}(z_{1+},z_{2+},z_3) + \Gamma_{\rm loc}(z_1,z_2,z_3)\, ,
}
where $ \Gamma_{\rm loc}(z_1,z_2,z_3)$ is a purely local contact term
which is necessary to ensure that the representation for the three point 
function for $L$ in \FFL\ is symmetric. Assuming superconformal invariance
$\Gamma_{\alpha\beta}(z_{1+},z_{2+},z_3)
=-\Gamma_{\beta\alpha}(z_{2+},z_{1+},z_3)$ is then determined by requiring
it  to be a three point function of the general form in  \OOO\ with 
${q_1=q_2 = {3\over 2}}, \, {\bq_1=\bq_2 =0}$ which  gives in this case
\eqn\GamFFL{
\Gamma_{\alpha\beta}(z_{1+},z_{2+},z_3) =  C_d \, 
{(\x_{1\3})_{\smash{\alpha\dal}} \,(\x_{2\3})_{\smash{\beta\dbe}} 
\over \big ( x_{\3 1}{}^{\! 2} \, x_{\3 2}{}^{\! 2} \big )^2}\, \vep^{\dal\dbe}
{\bTh_3{}^{\! 2} \over 2X_3{}^{\up{\! 2}}}  \, , \qquad
{\bTh_3{}^{\! 2} \over X_3{}^{\up{\! 2}}} = 
{\bTh_3{}^{\! 2} \over {\bar X}_3{}^{\up{\! 2}}} \, .
}
The overall coefficient is chosen so that, using \Dcon, \FFL\ gives the
symmetric form
\eqn\LLLpr{
\l L(z_1)  L(z_2) L(z_3) \r = C_d \, \bigg (
{1\over x_{\1 3}{}^{\! 2}\, x_{\3 2}{}^{\! 2} \, x_{\2 1}{}^{\! 2} }
+ {1\over x_{\3 1}{}^{\! 2} \,  x_{\2 3}{}^{\! 2} \,  x_{\1 2}{}^{\! 2}}
\bigg ) \, ,
}
at non coincident points. The singularities which appear in the expression
\GamFFL\ for $\Gamma_{\alpha\beta}(z_{1+},z_{2+},z_3)$
at coincident points are integrable and hence possible ambiguities proportional
to derivatives of $\de$-functions, which arise for \LLLpr, are not present. 
By its construction the representation \FFL\ ensures that anomalies arising
from the first term on the r.h.s. are
confined to the action of ${\bar D}_3{}^{\! 2}$ and $D_3{}^{\! 2}$.
To obtain the anomalies explicitly it is convenient to 
rewrite \GamFFL\ in the alternative form \OOOt
\eqn\GamFFL{
\Gamma_{\alpha\beta}(z_{1+},z_{2+},z_3) = -  C_d \, 
{(\x_{2\1})_{\smash{\beta\dbe}} \over \big ( x_{\1 2}{}^{\! 2} \big )^2} \,
{1 \over x_{\1 3}{}^{\! 2} \, x_{\3 1}{}^{\! 2}} \, 
\vep^{\dal\dbe} \, \bX_{1 \alpha \dal}
{\bTh_1{}^{\! 2} \over 2 \big ( {\bar X}_1^{\, 2} \big )^2} \, ,
}
so that, using the analogous result to \Dsqcon\ for $D_3{}^{\! 2}$, reduces
calculating the action of $D_3{}^{\! 2}$ to
${\bar \D}_1{}^{\! 2}\bTh_1{}^{\! 2}  = - 4 $ and hence
\eqn\DconFFT{
D_3{}^{\! 2} \Gamma_{\alpha\beta}(z_{1+},z_{2+},z_3) = - 2C_d\,
{(\x_{1\3})_{\smash{\alpha\dal}} \,(\x_{2\3})_{\smash{\beta\dbe}}
\over \big ( x_{\3 1}{}^{\! 2} \, x_{\3 2}{}^{\! 2} \big )^2}\, \vep^{\dal\dbe}
\, .
}
With the aid of
\eqn\derivR{
{\tD}_1{}^{\!\alpha} {(\x_{1\3})_{\alpha\dal} \over \big ( x_{\3 1}{}^{\! 2}
\big )^2} = 2\pi^2  \bD_{1\dal} \de_-^6(z_1-z_3) \, ,
}
and similarly for ${\tD}_2{}^{\!\beta}$, \FFL\ gives
\eqn\anomD{
D_3{}^{\! 2} \big ( \tD_2{}^{\!\beta}\tD_1{}^{\!\alpha}
\Gamma_{\alpha\beta}(z_{1+},z_{2+},z_3)\big ) = 8\pi^4 C_d \, 
\bD_{1\dal} \de_-^6(z_1-z_3) \, {\tilde\bD}_2{}^{\!\dal} \de_-^6(z_2-z_3) \, .
}

A similar calculation for ${\bar D}_3{}^{\! 2} 
\Gamma_{\alpha\beta}(z_{1+},z_{2+},z_3)$ naively gives zero but in this
case it is necessary to be more careful in the treatment of singularities
at coincident points. If we modify \FFL\ to
\eqn\GamFFLreg{
\!\! \Gamma_{\alpha\beta}(z_{1+},z_{2+},z_3)_{\lambda_1,\lambda_2} = C_d \,
{(\x_{1\3})_{\smash{\alpha\dal}} \,(\x_{2\3})_{\smash{\beta\dbe}}
\over  x_{\3 1}{}^{\up{\! 2(2+\lambda_1)}} \, 
x_{\3 2}{}^{\up{\! 2(2+\lambda_2)}} }\, \vep^{\dal\dbe}
{\bTh_3{}^{\! 2} \over 2X_3{}^{\up{\! 2}}}  \, ,
}
then
\eqn\derivreg{
{\bar D}_3{}^{\!2}\Gamma_{\alpha\beta}(z_{1+},z_{2+},z_3)_{\lambda_1,\lambda_2}
= 8 C_d (\lambda_1+\lambda_2)(\lambda_1+\lambda_2+1)
\theta_{13}^{\vphantom g}{}^{\! 2} \theta_{23}^{\vphantom g}{}^{\! 2}\,
{(\x_{13})_{\alpha\dal} (\x_{23})_{\smash{\beta\dbe}}\,\vep^{\dal\dbe}\over
x_{12}{}^{\up{\! 2}} x_{13}{}^{\up{\! 2(2+\lambda_1)}} 
x_{23}{}^{\up{\! 2(2+\lambda_1)}}} \, ,
}
where here $x_{12}= x_{1+} - x_{2+}$. As $\lambda_1,\lambda_2\to 0$ the
factor on the r.h.s. of \derivreg\ depending on $x_{12},x_{13},x_{23}$ 
generates a pole in $\lambda_1+\lambda_2$ with a residue $
\propto \de^4(x_{13})\de^4(x_{23})$ so that
\eqn\derivD{
{\bar D}_3{}^{\!2}\Gamma_{\alpha\beta}(z_{1+},z_{2+},z_3)
= 8\pi^4 C_d \, \vep_{\alpha\beta} \, \de_+^6(z_1-z_3) \de_+^6(z_2-z_3) \, ,
}
and hence, similar to \anomD,
\eqn\anomDbar{
{\bar D}_3{}^{\! 2} \big ( \tD_2{}^{\!\beta}\tD_1{}^{\!\alpha}
\Gamma_{\alpha\beta}(z_{1+},z_{2+},z_3)\big ) =  8\pi^4 C_d \,
\tD_1{}^{\!\alpha} \de_+^6(z_1-z_3)\, D_{2\alpha} \de_+^6(z_2-z_3) \, .
}

To obtain a suitable expression for $\Gamma_{\rm loc}(z_1,z_2,z_3)$ we first
define
\eqn\deff{\eqalign{
16&  f(z_1,z_2,z_3) \cr
&{} = \bD_{3\dal}\de^8(z_3-z_1) \, D_3{}^{\!2}
{\tilde \bD}_3{}^{\!\dal} \de^8(z_3-z_2) + \tD_3{}^{\!\alpha} \big (
\de^8(z_3-z_1) \, {\bD}_3{}^{\!2} D_{3\alpha} \de^8(z_3-z_2) \big ) \cr
{}& =  \bD_{3\dal} \big ( \de^8(z_3-z_1) \, D_3{}^{\!2} 
{\tilde \bD}_3{}^{\!\dal} \de^8(z_3-z_2) \big )
+ \tD_3{}^{\!\alpha} \de^8(z_3-z_1) \,
{\bD}_3{}^{\!2} D_{3\alpha} \de^8(z_3-z_2) \, . \cr}
}
which has the properties
\eqn\fprop{\eqalign{
D_3{}^{\! 2}  f(z_1,z_2,z_3) ={}&
\bD_{1\dal} \de_-^6(z_1-z_3) \, {\tilde\bD}_2{}^{\!\dal} \de_-^6(z_2-z_3)\, ,\cr
{\bar D}_3{}^{\! 2}f(z_1,z_2,z_3) = {}&
\tD_1{}^{\!\alpha} \de_+^6(z_1-z_3)\, D_{2\alpha} \de_+^6(z_2-z_3) \, , \cr
D_2{}^{\! 2}  f(z_1,z_2,z_3) ={}&{\bar D}_2{}^{\! 2}f(z_1,z_2,z_3) = 0 \, ,
\qquad  f(z_1,z_2,z_3) = - f(z_3,z_2,z_1) \, . \cr}
}
If we take
\eqn\loc{
\Gamma_{\rm loc}(z_1,z_2,z_3) = - {\ts{8\over 3}}\pi^4 C_d \big (
f(z_1,z_2,z_3) + f(z_2,z_1,z_3) \big ) \, ,
}
then \fprop\  and \anomD, \anomDbar\ give
\eqn\anom{\eqalign{
D_3{}^{\! 2} \l L(z_1)  L(z_2) L(z_3) \r ={}& {\ts{8\over 3}}\pi^4 C_d\,
\bD_{1\dal} \de_-^6(z_1-z_3) \, {\tilde\bD}_2{}^{\!\dal} \de_-^6(z_2-z_3)\, ,\cr
{\bar D}_3{}^{\! 2} \l L(z_1)  L(z_2) L(z_3) \r ={}& {\ts{8\over 3}}\pi^4 C_d\,
\tD_1{}^{\!\alpha} \de_+^6(z_1-z_3)\, D_{2\alpha} \de_+^6(z_2-z_3) \, , \cr}
}
and also the corresponding results required by symmetry of
$\l L(z_1)  L(z_2) L(z_3) \r$.

If an external real superfield $V$ is coupled to $L$ through an additional
term in the action $S_V = 2\int \! \d^8 z \, LV$ then the results in
\anom\ can be summarised through the operator equations
\eqn\anomL{
\bD^2 \l L \r_V^{\vphantom g} = - {\ts{16\over 3}}\pi^4 C_d\, W^2 \, , \qquad 
D^2 \l L \r_V^{\vphantom g} = - {\ts{16\over 3}} \pi^4 C_d\, \bW{}^2 \, ,
}
where $W,\bW$ are as in \SV. For the associated current $J_a$, defined
by $\J_{\alpha\dal}= - \half [D_\alpha, \bD_\dal] L$, then \anom\ gives
\eqn\anomJ{
\pr_a \l J^a \r_V^{\vphantom g} = {\ts{1\over 16}}i [D^2,\bD^2] 
\l L \r_V^{\vphantom g} = 
- {\ts{1\over 3}} \pi^4 C_d i\,( D^2 W^2 - \bD^2 \bW{}^2 ) \, ,
}
which reduces to the standard form for the anomaly  of the axial current
in a $U(1)$ gauge field background.

\newsec{Supercurrent Correlation Functions}

We here apply the general results of section 3 to a couple of particular non
trivial cases involving the supercurrent. First
we consider the three point function involving two supercurrents and a scalar 
superfield with $q=\bq$. Adapting the general form \OOO\ to this case we have
\eqn\TTO{
\l \T_{\alpha\dal}(z_1) \T_{\smash{\beta\dbe}}(z_2) O(z_3) \r = 
{(\x_{1\3})_{\smash{\alpha\dga}} \, (\x_{3\1})_{\gamma\dal} \,
(\x_{2\3})_{\smash{\beta\dde}} \,  (\x_{3\2})_{\smash{\de\dbe}} \over \big (
x_{\3 1}{}^{\! 2} \, x_{\1 3}{}^{\! 2} \, 
x_{\3 2}{}^{\! 2} \, x_{\2 3}{}^{\! 2} \big )^2  } \, t^{\dga\gamma,\dde\de}
(X_3, {\bar X_3}) \, .
}
It remains to determine the form of $t_{ab}(X,\Xb)= \quar (\si_a)_{\alpha\dal}
(\si_b)_{\smash{\beta\dbe}} t^{\dal\alpha,\dbe\beta}(X,\Xb)$, using
4-vector notation for
convenience, which is homogeneous of degree $2(q-3)$. From the invariance of
\TTO\ under $z_1 \leftrightarrow z_2$, when $X_3 \leftrightarrow - \Xb_3$,
and $\alpha\dal \leftrightarrow \beta\dbe$ this satisfies the symmetry
condition
\eqn\symt{
t_{ab}(X,\Xb) = t_{ba}(-\Xb,-X) \, ,
}
and also the reality constraint
\eqn\realt{
t_{ab}(X,\Xb)^* = t_{ab}(\Xb,X) \, .
}
A general form compatible with \symt\ and \realt\ is
\eqn\tab{\eqalign{
t_{ab}(X,\Xb) ={}& {\eta_{ab}\over (X{\cdot \Xb})^{\up{3-q}}}\bigg (A + B \, {P^2\over
X{\cdot \Xb}}\bigg )  + X_{(a}\Xb_{b)} \, {1\over (X{\cdot \Xb})^{\up{4-q}}} 
\bigg ( C + D \, {P^2\over X{\cdot \Xb}} \bigg ) \cr
{}& + E\, i \ep_{abcd}X^c \Xb^d \, {1\over (X{\cdot \Xb})^{\up{4-q}}} \, , \cr}
}
where $A,B,C,D,E$ are real coefficients and we have defined
\eqn\defP{
\Xb_a - X_a = i P_a \, , \qquad P_a P_b = \quar \, \eta_{ab} P^2 \, , \quad
P^2 = - 8 \Theta^2 \bTh^2 \, .
}
Using the results in \Dcon\ and \calD\ the conservation equation following from
applying ${\tilde \bD}_1{}^{\!\dal}$ to \TTO\ leads to
\eqn\cont{
\Theta \si^c \tsi^a {\pr \over \pr \Xb^c} \, t_{ab}(X,\Xb) = 0 \, .
}
The terms resulting from \cont\ which are ${\rm O}(P^0)$ give
\eqn\ACE{
C = - {3-q \over {\ts{3\over 2}}-q  } \, A \, , \qquad E = - \half (1-q) C \, ,
}
while, using $\Theta P^2 = 0$ and $\Theta P_a = \Theta^2 {\tilde \bTh} \tsi_a$,
the ${\rm O}(P)$ terms determine $B$ and $D$,
\eqn\BD{
B =  {\ts{1\over 8}}(3-q)(4-q)\, A \, , \qquad
D =  {\ts{1\over 8}}(3-q)(4-q)\, C \, .
}
The remaining conservation equations follow automatically as a consequence of the
symmetry and reality conditions \symt\ and \realt. Thus the three point function
\TTO\ is uniquely determined up to an overall constant although there are no
Ward identities in this case which allow the constant to be determined.

It is also convenient to rewrite the result \TTO\ in the form given by \OOOt\ so
that
\eqn\TTOt{
\l \T_{\alpha\dal}(z_1) \T_{\smash{\beta\dbe}}(z_2) O(z_3) \r = -
{(\x_{2\1})_{\smash{\beta\dga}} \, (\x_{1\2})_{\smash{\gamma\dbe}} \over \big (
x_{\1 2}{}^{\! 2} \, x_{\2 1}{}^{\! 2} \big )^2  } \, 
{1\over \big ( x_{\3 2}{}^{\! 2} \, x_{\1 3}{}^{\! 2} \big )^q  } \,
{\tilde t}_{\alpha\dal,}{}^{\dga\gamma} (X_1, {\bar X_1}) \, .
}
To obtain ${\tilde t}_{ab}(X,\Xb)$ we first use  \invt\ with the result \IX\
for $X^I,\Xb^I$ and $\det I = -1$ to give
\eqn\tbar{
{\bar t}_{ab}(X^I,\Xb^I) = I_a{}^c (\Xb,X) I_b{}^d (\Xb,X) \, t_{cd} (X,\Xb )
= t_{ab}(X,\Xb) \, ,
}
and then applying \IIt\
\eqn\tilt{
{\tilde t}_{ab}(X,\Xb) = {1\over (X^2 \Xb^2)^{\up{q-{3\over2}}}}\, 
I^c{}_b(\Xb,X) \, t_{ac}(X,\Xb) \, .
}
Using the explicit form for $I_{ab}$,
\eqn\Iex{
I_{ab}(\Xb,X) = {1\over (X^2 \Xb^2)^{\up{{1\over 2}}}}\big ( \eta_{ab} \, X{\cdot \Xb}
- 2 X_{(a}\Xb_{b)}
- i \ep_{abcd}X^c \Xb^d \big ) \, ,
}
and
\eqn\XbX{
{1\over (X^2 \Xb^2)^{\up \rho}} = {1\over (X{\cdot \Xb})^{\up {2\rho}}} \bigg (
1 + {\ts{3\over 4}}\rho \, {P^2 \over X{\cdot \Xb}} \bigg ) \, ,
}
we find that if
\eqn\tiltab{\eqalign{
{\tilde t}_{ab}(X,\Xb) ={}& 
{\eta_{ab}\over (X{\cdot \Xb})^{\up{q}}}\bigg ({\tilde A} + {\tilde B} \, {P^2\over
X{\cdot \Xb}}\bigg )  + X_{(a}\Xb_{b)} \, {1\over (X{\cdot \Xb})^{\up{q+1}}}
\bigg ( {\tilde C} + {\tilde D} \, {P^2\over X{\cdot \Xb}} \bigg ) \cr
{}& + {\tilde E}\, i\ep_{abcd}X^c \Xb^d \, {1\over (X{\cdot \Xb})^{\up{q+1}}}\, , \cr}
}
then \tilt\ implies
\eqn\ACEt{
{\tilde A} = A \, , \quad {\tilde C} = - C - 2A = - {q\over q - \ts{3\over 2}} A\, ,
\quad {\tilde E} = E - A = \half (q-2) {\tilde C} \, ,
}
and
\eqn\BDt{\eqalign{
{\tilde B} = {}& B +{\ts{3\over 4}}(q-1)A + {\ts{1\over 8}} C - \half E =
{\ts{1\over 8}} q(q+1) A \, , \cr
{\tilde D}  = {}& - D - \quar(3q-2)C - 2B - {\ts{3\over 2}}(q-1)A + \half E =
{\ts{1\over 8}} q(q+1) {\tilde C} \, . \cr}
}
The results \ACEt\ and \BDt\ are similar in form to \ACE\ and \BD\ with
$q \leftrightarrow 3-q$ which is necessary for the analogous conservation
equation to \cont\
to be satisfied by ${\tilde t}_{ab}(X,\Xb)$. When $q=0$, ${\tilde t}_{ab} =
A \eta_{ab}$, or ${\tilde t}_{\alpha\dal}{}^{\dbe\beta} = - 2A \,
\de_\alpha{}^{\! \beta}\de^\dbe{}_{\!\dal}$,
and \TTOt\ reduces to the form in \TT, as expected since
$O$ is then the  identity operator. We may also verify that \tiltab, with
${\tilde B},{\tilde D}$ determined by \BDt, obeys
$\D^2 {\tilde t}_{ab}(X,\Xb) = -4\bTh^2 \pr_X{}^{\!2} {\tilde t}_{ab}(X,\Xb) =0$
which, for $q=1$ and using the corresponding equation to \Dsqcon, ensures that 
\TTOt\ satisfies the conservation equations in \consL\ which become necessary
if we let $O \to L$.

Using the general formula \OPE\ the results for the three point function \TTO\
are equivalent to determination of the coefficient of the contribution of
the scalar superfield $O$ to the operator product expansion of two supercurrents
\eqn\OPETTO{
T_a(z_1) T_b(z_2) \sim {1\over C_O} \, {\bar t}_{ab}(x_{2\1}, x_{\2 1}) \,
O(z_2) \, ,
}
where ${\bar t}_{ab}$ is defined by $\tbar$ and is given by the same
solution of the constraints as $t_{ab}$ but with $E \to -E$.

Following a similar analysis we turn to the more intricate case of the
three point function of the supercurrent by itself.\foot{This case was
also investigated in \Park\ but with different conclusions.}
The general result \OOO\
now requires
\eqn\TTT{
\l \T_{\alpha\dal}(z_1) \T_{\smash{\beta\dbe}}(z_2) 
\T_{\smash{\gamma\dga}}(z_3)\r 
= {(\x_{1\3})_{\smash{\alpha\dep}} \, (\x_{3\1})_{\ep\dal} \,
(\x_{2\3})_{\smash{\beta\dta}} \,  (\x_{3\2})_{\smash{\eta\dbe}} \over \big (
x_{\3 1}{}^{\! 2} \, x_{\1 3}{}^{\! 2} \,
x_{\3 2}{}^{\! 2} \, x_{\2 3}{}^{\! 2} \big )^2  } \, t^{\dep\epsilon,\dta\eta}
{}_{\! \smash{, \gamma\dga}} (X_3, {\bar X_3}) \, ,
}
where it remains to determine $t_{abc}(X,\Xb) =
- {1\over 8}(\si_a)_{\alpha\dal}(\si_b)_{\smash{\beta\dbe}}
(\tsi_c)^{\dga\gamma}t^{\dal\alpha,\dbe\beta}{}_{\! \smash{, \gamma\dga}}
(X,\Xb)$, which is homogeneous of degree $-3$. 
As well as $P_a$ given by \defP\ it is convenient to define also
\eqn\defQs{
Q_a = \half (\Xb_a + X_a ) \, ,
}
so that under inversion following \IX\ they transform as
\eqn\invQP{
Q_{a}{}^{\!\! I} = I_a{}^b(\Xb,X) Q_b = - Q_a \, , \qquad
P_{a}{}^{\!\! I} = I_a{}^b(\Xb,X) P_b = P_a - 2{Q{\cdot P}\over Q^2} \, Q_a \, 
}
Assuming the symmetry condition
\eqn\tsym{
t_{abc}(X,\Xb) = t_{bac}(-\Xb,-X) \, ,
}
we can write a general expression, depending on 9 coefficients, for it as
\eqnn\tgen
$$\eqalignno{ \!\!\!
t_{abc}(X,\Xb &){}  = {1\over (X{\cdot \Xb})^2} \, \ep_{abcd} Q^d
\bigg (A + B \, {P^2\over X{\cdot \Xb}}\bigg ) 
+  {1\over (X{\cdot \Xb})^2}\, \eta_{ab}\bigg( C P_c + D \, {P{\cdot Q}
\over (X{\cdot \Xb})} \, Q_c \bigg ) \cr
{}& + {1\over (X{\cdot \Xb})^2} \bigg( E P_{(a} + F  \, {P{\cdot Q}
\over (X{\cdot \Xb})} \, Q_{(a}  \bigg ) \eta_{b)c} 
+  {1\over (X{\cdot \Xb})^3} \, Q_a Q_b \bigg( G P_c + H \, {P{\cdot Q}
\over (X{\cdot \Xb})} \, Q_c \bigg ) \cr
{}& +  {1\over (X{\cdot \Xb})^3} \, J \big ( Q_a P_b + Q_b P_a \big ) Q_c \, . 
& \tgen \cr}
$$
{}From \invt\ we let
\eqn\tprime{
t'{}_{\! abc}(X,\Xb ) =
{\bar t}_{abc}(X^I,\Xb^I) = I_a{}^e (\Xb,X) I_b{}^f (\Xb,X) I_c{}^g (\Xb,X) \, 
t_{efg} (X,\Xb ) \, ,
}
where, using \invQP, ${\bar t}_{abc}(X,\Xb ) = t_{bac}(X,\Xb ) $ and
$t'{}_{\! abc}(X,\Xb )$ has the same form as
$t_{abc}(X,\Xb)$ in \tgen\ but with
\eqn\AA{ \eqalign{
& (A',B',C',E',G',J')=(A,B,C,E,G,J) \, , \cr &  D' = - D -2C \, , \quad
F'= -F - 2E \, , \quad H' = - H -2G -4J \, . \cr}
}
Since the three point function in \TTT\ is totally symmetric we must now
impose, in addition to \tsym, by virtue of \OOOt\ and \IIt,
\eqn\tsymm{
t_{bca}(X,\Xb) = I^e{}_b(\Xb,X) t'{}_{\! aec}(X,\Xb ) \, .
}
By explicit calculation
\eqnn\Itp
$$\eqalignno{
 I^e{}_b(\Xb,X)& t'{}_{\! aec}(X,\Xb ) = 
{1\over (X{\cdot \Xb})^2} \,  \ep_{abcd} Q^d \bigg ( A' + \big (
B' + \quar A' - \quar C' + {\ts{1\over 8}} E ' \big ) 
{P^2\over X{\cdot \Xb}}\bigg ) \cr
{}& + {1\over (X{\cdot \Xb})^2}\, \big ( (C'-A')  \eta_{ab}  P_c 
+ \half E' \eta_{ac} P_b + (A'+\half E') \eta_{bc} P_a \big ) \cr
{}& + {1\over (X{\cdot \Xb})^3}\, P{\cdot Q} \,\big ( (A'+D')  \eta_{ab}  Q_c
- (E'+\half F') \eta_{ac} Q_b - (A'-\half F') \eta_{bc} Q_a \big ) \cr
{}& + {1\over (X{\cdot \Xb})^3}\, \big ( (A'-2C'-G') Q_a Q_b P_c + J'
Q_a Q_c P_b - ( A' + E' + J' ) Q_b Q_c P_a \big ) \cr
{}& - {1\over (X{\cdot \Xb})^4}\,  P{\cdot Q} \,Q_a Q_b  Q_c \big ( 2D' + F ' 
+ H' + 2J' \big ) \, , & \Itp \cr }
$$
and then, using \AA, it is easy to read off the conditions necessary to
satisfy \tsymm
\eqn\AAsym{
E = 2(C-A) \, , \qquad G+J = D+ \half F = A-2C \, .
}

For the conservation equations it is sufficient to impose just
\eqn\contt{
\Theta \si^e \tsi^a {\pr \over \pr \Xb^e} \, t_{abc}(X,\Xb) = 0 \, .
}
Inserting \tgen\ the equations split into those which are ${\rm O}(P^0)$,
\eqn\conPzero{
E= 2(C-A) \, , \quad F = -2C-5E \, , \quad G = -2A + \half F \, , \quad
J = 2A + D \, , \quad H = -2G - 6J \, ,
}
and also those which arise from terms which are  ${\rm O}(P)$,
\eqn\conPone{
4C + D + 6E + {\ts{3\over 2}} F + G +J = 0 \, , \quad H = 4E + F - 2J \, , \quad
8B = 4A - 4E - F - J \, . 
} 
In fact \conPzero\ and \conPone\ together imply \AAsym\ and there remain two
independent parameters which may be taken as $A,C$ so that we may determine
\eqn\BDEFGHJ{
B = \half A \, , \quad \half D = E = 2(C-A) \, , \quad F = 10A -12 C \, , \quad
G = \half H = -  {\ts{3\over 2}} J = 3 (A-2C) \, .
}
In consequence there are two linearly independent superconformal covariant
forms for the three point
function for the supercurrent \TTT. In contrast for the three point function
of the energy momentum tensor in conformal field theories there are in
general three linearly independent forms \refs{\hughone,\hughtwo}.

The result \tgen\ with \BDEFGHJ\ is not very transparent but it can be
recast more simply as
\eqn\ttau{
t_{abc}(X,\Xb)  = \tau_{abc}(X,\Xb) + \tau_{bac}(-\Xb,-X) \, ,
}
where
\eqn\spec{\eqalign{ \!\!\!\!
\tau_{abc}(X,\Xb) & =  -\half i A {1\over (X^2)^2}\big ( X_a \eta_{bc} +
X_b \eta_{ac} - X_c \eta_{ab} + i \ep_{abcd} X^d \big ) \cr
& \ {} + \half (2C-A) {1\over (X^2)^3} \Big ( 2 (X_a P_b + P_a X_b ) X_c
- 3 X_a X_b P_c - 6 {P{\cdot X}\over X^2} \, X_a X_b X_c \cr
&\  
{}-P{\cdot X}\big ( 3(X_a \eta_{bc} + X_b \eta_{ac}) - 2 X_c \eta_{ab}\big) 
+ \half X^2 \big ( P_a \eta_{bc} + P_b \eta_{ac} + P_c \eta_{ab} \big )\Big )
\, . \cr}
}

As in previous cases we may relate the results for the three point function
to the relevant coefficient in an associated operator product expansion
\eqn\OPETT{
\T_{\alpha\dal}(z_1) \T_{\smash{\beta\dbe}}(z_2) \sim - {1\over 2C_T} \, 
{\bar t}_{\smash{\alpha\dal,\beta\dbe}}{}^{\dga\gamma} (x_{2\1}, x_{\2 1}) \,
\T_{\smash{\gamma\dga}} (z_2) \, ,
}
with, from \tprime,
${\bar t}_{\smash{\alpha\dal,\beta\dbe}}{}^{\dga\gamma} (X,\Xb)
= (\si^a)_{\alpha\dal} (\si^b)_{\smash{\beta\dbe}}
(\tsi^c)^{\dga\gamma} {t}_{bac}(X,\Xb)$. Explicitly,
with a similar decomposition to \ttau,
\eqn\coeff{\eqalign{ \!\!\!\!
{\bar \tau}_{\smash{\alpha\dal,\beta\dbe}}& {}^{\dga\gamma} (X,\Xb) =  
2i A {1\over (X^2)^2}\X_{\beta\dal} \, \de_\alpha{}^{\!\gamma}\de^\dga
{}_{\smash{\! \dbe}} \cr
& \ {} + \half (2C-A) {1\over (X^2)^3} \bigg ( 2 \big (\X_{\alpha\dal} 
\P_{\smash{\beta\dbe}} + \P_{\alpha\dal}\X_{\smash{\beta\dbe}}\big ) 
\tX^{\dga\gamma}
- 3 \X_{\alpha\dal} \X_{\smash{\beta\dbe}}\Big (  {\tilde \P}^{\dga\gamma} 
+ 2 {P{\cdot X}\over X^2} \, \tX^{\dga\gamma} \Big ) \cr
{}& \qquad \qquad \qquad\quad {}
+2\big ( P{\cdot X} \, \X_{\alpha\dal} - X^2 \P_{\alpha\dal} \big )
\de_\beta{}^{\!\gamma}\de^\dga {}_{\smash{\! \dbe}}
+2\big (P{\cdot X}\,\X_{\smash{\beta\dbe}} - X^2 \P_{\smash{\beta\dbe} }\big )
\de_\alpha{}^{\!\gamma}\de^\dga {}_{\smash{\! \dal}} \cr
{}& \qquad \qquad \qquad\quad {}
+\big (4 P{\cdot X} \, \X_{\smash{\alpha\dbe}} 
+ X^2 \P_{\smash{\alpha\dbe}} \big )
\de_\beta{}^{\!\gamma}\de^\dga {}_{\smash{\! \dal}}
+\big (4 P{\cdot X}\,\X_{\smash{\beta\dal}}
+ X^2 \P_{\smash{\beta\dal} }\big )
\de_\alpha{}^{\!\gamma}\de^\dga {}_{\smash{\! \dbe}} \bigg ) 
\, . \cr}
}
The terms in \coeff\ with coefficient $2C-A$ have an ${\rm O}(X^{-4})$ 
singularity but closer analysis shows that $(X^2)^{-\lambda}
{\bar \tau}_{\smash{\alpha\dal,\beta\dbe}}{}^{\dga\gamma} (X,\Xb)$ has
no pole as $\lambda\to 0$ so this is integrable.
The construction of $t_{abc}(X,\Xb)$ guarantees that 
${\bar \tau}_{\smash{\alpha\dal,\beta\dbe}}{}^{\dga\gamma}(x_{2\1}, x_{\2 1})
+{\bar\tau}_{\smash{\beta\dbe,\alpha\dal}}{}^{\dga\gamma}(-x_{\2 1},-x_{2\1})$
satisfies the constraints obtaining from the conservation equations \consT\ 
for $z_1\ne z_2$ (in \coeff\ $P\to - 2 \theta_{21}\si \bth_{21}$ for this
case). To take account of singularities at $z_1 = z_2$ we first note that
\eqn\conTT{
\tD_1{}^{\!\alpha} 
{\bar \tau}_{\smash{\alpha\dal,\beta\dbe}}{}^{\dga\gamma} (x_{2\1}, x_{\2 1})
= 0 \, , \qquad
{\tilde \bD}_2{}^{\!\dbe}{\bar \tau}_{\smash{\alpha\dal,\beta\dbe}}
{}^{\dga\gamma} (x_{2\1}, x_{\2 1}) = 0 \, ,
}
without any $\de$-function contributions. However calculating the action
of ${\tilde \bD}_1{}^{\!\dal}$ requires a more careful treatment. Modifying
the singularity in \coeff\ as in \derivS\ we find
\eqn\conTTbar{\eqalign{
{\tilde \bD}_1{}^{\!\dal} \Big ( {1\over x_{\1 2}{}^{\! 2\lambda}} &
{\bar \tau}_{\smash{\alpha\dal,\beta\dbe}}{}^{\dga\gamma} (x_{2\1}, x_{\2 1})
\Big ) =  8A \, \lambda \,
{1\over (x_{\1 2}{}^{\! 2})^{2+\lambda}} (\tth_{12})_\beta \, \de_\alpha
{}^{\! \gamma} \de^\dga{}_{\smash{\!\dbe}} \cr
& {} - 20i(2C-A) \, \lambda \, {\theta_{12}{}^{\! 2}\over
(x_{\1 2}{}^{\! 2})^{3+\lambda}} \Big ( (\x_{2\1} \bth_{12})_\beta
\de_\alpha {}^{\! \gamma} \de^\dga{}_{\smash{\!\dbe}} + (\x_{2\1})_{\smash
{\alpha\dbe}} \de_\beta {}^{\! \gamma} (\bth_{12})^\dga \cr
& \qquad \qquad \qquad\qquad \qquad \qquad {}
+ {1\over x_{\1 2}{}^{\! 2}} (\x_{2\1} \bth_{12})_\alpha 
(\x_{2\1})_{\smash {\beta\dbe}} (\tx_{\1 2})^{\dga\gamma} \Big ) \, .
\cr}
}
Taking the limit $\lambda\to 0$ then reveals the local contributions
with support when $z_1=z_2$
\eqnn\conTde
$$ \eqalignno{ \!\!\!\!\!\!\!\!\!\!
{\tilde \bD}_1{}^{\!\dal}
{\bar \tau}_{\smash{\alpha\dal,\beta\dbe}}{}^{\dga\gamma}& (x_{2\1}, x_{\2 1})
= 8\pi^2 i A \, (\tth_{12})_\beta \de^4(x_{2\1}) \de_\alpha{}^{\! \gamma} 
\de^\dga{}_{\smash{\!\dbe}} \cr
{}&\qquad \qquad \quad{}  - {\ts {40\over 3}}\pi^2 (2C-A) \,
\theta_{12}{}^{\! 2} \de_{(\alpha}{}^{\!\gamma}\pr_{\smash{2\beta)(\dbe}}
\de^4(x_{2\1}) \de^\dga{}_{\smash{\dde)}} (\bth_{12})^\dde \cr
=   4\pi^2iA & D_{1\beta} \de_+^6(z_1-z_2)  \de_\alpha{}^{\! \gamma}
\de^\dga{}_{\smash{\!\dbe}} +  {\ts {20\over 3}}\pi^2 (2C-A) \,
\de_{(\alpha}{}^{\!\gamma}\pr_{\smash{1\beta)(\dbe}}
\de^\dga{}_{\smash{\dde)}} {\tilde \bD}_1{}^{\!\dde} \de^8(z_1-z_2) \, .  
& \conTde \cr}$$
A similar result may also be derived for $\tD_2{}^{\!\beta}
{\bar \tau}_{\smash{\alpha\dal,\beta\dbe}}{}^{\dga\gamma}(x_{2\1}, x_{\2 1})$.

The association of the supercurrent with superconformal transformations allows
the derivation of Ward identities which constrain one linear combination
of the parameters in ${t}_{abc}$. To derive these from \WardT\
we need to define $\de_{h,{\bar h}} T_a$. In the superconformal 
case, given by \hh, this must reduce to the particular case of \supertr\
appropriate for the supercurrent and therefore, based on particular
examples, we postulate the form\foot{For an alternative approach see
\Johanna.}
\eqn\tranT{\eqalign{
\de_{h,{\bar h}} \T_{\alpha\dal} = {}& - 
\big ( \half (h^a + \bh^a ) \pr_a + \lambda^\alpha D_\alpha
+ {\tilde\bla}_\dal {\tilde\bD}{}^\dal + 3(\si_h^{\vphantom g} +
 \bsi_{\bar h} ) \big )  \T_{\alpha\dal} \cr
{}&  + \omega_{h}^{\vphantom g}{}_\alpha{}^\beta \T_{\beta\dal}
- \T_{\smash{\alpha\dbe}}\,\bom_{\bh}{}^{\! \dbe}{}_{\dal} 
+ \big ( \D (h-{\bar h}) \big )^I{}_{\!\!\! \alpha\dal} O_I \, ,\cr}
}
where $\D$ represents the action of various derivatives and $O_I$ are a basis
of superfield operators in the theory. These terms are model dependent, 
the results for free theories are given in an appendix. Nevertheless the fields
which contribute may include the supercurrent itself so that these terms are
relevant for Ward identities applied to the three point function of
the supercurrent by itself. Using the prepotentials given in \defL\ we have
\eqn\prep{
\omega_{h}^{\vphantom g}{}_\alpha{}^\beta - 3 \de_{\alpha}{}^{\!\beta}
\si_h^{\vphantom g} = - \quar i \, \bD{}^2 D_\alpha  L^\beta \, , \qquad
\bom_{\bh}{}^{\! \dbe}{}_{\dal} + 3 \de^{\dbe}{}_{\! \dal}\,
\bsi_{\bar h} = - {\ts{1\over 4}} i \, D^2 \bD_\dal  \bL^\dbe \, ,
}
we may obtain from \WardT
\eqn\WardTT{\eqalign{
{\tilde \bD}_1{}^{\!\dal} 
\l \T_{\alpha\dal}&(z_1) \T_{\smash{\beta\dbe}}(z_2) \dots \r\cr
{}& = \big ( 2i D_{1\beta} \de^6_+(z_1 - z_2 )\de_\alpha{}^{\! \gamma}
\de^\dga{}_{\smash{\!\dbe}} + {\tilde \bD}_1{}^{\!\dal} 
\chi_{\smash{\alpha\dal,\beta\dbe}}{}^{\dga\gamma} (z_{12}) \big ) 
\l  \T_{\smash{\gamma\dga}}(z_2) \dots \r
+ \dots \, , \cr
{\tD}_1{}^{\!\alpha} 
\l \T_{\alpha\dal}&(z_1) \T_{\smash{\beta\dbe}}(z_2) \dots \r\cr
{}& = \big ( 2i \bD_{\smash{1\dbe}} \de^6_-(z_1 - z_2 )\de_\beta{}^{\! \gamma}
\de^\dga{}_{\smash{\!\dal}} + {\tD}_1{}^{\!\alpha} 
\chi_{\smash{\alpha\dal,\beta\dbe}}{}^{\dga\gamma} (z_{12}) \big ) 
\l  \T_{\smash{\gamma\dga}}(z_2) \dots \r
+ \dots \, ,
\cr}
}
where additional terms representing contributions from other operators in the
correlation function and also less singular terms, involving derivatives
of $\T_{\smash{\gamma\dga}}(z_2)$, are not shown. The terms involving
$\chi_{\smash{\alpha\dal,\beta\dbe}}{}^{\dga\gamma} (z_{12})$, where $z_{12}$
is defined in \yint\ and which
is restricted to be a linear combination constructed
from $\bD_{\smash \dde}D_\de \de^8(z_1-z_2) = -4\, \de^4(x_{1\2}) \tth_{12\de}
{\tilde \bth}_{\smash{12\dde}}$ and $D_\de\bD_{\smash \dde}
\de^8(z_1-z_2) = 4\,\de^4(x_{\1 2})\tth_{12\de}{\tilde \bth}_{\smash{12\dde}}$, 
arise from the model dependent
$h-\bh$ contributions in \tranT. However such terms may also be viewed as
a reflection of the arbitrariness of the operator product coefficient 
${\bar t}_{\smash{\alpha\dal,\beta\dbe}}{}^{\dga\gamma} (x_{2\1}, x_{\2 1})$
in \OPETT\ up to purely local $\de$-function contributions and in
consequence it is therefore possible to redefine it so as to remove the
$\chi_{\smash{\alpha\dal,\beta\dbe}}{}^{\dga\gamma} (z_{12})$ terms from
\WardTT.

To apply the results in \conTT\ and \conTde\ to the Ward identity we
first introduce
\eqn\comph{ \eqalign{
h_{\smash{\alpha\dal,\beta\dbe}}{}^{\dga\gamma}(z_{12})
= {}& h_{\smash{\beta\dbe,\alpha\dal}}{}^{\dga\gamma}(z_{21}) \cr
= {}& [ D_\alpha , \bD_\dal ] \de^8 ( z_1-z_2)\,  \de_\beta{}^{\!\gamma}
\de^\dga{}_{\smash{\!\dbe}} +
[ D_\beta , \bD_{\smash{\dbe}} ] \de^8 ( z_1-z_2)\,  \de_\alpha{}^{\!\gamma}
\de^\dga{}_{\smash{\!\dal}} \cr
{}& - 2 [ D_\alpha ,\bD_{\smash{\dbe}}]\de^8 ( z_1-z_2)\, \de_\beta{}^{\!\gamma}
\de^\dga{}_{\smash{\!\dal}} 
- 2 [ D_\beta , \bD_\dal ] \de^8 ( z_1-z_2)\,  \de_\alpha{}^{\!\gamma}
\de^\dga{}_{\smash{\!\dbe}} \cr
{}& + 6i \big ( \pr_{\beta\dal} \de^8 ( z_1-z_2) \, \de_\alpha{}^{\!\gamma}
\de^\dga{}_{\smash{\!\dbe}} 
- \pr_{\smash{\alpha\dbe}} \de^8 ( z_1-z_2) \,
\de_\beta{}^{\!\gamma} \de^\dga{}_{\smash{\!\dal}} \big ) \, , \cr}
}
which satisfies
\eqn\derivh{ \eqalign{\!\!\!\!\!\!
{\tilde \bD}_1{}^{\!\dal} 
h_{\smash{\alpha\dal,\beta\dbe}}{}^{\dga\gamma}(z_{12}) = {}&
12 D_{1\beta} \de^6_+(z_1 - z_2 )\, \de_\alpha{}^{\! \gamma}
\de^\dga{}_{\smash{\!\dbe}} - 8i 
\de_{(\alpha}{}^{\!\gamma}\pr_{\smash{1\beta)(\dal}}
\de^\dga{}_{\smash{\dbe)}} {\tilde \bD}_1{}^{\!\dal} \de^8(z_1-z_2) \, , \cr
\!\!\!\!\!\! {\tD}_1{}^{\!\alpha} 
h_{\smash{\alpha\dal,\beta\dbe}}{}^{\dga\gamma}(z_{12}) = {}&
12 \bD_{\smash{1 \dbe}} \de^6_- (z_1 - z_2 )\,  \de_\beta{}^{\! \gamma}
\de^\dga{}_{\!\dal}
+ 8i \de_{(\alpha}{}^{\!\gamma}\pr_{\smash{1\beta)(\dal}} 
\de^\dga{}_{\smash{\dbe)}} \tD_1{}^{\!\alpha} \de^8(z_1-z_2) \, . \cr}
}
Redefining the operator product coefficient in \OPETT\ to be
\eqn\OPETTn{\eqalign{
{\bar t}_{\smash{\alpha\dal,\beta\dbe}}{}^{\dga\gamma}(x_{2\1}, x_{\2 1}) ={}&
{\bar \tau}_{\smash{\alpha\dal,\beta\dbe}}{}^{\dga\gamma}(x_{2\1}, x_{\2 1})
+{\bar\tau}_{\smash{\beta\dbe,\alpha\dal}}{}^{\dga\gamma}(-x_{\2 1},-x_{2\1})\cr
{}& - {\ts{5\over 6}}\pi^2 i (2C-A) \,
h_{\smash{\alpha\dal,\beta\dbe}}{}^{\dga\gamma}(z_{12}) \, , \cr}
}
we then obtain
\eqn\OPEd{
{\tilde \bD}_1{}^{\!\dal} 
{\bar t}_{\smash{\alpha\dal,\beta\dbe}}{}^{\dga\gamma}(x_{2\1}, x_{\2 1})
=  2\pi^2 i \big ( 2A - 5(2C-A) \big ) D_{1\beta} \de^6_+(z_1 - z_2 ) \,
\de_\alpha{}^{\! \gamma}\de^\dga{}_{\smash{\!\dbe}} \, .
}
The additional term in \OPETTn\ is equivalent to setting
$\chi_{\smash{\alpha\dal,\beta\dbe}}{}^{\dga\gamma} (z_{12}) = 0$ in
\WardTT\ and hence, using \OPEd, we therefore find from the Ward identity
the constraint
\eqn\constraint{
\pi^2(10C-7A) = 2 C_T \, .
}
Thus the general superconformal three point function for the supercurrent
contains one new parameter beyond the coefficient $C_T$ for the two point
function.

\newsec{Free Fields}

The trivial realisations of superconformal field theories are given by
free fields, i.e. for the chiral scalar superfield theory defined by the
action \Sphi\ or the abelian gauge theory described by \SV. As a consistency
check we give the reduction of some of our general results to these cases.
{}From \Sphi\ the two point function is easily found,
\eqn\twophi{
\l \phi(z_{1+}) \bph(z_{2-})\r = {1\over 4\pi^2 x_{\2 1}{}^{\! 2}} \, .
}
For gauge superfield $V$, and with covariant gauge fixing parameterised by 
$\xi$, the normalisations in \SV\ require
\eqn\twoV{
\l V(z_1) V(z_2) \r = {1\over 16\pi^2} \bigg ( \Big ( 1 - {1\over \xi}
\Big ) \ln y_{12}{}^{\! 2} - \Big ( 1 + {1\over \xi} \Big ) 
{1\over y_{12}{}^{\! 2}} \, \theta_{12}{}^{\! 2} \bth_{12}{}^{\! 2}\bigg ) \, ,
}
where $y_{12}$ is given by \yint. With the definitions in \SV\ we may then
show that
\eqn\twoW{
\l W_\alpha (z_{1+}) \bW_\dal (z_{2-})\r = i \, {(\x_{1\2})_{\alpha\dal}
\over 2\pi^2 (x_{\2 1}{}^{\! 2})^{\up 2}} \, ,
}
while $\l W_\alpha (z_{1+}) W_\beta (z_{2+}) \r = 0$. Clearly \twophi\ and
\twoW\ are in accord with the general superconformal expression \OO.

With these results and the explicit forms \Tphi\ and \TV\ we may evaluate
the coefficient of the two point function for the supercurrent, defined
by \TT, directly to be
\eqn\TTpV{
C_{T,\phi} = {1\over 6\pi^4} \, , \qquad \quad
C_{T,V} = {1\over 2\pi^4} \, .
}
The form required by \TT\ is trivial to obtain in the $V$ case, using
\twoW, but emerges in the $\phi$ case after lengthy calculation.
To obtain the coefficients in the supercurrent three point function it is
sufficient to find the leading contributions to the operator product
$\T_{\alpha\dal}(z_1) \T_{\smash{\beta\dbe}}(z_2)$. This is easy in the
$V$ case since it is evident from \TV\ and \twoW\ that the first
operator term in the short distance expansion is the supercurrent itself.
This then gives
\eqn\ACV{
A_V = 2C_V = -{1\over \pi^2} C_{T,V} \, .
}
In the operator product expansion for two supercurrents formed from free
chiral fields, as in \Tphi, the leading term is $\phi\bph$ and it is
necessary to remove this and its derivatives after using Taylor expansions
in the form
\eqn\taylor{\eqalign{
\phi(z_{2+}) = {}& \phi(z_{1+}) + ( x_{2\1}{\cdot \pr}_1 + 
\theta_{21}{}^{\!\!\alpha}D_{1\alpha} )  \phi(z_{1+}) + \dots\, , \cr
\bph(z_{2-}) = {}& \bph(z_{1-}) + ( x_{\2 1}{\cdot \pr}_1 +
\bth_{21}{}^{\!\!\dal}\bD_{1 \dal} )  \bph(z_{1-}) + \dots\, . \cr}
}
Noting that $\half [D_\alpha, \bD_\dal]\phi\bph =  D_\alpha \phi \, 
\bD_\dal \bph - i \, \phi  \olr {\pr}_{\alpha\dal} \bph $ we then find
\eqn\ACphi{
A_S = {\ts {2\over 5}} C_S = {1\over 9\pi^2} C_{T,\phi} \, .
}
Of course both \ACV\ and \ACphi\ are in accord with the Ward identity 
\constraint.\foot{The results for free fields may be used to relate
the parameters $A,C$ for the general superconformal supercurrent three
point function to those specifying the conformal energy momentum tensor
three point function. In terms of the parameters $r,s,t$ in \hughone\
we have $2r = 29A-9C, \, 4s = -45A + 24 C, \, t=4A$.} 

As an example of a current superfield we consider $L_i = \bph t_i\phi$
for free chiral scalar superfields where $t_i$ are hermitian matrices
obeying the Lie algebra $[t_i , t_j ] = if_{ijk} t_k$ and we take
$\tr(t_i t_j) = T \de_{ij}$. It is very easy to see that this gives the
results \LLij\ and \LLLf\ for the two and three point functions with
\eqn\CL{
C_L = {T\over (4\pi^2)^2} \, , \qquad \quad C_f = {T\over 2(4\pi^2)^3} \, .
}
Manifestly these satisfy \Cf. If $\half \tr(\{t_i,t_j\}t_k) = d_{ijk}$
then the symmetric form \LLLd\ is also given by
\eqn\Cd{
C_d = {1\over (4\pi^2)^3} \, ,
}
and this is appropriate in \anomL\ or \anomJ\ to give the standard one
loop anomaly result.

\newsec{Superconformal Integrals}

It was realised long ago \DEPP\ that integrals such as those appearing in field
theoretic calculations may be significantly simplified in special
cases as a consequence of the restrictions of conformal invariance. As
general discussion was given by Symanzik \Symanzik\ and we extend this
here to particular superconformal examples.

If we define
\eqn\defxx{
x_{i}{}^{\!2} = (x_{i+} -2i \theta_i \si \bth - x)^2 \, ,
}
then the integrations over anti-chiral superspace we consider here are
\eqn\SN{
S_N = i \int \! \d^4 x \d^2 \bth \, \prod_{i=1}^N 
{1\over (x_{i}{}^{\!2})^{\up{q_i}}} \, , \qquad 
\sum_i q_i = 3 \, .
}
Under a superconformal transformation $z\to z'$, $x'{}_{\!i}{}^{2} =
x_{i}{}^{\!2} /{\Omega}(z_i) {\bar \Omega}(z)$, from \vars, while
the measure $\d^6 z'{}_{\! -} = \d^6 z_- /{\bar\Omega}^3(z)$ so that the 
condition on $\sum_i q_i$ in \SN\ defines a superconformal covariant
function. Using the standard result
\eqn\rep{
{1\over (x^2+i\epsilon)^\alpha} = {e^{-i {\pi\over 2}\alpha}\over \Gamma
(\alpha)}\int_0^\infty \!\!\!\!
\d\lambda \, \lambda^{\alpha-1} e^{i\lambda x^2} \, ,
}
then we find
\eqn\SNa{
S_N = {\pi^2 \over \prod_i  \Gamma(q_i)}  \int_0^\infty \!\!\!\! 
{\scriptstyle{\prod}}_{i} \d\lambda_i \lambda_i{}^{\! q_i -1} 
{1\over \Lambda^2}
\int \! \d^2\bth \, e^{-{1\over \Lambda}\sum_{i<j} \lambda_i \lambda_j
X_{ij}^{2}} \, , \quad
X_{ij} = x_{i+}-x_{j+} - 2i \theta_{ij} \si \bth \, ,
}
where $\Lambda = \sum_i \lambda_i$. Expanding the exponential allows the
$\bth$ integration to be performed giving
\eqn\SNb{
S_N = - {\pi^2 \over \prod_i  \Gamma(q_i)}  \int_0^\infty \!\!\!\!
{\scriptstyle{\prod}}_{i} \d\lambda_i \lambda_i{}^{\! q_i -1} 
{1\over \Lambda^2} 
\sum_{jk} {\theta}_j \si{\cdot \pr}_j \, \tsi{\cdot \pr}_k \tth_k \,
e^{-{1\over \Lambda}\sum_{i<j} \lambda_i \lambda_j x_{ij+}^{\, \,2}}  ,
}
for $x_{ij+} = x_{i+}- x_{j+}$.
In \SNa\ and \SNb\ we have assumed that the integrals are initially defined
in a region where $x_{ij+}^{\, \, 2}>0$.

The crucial observation of Symanzik is that in an integral
\eqn\Syn{
\int_0^\infty \!\!\! {\scriptstyle{\prod}}_{i} \d\lambda_i 
\lambda_i{}^{\! \delta_i -1} {1\over\Lambda{}^{\up p}} \,
e^{-{1\over \Lambda}\sum_{i<j} \lambda_i \lambda_j u_{ij}}\, ,
\qquad u_{ij}=u_{ji} \, , \quad \sum_i \delta_i = 2p \, , 
}
it is possible to transform $\Lambda$ to $\Lambda= \sum_i \kappa_i \lambda_i$
with arbitrary $\kappa_i \ge 0 , \, \sum \kappa_i > 0$. This then allows
the choice $\Lambda = \lambda_i$ for some $i$ and the integral, using
contour techniques, written in terms of the conformal invariant cross
ratios $u_{ij}u_{kl}/u_{ik}u_{jl}$. For the case $i=1,2,3$, when there
are no invariants, the integral is easily determined to be
\eqn\three{
{\Gamma(p-\de_1) \Gamma(p-\de_2) \Gamma(p-\de_3) \over 
u_{12}^{\, p-\de_3} \, u_{23}^{\, p-\de_1} \, u_{31}^{\, p-\de_2}} \, .
}

There are various alternative ways of writing \SNb\ in the desired form.
One convenient representation is
\eqn\SNc{\eqalign{
S_N = {4\pi^2 \over \prod_i  \Gamma(q_i)}&  \int_0^\infty \!\!\!\!
{\scriptstyle{\prod}}_{i} \d\lambda_i \lambda_i{}^{\! q_i -1}
{1\over \Lambda^3}\, 
e^{-{1\over \Lambda}\sum_{i<j} \lambda_i \lambda_j x_{ij+}^{\, \, 2}} \cr
& {}\times
\bigg( \sum_{jkl}\lambda_j\lambda_k \lambda_l\, {\theta}_j^{\vphantom g}
\, \x_{jl+}^{\vphantom g}\,  \tx_{lk+}^{\vphantom g}\, \tth_k^{\vphantom g} 
- \half \sum_{jk}  \lambda_j \lambda_k \, x_{jk+}^{\,\,2} \sum_l \lambda_l \,
\theta_l^{\, 2} \bigg) \, , \cr}
}
which has the form required by \Syn\ with $p=3$. For $N>3$ the integral
involves non trivial functions of superconformal invariants but here we
consider just $N=3$ when the result by applying \three\ is
\eqn\SNd{
S_3 = - 4\pi^2 \prod_i{\Gamma(2-q_i)\over \Gamma(q_i)} \,
{{\theta}_{12}^{\vphantom g} \tth_{13}^{\vphantom g}\, 
x_{23+}^{\,\, 2} + {\theta}_{23}^{\vphantom g} \tth_{21}^{\vphantom g}\, 
x_{31+}^{\,\, 2}  + {\theta}_{31}^{\vphantom g} \tth_{32}^{\vphantom g}\,
x_{12+}^{\,\, 2} \over
(x_{12+}^{\,\, 2})^{\up{2-q_3}} ( x_{23+}^{\,\, 2})^{\up{2-q_1}} 
(x_{31+}^{\,\, 2})^{\up{2-q_2}}} \, .
}
It is straightforward to reexpress this in the manifestly superconformal form
expected from \phphph.

\newsec{Superconformal Invariants}

For analysis of higher point correlation functions it is necessary to
understand what conformal invariants may occur. For an $N$-point function
depending on $z_r  \in \bR^{4|4}$, $r=1,\dots N$ we may use supertranslations
to set $z_1=0$ and then superconformal transformations to set
$x_2 = \infty$ (in a suitable compactification) and also $\theta_2, \bth_2=0$.
Superconformal invariants are then given by those scalars formed from
$z_r $, $r=3\dots N$ which are invariant under the residual symmetry group
$O(3,1)\times D \times U(1)_R$, where $D$ denotes the group of scale
transformations. For the $x_r$ coordinates we may consider the
$\half (N-2)(N-1) - 1 = \half N(N-3)$ scalars $x_r {\cdot x}_s / x_3{}^{\!2}$,
$ s\ge r>3 , \, s>r=3$ and with the Grassmann coordinates we may define
the $(N-2)^3$ invariants $\theta_s\, \x_r\, \bth_t / x_r{}^{\! 2}$. However for
$N>6$ the $x_r$ are linearly dependent and we may restrict $r = 3,4,5,6$,
giving $4N-15$ $c$-number invariants (in this case the ordinary
$O(4,2)$ conformal group is acting transitively)
and $4(N-2)^2$ invariants formed from
$\theta, \bth$. When $N=3$ there is clearly one Grassmann invariant which
corresponds to $J$ defined in \invJ.

To proceed we extend the definition of $X$ for three points $z_1,z_2,z_3$
in \defX\ to a similar expression formed from $z_r,z_s,z_t$
\eqn\Xdef{
\X_{r(st)} = {\x_{r\bs}\, \tx_{\bs t}\, \x_{t\br}\over x_{\bs r}{}^{\! 2}
\, x_{\br t}{}^{\! 2}} \, , \qquad
\Xb_{r(st)} = - X_{r(ts)} \, ,
}
and also, extending \defTh,
\eqn\Thdef{
\Theta_{r(st)} = i\bigg ( {1\over x_{\bs r}{}^{\! 2}}\, 
{\tilde\bth}_{rs}\,\tx_{\bs r} - {1\over x_{\bt r}{}^{\! 2}}\, 
{\tilde\bth}_{rt}\,\tx_{\smash{\bt r}}\bigg ) \, , \qquad
{\bTh}_{r(st)} = i\bigg ( {1\over x_{\br s}{}^{\! 2}}\, \tx_{\br s}\,\tth_{rs}
- {1\over x_{\br t}{}^{\! 2}}\, \tx_{\br t}\,\tth_{rt} \bigg ) \, .
}
These functions of $z_r,z_s,z_t$, $r\ne s \ne t$, transform homogeneously at 
$z_r$ according to \varX\ and \varTh. Trivially from \Xdef\ we have
\eqn\diffTh{
\Theta_{r(su)} = \Theta_{r(st)} + \Theta_{r(tu)} \, , \qquad
\bTh_{r(su)} = \bTh_{r(st)} + \bTh_{r(tu)} \, ,
}
and 
\eqn\diffX{
X_{r(su)} = X_{r(st)} + X_{r(tu)} - 2i \Theta_{r(st)} \si \bTh_{r(tu)} \, .
}
As special case \diffX\ reduces for $u=s$ to
$ X_{r(st)} + X_{r(ts)} = - 2i \Theta_{r(st)} \si \bTh_{r(st)}$, 
which is equivalent to \XTh. 

In the limiting situation considered above, $z_1 = (0,0,0) , \, 
z_2 = (\infty, 0,0)$, we have $\X_{1(2r)} = \x_{r+}/x_{r+}^{\, \, 2}, \, 
\Theta_{1(2r)} = i{\tilde \bth}_r \, \tx_{r-}/ x_{r-}^{\, \, 2}, \,
\bTh_{1(2r)} = - i\tx_{r+} \tth_r / x_{r+}^{\, \, 2}$. Hence it is natural
to construct a basis of superconformal invariants in terms of
$X_{1(2r)}, \, \Theta_{1(2r)} , \, \bTh_{1(2r)}$ for $r=3, \dots N$.
Thus we may define a set of bosonic invariants formed in this fashion, in
which the points $z_1,z_2$ play a privileged role, by
\eqn\defu{
u_r = {X_{1(2r)}^{\,\, 2} \over X_{1(23)}^{\,\, 2} }
= \det \big ( \X_{1(23)}{}^{\!\!\! -1} \X_{1(2r)} \big ) \, , \quad r>3 \, ,
}
and
\eqn\defv{
v_{rs} = {X_{1(2r)}\,{\cdot \, X_{1(2s)}} \over X_{1(2r)}^{\,\, 2}} =  \half \,
\tr \big ( \X_{1(2r)}{}^{\!\!\! -1} \X_{1(2s)} \big ) \, , \quad s>r\ge 3  \, .
}
The definitions \defu\ and \defv\ are special cases of an extension to
the superconformal case of the  usual
invariant cross ratios and a related invariant trace given by
\eqn\defuv{
u_{rs,tu} = {x_{\br t}{}^{\! 2}\,
x_{\bs u}{}^{\! 2} \over x_{\br u}{}^{\! 2}\, x_{\bs t}{}^{\! 2} } \, , \qquad
v_{rs,tu} =  \half \, \tr \big ( \tx_{\br t}\, \tx_{\bs t}{}^{\!\!\! -1} 
\tx_{\bs u} \, \tx_{\br u}{}^{\!\!\! -1} \big ) \, ,
}
since it is easy to see, from the definitions \defu\ and \defv\ together
with \Xdef, that $u_r = u_{12,3r} , \, v_{rs} =  v_{12,rs}$. From
\defuv\ it follows that
\eqn\tranuv{
u_{rs,tu} = u_{sr,ut} = {1\over u_{sr,tu}}  \, , \qquad 
v_{rs,tu} = v_{sr,ut} = {v_{sr,tu} \over u_{sr,tu}}  \, ,
}
which may also be derived from the definitions in \defu\ and \defv\ using the
relation $\tx_{\bs r} \X_{r(st)} \tx_{\br s} = \X_{s(rt)}{}^{\!\!\! -1}$
which follows from \Xtran.

Restricting to functions of four points the above discussion also suggests, in
addition  to those in \defu\ and \defv, introducing
an associated set of Grassmann invariants given by 
$Q_{12,rs}, \, {\bar Q}_{12,rs}, \ r,s \ge 3$ where in general we define
\eqn\defQ{
{\bar Q}_{rs,tu}=  4i \Theta_{r(su)} \, \tX_{r(st)}{}^{\!\! -1} \, 
\bTh_{r(st)} \, ,  \qquad 
Q_{rs,tu}=  4i \Theta_{r(st)} \, \tX_{r(ts)}{}^{\!\! -1} \,
\bTh_{r(su)} \,  .
}
Using the transformation relations in \Thtran\ as well as \Xtran\ we have
\eqn\tranQ{
Q_{sr,tu} = {\bar Q}_{rs,ut} \, .
}
From \diffX\ we may show that
\eqn\XQ{
{X_{r(st)}^{\,\, 2} \over X_{r(ts)}^{\,\, 2} }=\big ( 1 + Q_{rs,tt} \big )^{-1}
= 1 + {\bar Q}_{rs,tt} \, ,
}
and, just in \invJ, we may define, using \tranQ\ and $Q_{rs,tt} = 
{\bar Q}_{rt,ss}$,
\eqn\JJ{
J_{rst} = - \half \big ( Q_{rs,tt} - {\bar Q}_{rs,tt} \big ) \, ,
}
as an totally antisymmetric invariant depending on $z_r,z_s,z_t$.

Other invariants formed from four points $z_r,z_s,z_t,z_u$ should be expressible
in general in terms of the basis described above. Alternatively the invariants
$u_{rs,tu}, \, v_{rs,tu}$ and $Q_{rs,tu}, \, {\bar Q}_{rs,tu}$ obey 
various relations, besides those given in \tranuv\ and \tranQ.
Using \diffTh\ and \diffX\ we may obtain
\eqn\uv{
u_{us,tr} = w_{rs,tu} \big ( 1 + {\bar Q}_{ru,ts} \big ) \, , \qquad 
w_{rs,tu} =  1 + u_{rs,tu} - 2v_{rs,tu} \, ,
}
which allows the trace invariants given by \defv\ to
be expressed in terms of the invariant cross ratios as given by \defu, and also
\eqn\Qrel{
1+ {\bar Q}_{rt,su} = \big ( 1 + {\bar Q}_{rs,tu} \big )
\big ( 1 + Q_{rs,tt} \big ) \, .
}
By combining \uv\ with $w_{rs,tu}=w_{sr,ut}$, which follows from \tranuv,
we may obtain
\eqn\uubar{
u_{tu,rs} = u_{rs,tu} {1+  {\bar Q}_{st,tu} \over 1+ {\bar Q}_{ur,ts}}
= u_{rs,tu} {1+  {\bar Q}_{rs,tt} \over 1+  {\bar Q}_{rs,uu}} \, ,
}
which also follows from \XQ\ and \Qrel. This results shows how $u_{rs,tu}$ 
is related to its conjugate ${\bar u}_{rs,tu}=u_{tu,rs}$.

The presence of the Grassmann invariants, such as given by \defQ, clearly
complicates the analysis of superconformal $N$ point functions for $N\ge 4$.
For chiral fields the necessary additional conditions are more restrictive. 
As an illustration we consider a 4 point function  for chiral scalar fields.
If we express it in the form
\eqn\phfour{
\l \phi_1(z_{1+}) \phi_2(z_{2+})  \phi_3(z_{3+}) \phi_4(z_{4+}) \r 
= { X_{1(23)}{}^{\!\!\!2(q_1-2)} \over x_{\1 2}{}^{\! 2q_2} 
x_{\1 3}{}^{\! 2q_3} x_{\1 4}{}^{\! 2q_4} } \, F_{12,34}(z_1,z_2,z_{3+},z_{4+})
\, ,
}
then superconformal invariance, if $\sum_i q_i = 3$, requires
\eqn\fourinv{
F_{12,34}(z'{}_{\!1},z'{}_{\!2},z'{}_{\!3+},z'{}_{\!4+}) =
\Omega(z_{1+})^{-2} {\bar \Omega}(z_{1-}) F_{12,34}(z_1,z_2,z_{3+},z_{4+}) \, .
}
$F_{12,34}$ may be expanded as
\eqn\Ffour{\eqalign{
F_{12,34}& (z_1,z_2,z_{3+},z_{4+}) \cr
= {}& A \bTh_{1(23)}{}^{\!\!\! 2} +  B \bTh_{1(24)}{}^{\!\!\! 2} + 
C {\tilde \bTh}_{1(23)}  \bTh_{1(24)} 
+  D {\tilde \bTh}_{1(23)} \X_{1(23)}{}^{\!\!\!\!\! -1}
\X_{1(24)} \bTh_{1(24)}  \, , \cr}
}
so that, if the coefficients $A,B,C,D$ are functions of just  the two 
invariants $u=u_{12,34}, {w= w_{12,34}}$, \phfour\
and \Ffour\ give a generalisation of the result displayed in \phphph\ for the 3
point function which has the required superconformal transformation properties
and further depends manifestly only on $z_{3+},z_{4+}$.
Interchanging $z_3$ and $z_4$ so that  in the corresponding expression to
\phfour\ we have $F_{12,43}(z_1,z_2,z_{4+},z_{3+})$ it is easy to see that
\eqn\Ftrch{
F_{12,43}(z_1,z_2,z_{4+},z_{3+}) = u^{-q_1+2} 
F_{12,34}(z_1,z_2,z_{3+},z_{4+}) \, .
}
Writing now
\eqn\Ffourtr{\eqalign{
F_{12,43}& (z_1,z_2,z_{4+},z_{3+}) \cr
= {}& {\bar A} \bTh_{1(24)}{}^{\!\!\! 2}+ {\bar B} \bTh_{1(23)}{}^{\!\!\! 2}
+ {\bar C} {\tilde \bTh}_{1(24)}  \bTh_{1(23)}
+ {\bar D} {\tilde \bTh}_{1(24)} \X_{1(24)}{}^{\!\!\!\!\! -1}
\X_{1(23)} \bTh_{1(23)} \, , \cr}
}
then this leads to
\eqn\ABtr{
{\bar A} = u^{-q_1+2} B \, , \quad  {\bar B} = u^{-q_1+2} A \, , \quad  
{\bar C}  = u^{-q_1+2} C \, , \quad   {\bar D} = u^{-q_1+3} D \, .
}
In a similar fashion if $z_1 \leftrightarrow z_2$ then we obtain
$F_{21,34}(z_2,z_1,z_{3+},z_{4+})$ which may be  expressed as
\eqn\Ftr{\eqalign{
F_{21,34}& (z_2,z_1,z_{3+},z_{4+}) \cr
= {}& A' \bTh_{2(13)}{}^{\!\!\! 2} + B' \bTh_{2(14)}{}^{\!\!\! 2} + 
C'  {\tilde \bTh}_{2(13)}  \bTh_{2(14)}
+ D' {\tilde \bTh}_{2(13)} \X_{2(13)}{}^{\!\!\!\!\! -1}
\X_{2(14)} \bTh_{2(14)}\, . \cr}
}
where the necessary relations for compatibility are
\eqn\chir{
A' = Au^{q_4} \, , \quad B'= B u^{q_4+1} \, , \quad C' = D u^{q_4+1} \, , \quad
D'  = C u^{q_4+1} \, .
}

Although in the expression given by
\phfour\ with \Ffour\ only $z_{3+}, z_{4+}$ appear
further conditions are necessary to ensure that the whole result
depends only on $z_{1+}, z_{2+}$. These may be obtained by considering
$z_2 \leftrightarrow z_3$ applied to \phfour,
\eqn\phfourn{
\l \phi_1(z_{1+}) \phi_2(z_{2+})  \phi_3(z_{3+}) \phi_4(z_{4+}) \r 
= { X_{1(32)}{}^{\!\!\!2(q_1-2)} \over x_{\1 2}{}^{\! 2q_2}
x_{\1 3}{}^{\! 2q_3} x_{\1 4}{}^{\! 2q_4} } 
\, E(z_1,z_2,z_3,z_{4+}) \, ,
}
where by  using \XQ\ we have
\eqn\FF{
E(z_1,z_2,z_3,z_{4+}) = 
(1+{\bar Q}_{12,33} )^{q_1-2} F_{12,34}(z_1,z_2,z_{3+},z_{4+})
= F_{13,24}(z_1,z_3,z_{2+},z_{4+}) + G \, ,
}
so that the dependence on $z_{2-}$  is  isolated in $G$.
To obtain an explicit form for $G$ we first rewrite $F_{12,34}$ as
\eqn\Fchtr{\eqalign{
F_{12,34} (z_1,z_2,z_{3+},z_{4+}) 
={}& {\cal A} \, \bTh_{1(32)}{}^{\!\!\! 2} + (B+ {\bar Q}_{12,33} D)
\bTh_{1(34)}{}^{\!\!\! 2} + {\cal C}\,  {\tilde \bTh}_{1(32)}  \bTh_{1(34)} \cr
{}& + (1+{\bar Q}_{12,33} ) D {\tilde \bTh}_{1(32)} \X_{1(32)}{}^{\!\!\!\!\! -1}
\X_{1(34)} \bTh_{1(34)}\, ,\cr}
}
where
\eqn\AC{
{\cal A} =  A + B + C + v D \, , \qquad {\cal C} = - ( 2B + C + D ) \, , \qquad
v= v_{12,34} \, .
}
The invariants $u,w$, on which ${\cal A},B,{\cal C},D$ depend may also be
re-expressed in terms of ${\hat u}=u_{13,24},{\hat w}=w_{13,24}$,
which depend only on $z_{2+}$, using \uv\ with \Qrel\ giving
\eqn\uvhat{
u = {\hat w}\, {1+{\bar Q}_{12,43}\over 1+{\bar Q}_{12,33}}\, , \qquad
w = {\hat u}\, {1\over ( 1+{\bar Q}_{13,42} )(1+{\bar Q}_{12,33})}\, .
}
Hence the dependence on $z_{2-}$ in \FF\ occurs only in terms involving
${\bar Q}_{12,33},{\bar Q}_{12,43},{\bar Q}_{13,42}$. By combining such
terms with \Fchtr\ in \FF\  $G$ may be reduced to the form
\eqn\Gch{\eqalign{
G = {}& \big (\alpha_1 {\bar Q}_{12,43} + \alpha_2 {\bar Q}_{21,43} \big ) 
\bTh_{1(32)}{}^{\!\!\! 2} 
+ \big (\beta_1 {\bar Q}_{12,43} + \beta_2 {\bar Q}_{21,43} \big ) 
\bTh_{1(34)}{}^{\!\!\! 2} \cr
&{} + \gamma \, {1\over X_{1(34)}{}^{\!\! 2}}
\Theta_{1(23)}{}^{\!\!\! 2}\bTh_{1(32)}{}^{\!\!\! 2} 
\bTh_{1(34)}{}^{\!\!\! 2}\, , \cr}
}
where $\alpha_1,\alpha_2,\beta_1,\beta_2,\gamma$ are linear in
${\cal A}, B ,  {\cal C} , D$. In consequence the dependence on $z_{2-}$ may be
eliminated by imposing the conditions that $\alpha_1,\alpha_2,\beta_1,
\beta_2,\gamma$ each vanish and hence in \phfourn\ $E\to F_{13,24}$ given by
\eqn\Ftr{\eqalign{
F_{13,24}& (z_1,z_3,z_{2+},z_{4+}) \cr
= {}& {\cal A} \bTh_{1(32)}{}^{\!\!\! 2} + B \bTh_{1(34)}{}^{\!\!\! 2} +
{\cal C}  {\tilde \bTh}_{1(32)}  \bTh_{1(34)}
+ D {\tilde \bTh}_{1(32)} \X_{1(32)}{}^{\!\!\!\!\! -1}
\X_{1(34)} \bTh_{1(34)}\, , \cr}
}
where in the arguments of ${\cal A}, B ,  {\cal C} , D$ $u,w \to {\hat w},
{\hat u}$. 
In a similar fashion the necessary dependence only on $z_{1+}$
may also be ensured. These conditions require relations between the 
functions $A,B,C,D$ but it is clear from
the results of the previous section  that there is at least
a single arbitrary function of $u,v$ remaining in the general solution.

\newsec{Conclusion}

In the above we have endeavoured to generalise the kinematic analysis in 
\refs{\hughone,\hughtwo} of conformal invariance and its implications in 
quantum field theory in general dimensions on flat space to the simplest case
of $\N=1$ supersymmetry in four dimensions. In the analysis of the two and
three point functions of the energy momentum tensor in four dimensions the 
coefficients which appeared in the two and three point functions (for the
latter there are in general three parameters which may be connected with
the three trivial free conformal field theories in four dimensions) can be
related to the coefficients which appear in the trace of the energy
momentum tensor when a conformal field theory is extended to a curved space
background. Here we describe the connections of the results obtained here
with the similar parameters which may be defined when a superconformal
theory is extended to a minimal $\N=1$ supergravity background.

In this case the theory includes a superfield $H^a(z)$, which contains the
metric, such that the expectation of the energy momentum tensor may
be defined by
\eqn\WT{
\l T_a \r = {\de W \over \de H^a} \, ,
}
where $W$ is the connected vacuum functional for the curved background (in our
conventions the functional integral gives $e^{iW}$). Assuming the theory
is defined to preserve the usual supergravity superspace reparameterisation
invariance we may extend the definitions in \deTh\ to obtain
\eqn\sga{
\half i \int \! \d^8 z  E^{-1}\, \big ( h^a - {\bar h}^a \big )
{\de W \over \de H^a} = 
\int \! \d^6 z_+{\hat \vphi}^3 \, {\hat \si}_h^{\vphantom g} {\hat \cT}
+ \int \! \d^6 z_- {\hat {\bar\vphi}}{}^3 \, 
{\hat {\bsi}}_{\bar h} {\hat {\bar \cT}} \, ,
}
where $\d^8 z  E^{-1}$ and $ \d^6 z_+{\hat \vphi}^3, \, \d^6 z_- 
{\hat {\bar\vphi}}{}^3$ are the appropriate invariant integration measures
on full superspace and its chiral, anti-chiral projections \Buch. 
$ h^a , {\bar h}^a$
satisfy supercovariant generalisations of \Dhh\ while $\cT, \, {\bar \cT}$
are covariantly chiral, anti-chiral scalars formed from $H^a$ (${\hat \cT}, \,
{\hat {\bar \cT}}$ are defined by transformation to a chiral representation
when they depend only on $z_+ , \, z_-$ respectively). $\cT, \, {\bar \cT}$
are formed from the supergravity curvatures,
the chiral superfields $W_{\alpha\beta\gamma}{=W_{(\alpha\beta\gamma)}}, \, R$,
and their anti-chiral conjugates 
${\bar W}_{\smash{\dal\dbe\dga}} , \, {\bar R}$, 
together with the real vector superfield $G_a$, and supercovariant derivatives.
The general form for $\cT$ can be written as \Bon
\eqn\trace{
8\pi^2\, \cT = c \, W^{\alpha\beta\gamma} W_{\alpha\beta\gamma} - a \, G
+ h ( {\bcD}{}^2 - 4R ) \D^2 R \, ,
}
with  $G$ a topological density whose chiral superspace integral is related to
the difference of the Euler and Pontryagin invariants,
\eqn\topden{
G =  W^{\alpha\beta\gamma} W_{\alpha\beta\gamma} - \quar
( {\bcD}{}^2 - 4R ) ( G^a G_a + 2{\bar R} R ) \, ,
}
and where $\D_\alpha, \bcD_\dal$ are supercovariant spinor derivatives.
In \trace\ $h$ is arbitrary since it may be varied at will by
adding a purely local term $\propto \int \! \d^8 z  E^{-1}\, {\bar R} R$
to $W$. The coefficients $c,a$ have a non trivial significance in any
superconformal theory \Gris\ and for $n_S , \, n_V$ free superfields,
as described by actions \Sphi, \SV, we have \Buch
\eqn\ca{
c = {\ts{1\over 24}}( 3n_V + n_S )\, , \qquad 
a  = {\ts{1\over 48}}( 9n_V + n_S ) \, .
}

There is a direct relation between $c,a$ and the parameters $A,C$
specifying the general superconformal supercurrent three point function
as found in section 7. In principle the relation may be found from \sga\
first by obtaining
\eqn\covT{
{\tilde \bcD}{}^\dal \l \T_{\alpha \dal} \r = {\ts {2\over 3}} \D_\alpha
\cT \, , \qquad {\tilde \D}^\alpha \l \T_{\alpha \dal} \r 
 = {\ts {2\over 3}} \bcD_\dal {\bar \cT} \, ,
}
and then taking two functional derivatives with respect to $H$
of both sides and restricting to flat space. This gives contributions to
${\tilde \bD}_1{}^{\!\dal} \l \T_{\alpha\dal}(z_1)
\T_{\smash{\beta\dbe}}(z_2)\T_{\smash{\gamma\dga}}(z_3)\r$
and ${\tD}_1{}^{\!\alpha} \l \T_{\alpha\dal}(z_1) \T_{\smash{\beta\dbe}}(z_2) 
\T_{\smash{\gamma\dga}}(z_3)\r$ which are
proportional to various derivatives acting on $\de^8(z_1-z_2)\de^8(z_1-z_3)$.
With careful regularisation the results from the r.h.s. of \covT,
depending on $c,a,h$ may be matched with results arising from explicit
calculation using the general form \TTT. However such an analysis is not
straightforward (an analogous investigation of the energy momentum tensor
three point function assuming conformal invariance was undertaken in \hughtwo)
although the necessary relations are easy to read off from the
results for free fields \ca\ and \ACphi, \ACV\ with \TTpV. This gives
\eqn\ACac{
A = {8\over 9\pi^6}( 3c-5a) \, , \qquad C = {4\over 9\pi^6} (6c-7a) \, .
}

As a consistency check we verify the relation between $c$ and the
coefficient $C_T$ of the supercurrent two point function. For this
it is sufficient to restrict to constant rescalings when we may take in
\sga\ $ {\hat \si}_h^{\vphantom g} = {\hat {\bsi}}_{\bar h} =1$. With
$\mu$ an arbitrary renormalisation scale we may write
\eqn\Wmu{
\mu{\pr \over \pr \mu} W = \int \! \d^6 z_+{\hat \vphi}^3 \, {\hat \cT}
+ \int \! \d^6 z_- {\hat {\bar\vphi}}{}^3 \, {\hat {\bar \cT}} \, .
}
In this the terms depending on $a$ are topological invariants while $h$
disappears since it is the coefficient of terms which are total derivatives. 
Furthermore the difference between the integrals of 
$W^{\alpha\beta\gamma} W_{\alpha\beta\gamma}$ and
${\bar W}_{\smash{\dal\dbe\dga}}{\bar W}^{\dal\dbe\dga}$ is also a topological
invariant so that we may write from \Wmu\ and \WT
\eqn\TTmu{
\mu{\pr \over \pr \mu} \l \T_{\alpha\dal}(z_1) \T_{\smash{\beta\dbe}}(z_2) \r
= 8c \, {\de^2 \over \de{\rm H}^{\dal\alpha}(z_1) \de {\rm H}^{\dbe \beta}(z_2)}
\int\! \d^6 z_+{\hat \vphi}^3\, W^{\alpha\beta\gamma} W_{\alpha\beta\gamma}
\bigg |_{\rm flat \ space} \, .
}
Using, to lowest order in expansion about flat space,
$\de W_{\alpha\beta\gamma} = - {1\over 16} \bcD{}^2 \D_{(\alpha}
{\tilde \bcD}{}^\dbe \D_\beta \de H_{\smash{\gamma)\dbe}}$ this may be
readily calculated giving
\eqn\TTmud{
\mu{\pr \over \pr \mu} \l \T_{\alpha\dal}(z_1) \T_{\smash{\beta\dbe}}(z_2) \r
= 4c \, \pr_{1\gamma\dal}D_{1\eta} \pr_{\smash{2\de\dbe}} D_{2\ep}
\de^6_+(z_1-z_2) \, \E_\alpha{}^{\gamma\ep}{}_{,\beta}{}^{\de\eta} \, ,
}
with $\E$ defined by \proj. The result \TTmud\ may be compared with that
obtained from the regularised version of \TTt\ using, with the definition
in \reg,
\eqn\regd{
\mu{\pr \over \pr \mu}\R\bigg({\theta_{12}^{\,\, 2}
 \over (x_{12+}^{\,\,\, 2})^{\up 2}}\bigg ) = 2\pi^2 \de^6_+(z_1-z_2) \, .
}
It is then evident that we must have
\eqn\CT{
C_T = {4 \over \pi^4} c \, ,
}
which is compatible with the Ward identity result \constraint\ and \ACac.

\newsec{Note Added}

A further consistency check may be found by reducing the results \TTT\ and \tgen\
to their $\theta,\bth$ independent forms. With $T^a(z)| = R^a(x)$, the $R$-symmetry
current, we have
\eqn\RRR{
\l R^a(x_1) R^b(x_2) R^c(x_3) \r = A \, {I^a{}_{\! e}(x_{13}) I^b{}_{\! f}(x_{23}) \over
\big ( x_{13}{}^{\! 2} \, x_{2 3}{}^{\! 2} \big )^2  } \, i \ep^{efcd} {X_{3 d}
\over \big (X_3{}^{\! 2}  \big )^2 } \, ,
}
where $I^a{}_{\! b}(x) = \de^a{}_{\! b} - 2 x^a x_b / x^2$ is the reduction of the
inversion tensor given by  \Iex.
Using the standard form for the anomaly, such as obtained in \hughtwo\ with symmetrisation
and transforming to Minkowski space,
\eqn\anomR{
\pr_{3c} \l R^a(x_1) R^b(x_2) R^c(x_3) \r = - A\, {\ts{1\over 6}}\pi^4 \pr_{1c} \pr_{2d}
\big ( \ep^{acbd} \de^4(x_{13}) \de^4(x_{23}) \big ) \, .
}
{}From \covT, for a general background, $i\D_a \l T^a\r = {1\over 6}( \D^2 \cT -
\bcD^2 {\bar \cT} )$ which for flat space becomes \Gris\ (adapting to the supergravity
conventions of \Buch), with $G_a(z)| = {4\over 3} A_a(x)$,
\eqn\conR{
\pr_a \l R^a \r = {1\over 54\pi^2} (3c-5a) \, \ep^{abcd} F_{ab} F_{cd} \, , \qquad
F_{ab} = \pr_a A_b - \pr_b A_a \, .
}
Since \WT\ now reduces to $\l R^a \r = {\de W / \de A_a}$ the compatibility of
\anomR\ and \conR\ gives the first of eqs.\ACac.

\bigskip\bigskip
\vfill\eject
\appendix{A}{}

We here describe how the assumed transformation rule \tranT\ for the
supercurrent is realised for free fields. For the scalar case with the
supercurrent given by \Tphi\ and the elementary chiral transforming as
in \tranphi\ with $q=\bq=1$ we have
\eqnn\tranTphi
$$\eqalignno{
\de_{h,{\bar h}} \T_{\alpha\dal} = {}& -
\big ( \half (h^a + \bh^a ) \pr_a + \lambda^\alpha D_\alpha
+ {\tilde\bla}_\dal {\tilde\bD}{}^\dal + 3(\si_h^{\vphantom g} +
\bsi_\bh  ) \big )  \T_{\alpha\dal} 
+ \omega_{h}^{\vphantom g}{}_\alpha{}^\beta \T_{\beta\dal}
- \T_{\smash{\alpha\dbe}}\,\bom_{\bh}{}^{\! \dbe}{}_{\dal} \cr
&{} - X{}_\alpha{}^\beta \T_{\beta\dal} 
+  \T_{\smash{\alpha\dbe}} {\bar X} {}^{\dbe}{}_{\dal}
+ (h^a - {\bar h}^a ) {\ts{1\over 6}} D_\alpha \phi \olr {\pr}_{a}
\bD_\dal \bph \cr
&{} - D_\alpha ( h^a - {\bar h}^a )\, {\ts{1\over 3}} \pr_a \phi \bD_\dal \bph
+ \bD_\dal ( h^a - {\bar h}^a )\, {\ts{1\over 3}} D_\alpha \phi \pr_a \bph  \cr
&{} + i \pr_{\alpha\dal} (  h^a - {\bar h}^a )\, {\ts{1\over 3}}  \pr_a 
(\phi \bph ) 
- i[D_\alpha, \bD_\dal] ( \th^{\dbe\beta} - {\tilde{\brh}} {}^{\dbe\beta}) \, 
{\ts{1\over 12}} i \phi \olr {\pr}_{\smash{\beta\dbe}}\bph\cr
&{}+ \big ( i \pr_{\alpha\dal} \lambda^\beta + \de_\alpha{}^{\!\beta}
\bD_\dal \bsi_\bh \big ) \, {\ts{2\over 3}} D_\beta \phi \bph
+  \big ( i \pr_{\alpha\dal} \bla^\dbe - \de^\dbe{}_{\!\dal} D_\alpha
\si_h^{\vphantom g} \big ) \, {\ts{2\over 3}} \bph \bD_{\smash \dbe} \bph \cr
&{}+ i \pr_{\alpha \dal} (\si_h^{\vphantom g} - \bsi_\bh ) \, 
{\ts{4\over 3}} \phi \bph \, , & \tranTphi \cr}
$$
where the extra terms depend only on $h-{\bar h}$ as a consequence of
\eqn\hh{\eqalign{
X{}_\alpha{}^\beta = {}& D_\alpha \lambda^\beta + 
\omega_{h}^{\vphantom g}{}_\alpha{}^\beta - \de_\alpha {}^{\! \beta}
( 2 \bsi_\bh  - \si_h^{\vphantom g} )\cr
= {}& {\ts{1\over 4}} i
\bD_\dal D_\alpha ( \th^{\dal\beta} - {\tilde{\brh}}{}^{\dal\beta})
- \de_\alpha {}^{\! \beta} {\ts{1\over 12}} i
( D_\gamma \bD_{\smash{\dga}} + 2 \bD_{\smash{\dga}} D_\gamma )
( \th^{\dga\gamma} - {\tilde{\brh}}{}^{\dga\gamma}) \, , \cr
{\bar X} {}^{\dbe}{}_{\dal}  = {}& \bD_\dal \bla^\dbe +
\bom_{\bh}{}^{\! \dbe}{}_{\dal} + \de^\dbe{}_{\! \dal} 
( 2 \si_h^{\vphantom g} - \bsi_\bh ) \cr
= {}& {\ts{1\over 4}} i D_\alpha \bD_\dal 
( \th^{\dbe\alpha} - {\tilde{\brh}}{}^{\dbe\alpha})
- \de^\dbe{}_{\! \dal}  {\ts{1\over 12}} i 
( \bD_{\smash{\dga}} D_\gamma + 2 D_\gamma \bD_{\smash{\dga}} )
( \th^{\dga\gamma} - {\tilde{\brh}}{}^{\dga\gamma})  \, , \cr
\!\!\!\!
i \pr_{\alpha\dal} \lambda^\beta + \de_\alpha{}^{\!\beta} \bD_\dal \bsi_\bh 
= {}& {\ts{1\over 16}} i \ep_{\smash{\dal\dga}} \bD^2 D_\alpha
( \th^{\dga\beta} - {\tilde{\brh}}{}^{\dga\beta}) \cr
{}& - {\ts{1\over 48}} \de_\alpha{}^{\!\beta}
( 4 \bD_\dal \pr_{\smash{\gamma\dga}} + i\ep_{\smash{\dal\dga}} \bD^2 D_\gamma )
( \th^{\dga\gamma} - {\tilde{\brh}}{}^{\dga\gamma})  \, , \cr
\!\!\!\!  i \pr_{\alpha\dal} \bla^\dbe - \de^\dbe{}_{\!\dal} D_\alpha 
\si_h^{\vphantom g} = {}& - {\ts{1\over 16}}i \ep_{\alpha\gamma}
D^2 \bD_{\smash \dga} ( \th^{\dbe\gamma} - {\tilde{\brh}}{}^{\dbe\gamma}) \cr
{}& -{\ts{1\over 48}}\de^\dbe{}_{\! \dal} (4 D_\alpha \pr_{\smash{\gamma\dga}}
- i \ep_{\alpha\gamma} D^2 \bD_{\smash \dga} )
( \th^{\dga\gamma} - {\tilde{\brh}}{}^{\dga\gamma})  \, , \cr
\pr_{\alpha\dal} ( \si_h^{\vphantom g} - \bsi_\bh ) = {}&
{\ts{1\over 48}} \big ( \bD_\dal D_\alpha ( \bD_{\smash \dbe} D_\beta
+ 2 D_\beta  \bD_{\smash \dbe} ) + 3 D_\beta \bD_\dal \bD_{\smash \dbe} 
 D_\alpha \big ) ( \th^{\dbe\beta} - {\tilde{\brh}}{}^{\dbe\beta} ) \, .   
\cr}
}
For the case of free vector fields we may use \TV\ and \tranW\ to easily
obtain
\eqnn\tranTV
$$\eqalignno{
\de_{h,{\bar h}} \T_{\alpha\dal} = {}& -
\big ( \half (h^a + \bh^a ) \pr_a + \lambda^\alpha D_\alpha
+ {\tilde\bla}_\dal {\tilde\bD}{}^\dal + 3(\si_h^{\vphantom g} +
\bsi_\bh  ) \big )  \T_{\alpha\dal}
+ \omega_{h}^{\vphantom g}{}_\alpha{}^\beta \T_{\beta\dal}
- \T_{\smash{\alpha\dbe}}\,\bom_{\bh}{}^{\! \dbe}{}_{\dal} \cr
& {} - (h^a - {\bar h}^a ) W_\alpha \olr {\pr}_{a} \bW_\dal \cr
&{} + {\ts {1\over 4}}i \ep_{\alpha\beta}\bD^2 \big ( \bW_{\smash\dbe}
( \th{}^{\dbe\beta} - {\tilde \brh}{}^{\dbe\beta}) \big ) \bW_{\smash \dal}
+ {\ts {1\over 4}}i \ep_{{\smash{\dal\dbe}}}  W_\alpha D^2  \big ( (
\th{}^{\dbe\beta} - {\tilde \brh}{}^{\dbe\beta}) W_\beta \big ) \, .
& \tranTV \cr}
$$
The extra terms in \tranTphi\ and \tranTV\ may both be decomposed into
quasi-primary operators and their derivatives. The crucial difference
is that in the case of chiral superfields this includes the supercurrent
itself. In this case they therefore contribute to the Ward identity
for the supercurrent three point function, giving a non zero
$\chi_{\smash{\alpha\dal,\beta\dbe}}{}^{\dga\gamma}$ in \WardT\ whereas such
terms were absent in the vector case when the parameters $A,C$ satisfied \ACV. 
\listrefs
\bye